\documentclass[12pt,a4paper]{article}
\usepackage{ifthen} %
\newboolean{pdflatex}
\setboolean{pdflatex}{true} %

\newboolean{articletitles}
\setboolean{articletitles}{true} %

\newboolean{uprightparticles}
\setboolean{uprightparticles}{false} %

\newboolean{inbibliography}
\setboolean{inbibliography}{false} %

\def\paperauthors{LHCb collaboration} %
\def\paperasciititle{LHCb-PAPER-2020-030} %
\def\papertitle{Measurement of the CKM angle $\gamma$\\
\vspace{0.15\baselineskip}
and \Bs-\Bsb mixing frequency
\\ \vspace{0.15\baselineskip}
 with
$\Bs \to \Dsmp h^\pm \pipm \pimp$
decays
}

\def\paperkeywords{{High Energy Physics}, {LHCb}} %
\def\papercopyright{\the\year\ CERN for the benefit of the LHCb collaboration} %
\def\paperlicence{CC BY 4.0 licence}
\def\paperlicenceurl{https://creativecommons.org/licenses/by/4.0/}

\usepackage[top=1in, bottom=1.25in, left=1in, right=1in]{geometry}

\usepackage{microtype}
\usepackage{lineno}  %
\usepackage{xspace} %
\usepackage{caption} %

\usepackage{graphicx}  %
\usepackage{color}
\usepackage{colortbl}
\graphicspath{{./figs/}} %
\usepackage{amsmath} %
\usepackage{amssymb}
\usepackage{amsfonts}
\usepackage{upgreek} %

\newcommand*\patchAmsMathEnvironmentForLineno[1]{%
\expandafter\let\csname old#1\expandafter\endcsname\csname #1\endcsname
\expandafter\let\csname oldend#1\expandafter\endcsname\csname
end#1\endcsname
 \renewenvironment{#1}%
   {\linenomath\csname old#1\endcsname}%
   {\csname oldend#1\endcsname\endlinenomath}%
}
\newcommand*\patchBothAmsMathEnvironmentsForLineno[1]{%
  \patchAmsMathEnvironmentForLineno{#1}%
  \patchAmsMathEnvironmentForLineno{#1*}%
}
\AtBeginDocument{%
\patchBothAmsMathEnvironmentsForLineno{equation}%
\patchBothAmsMathEnvironmentsForLineno{align}%
\patchBothAmsMathEnvironmentsForLineno{flalign}%
\patchBothAmsMathEnvironmentsForLineno{alignat}%
\patchBothAmsMathEnvironmentsForLineno{gather}%
\patchBothAmsMathEnvironmentsForLineno{multline}%
\patchBothAmsMathEnvironmentsForLineno{eqnarray}%
}

\usepackage{hyperxmp}

\usepackage[pdftex,
            pdfauthor={\paperauthors},
            pdftitle={\paperasciititle},
            pdfkeywords={\paperkeywords},
            pdfcopyright={Copyright (C) \papercopyright},
            pdflicenseurl={\paperlicenceurl}]{hyperref}
\usepackage[colorinlistoftodos,textsize=scriptsize]{todonotes}

\usepackage[bottom,flushmargin,hang,multiple]{footmisc}

\usepackage[all]{hypcap} %

\usepackage{xspace} 
\usepackage{upgreek}

\def\lhcb   {\mbox{LHCb}\xspace}

\def\MagUp {\mbox{\em Mag\kern -0.05em Up}\xspace}

\ifthenelse{\boolean{uprightparticles}}%
{

 \def\Ppi         {\ensuremath{\uppi}\xspace}

 \def\Ppsi        {\ensuremath{\uppsi}\xspace}

 \def\PDelta      {\ensuremath{\Delta}\xspace}                 
 \def\PXi         {\ensuremath{\Xi}\xspace}                 
 \def\PLambda     {\ensuremath{\Lambda}\xspace}                 
 \def\PSigma      {\ensuremath{\Sigma}\xspace}                 
 \def\POmega      {\ensuremath{\Omega}\xspace}                 
 \def\PUpsilon    {\ensuremath{\Upsilon}\xspace}

 \def\PB      {\ensuremath{\mathrm{B}}\xspace}                 
                  
 \def\PD      {\ensuremath{\mathrm{D}}\xspace}

 \def\PJ      {\ensuremath{\mathrm{J}}\xspace}                 
 \def\PK      {\ensuremath{\mathrm{K}}\xspace}

 \def\Pb      {\ensuremath{\mathrm{b}}\xspace}                 
 \def\Pc      {\ensuremath{\mathrm{c}}\xspace}

 \def\Pi      {\ensuremath{\mathrm{i}}\xspace}

 \def\Pp      {\ensuremath{\mathrm{p}}\xspace}

 \def\Ps      {\ensuremath{\mathrm{s}}\xspace}                 
                  
 \def\Pu      {\ensuremath{\mathrm{u}}\xspace}

 \def\thebaroffset{0.0em}
}
{

 \def\Ppi         {\ensuremath{\pi}\xspace}

 \def\Ppsi        {\ensuremath{\psi}\xspace}                 
                  
 \mathchardef\PDelta="7101
 \mathchardef\PXi="7104
 \mathchardef\PLambda="7103
 \mathchardef\PSigma="7106
 \mathchardef\POmega="710A
 \mathchardef\PUpsilon="7107
                  
 \def\PB      {\ensuremath{B}\xspace}                 
                  
 \def\PD      {\ensuremath{D}\xspace}

 \def\PJ      {\ensuremath{J}\xspace}                 
 \def\PK      {\ensuremath{K}\xspace}

 \def\Pb      {\ensuremath{b}\xspace}                 
 \def\Pc      {\ensuremath{c}\xspace}

 \def\Pi      {\ensuremath{i}\xspace}

 \def\Pp      {\ensuremath{p}\xspace}

 \def\Ps      {\ensuremath{s}\xspace}                 
                  
 \def\Pu      {\ensuremath{u}\xspace}

 \def\thebaroffset{0.18em}
}
\newcommand{\offsetoverline}[2][\thebaroffset]{\kern #1\overline{\kern -#1 #2}}%

\makeatletter
\ifcase \@ptsize \relax%
  \newcommand{\miniscule}{\@setfontsize\miniscule{4}{5}}%
\or%
  \newcommand{\miniscule}{\@setfontsize\miniscule{5}{6}}%
\or%
  \newcommand{\miniscule}{\@setfontsize\miniscule{5}{6}}%
\fi
\makeatother

\DeclareRobustCommand{\optbar}[1]{\shortstack{{\miniscule (\rule[.5ex]{1.25em}{.18mm})}
  \\ [-.7ex] $#1$}}

\def\uquark    {{\ensuremath{\Pu}}\xspace}

\def\squark    {{\ensuremath{\Ps}}\xspace}

\def\cquark    {{\ensuremath{\Pc}}\xspace}

\def\bquark    {{\ensuremath{\Pb}}\xspace}

\def\pion   {{\ensuremath{\Ppi}}\xspace}
\def\piz    {{\ensuremath{\pion^0}}\xspace}
\def\pip    {{\ensuremath{\pion^+}}\xspace}
\def\pim    {{\ensuremath{\pion^-}}\xspace}
\def\pipm   {{\ensuremath{\pion^\pm}}\xspace}
\def\pimp   {{\ensuremath{\pion^\mp}}\xspace}

\def\kaon    {{\ensuremath{\PK}}\xspace}
\def\Kbar    {{\ensuremath{\offsetoverline{\PK}}}\xspace}
\def\Kb      {{\ensuremath{\Kbar}}\xspace}
\def\KorKbar {\kern \thebaroffset\optbar{\kern -\thebaroffset \PK}{}\xspace}

\def\Kp      {{\ensuremath{\kaon^+}}\xspace}
\def\Km      {{\ensuremath{\kaon^-}}\xspace}
\def\Kpm     {{\ensuremath{\kaon^\pm}}\xspace}
\def\Kmp     {{\ensuremath{\kaon^\mp}}\xspace}
\def\KS      {{\ensuremath{\kaon^0_{\mathrm{S}}}}\xspace}

\def\D       {{\ensuremath{\PD}}\xspace}

\def\DorDbar {\kern \thebaroffset\optbar{\kern -\thebaroffset \PD}\xspace}
\def\Dz      {{\ensuremath{\D^0}}\xspace}

\def\Dp      {{\ensuremath{\D^+}}\xspace}
\def\Dm      {{\ensuremath{\D^-}}\xspace}

\def\DpDm    {\ensuremath{\Dp {\kern -0.16em \Dm}}\xspace}

\def\Dsp     {{\ensuremath{\D^+_\squark}}\xspace}
\def\Dsm     {{\ensuremath{\D^-_\squark}}\xspace}
\def\Dspm    {{\ensuremath{\D^{\pm}_\squark}}\xspace}
\def\Dsmp    {{\ensuremath{\D^{\mp}_\squark}}\xspace}

\def\Dssm    {{\ensuremath{\D^{*-}_\squark}}\xspace}

\def\B       {{\ensuremath{\PB}}\xspace}
\def\Bbar    {{\ensuremath{\offsetoverline{\PB}}}\xspace}
\def\Bb      {{\ensuremath{\Bbar}}\xspace}
\def\BorBbar {\kern \thebaroffset\optbar{\kern -\thebaroffset \PB}\xspace}
\def\Bz      {{\ensuremath{\B^0}}\xspace}

\def\Bd      {{\ensuremath{\B^0}}\xspace}

\def\BdorBdbar {\kern \thebaroffset\optbar{\kern -\thebaroffset \Bd}\xspace}
\def\Bu      {{\ensuremath{\B^+}}\xspace}

\def\Bp      {{\ensuremath{\Bu}}\xspace}

\def\Bs      {{\ensuremath{\B^0_\squark}}\xspace}
\def\Bsb     {{\ensuremath{\Bbar{}^0_\squark}}\xspace}
\def\BsorBsbar {\kern \thebaroffset\optbar{\kern -\thebaroffset \Bs}\xspace}

\def\Bdsb    {{\ensuremath{\Bbar{}_{(\squark)}^0}}\xspace}

\def\jpsi     {{\ensuremath{{\PJ\mskip -3mu/\mskip -2mu\Ppsi}}}\xspace}
\def\psitwos  {{\ensuremath{\Ppsi{(2S)}}}\xspace}

\def\Y#1S{\ensuremath{\PUpsilon{(#1S)}}\xspace}

\def\proton      {{\ensuremath{\Pp}}\xspace}

\def\Lz          {{\ensuremath{\PLambda}}\xspace}

\def\LorLbar     {\kern \thebaroffset\optbar{\kern -\thebaroffset \PLambda}\xspace}

\def\Lb           {{\ensuremath{\Lz^0_\bquark}}\xspace}

\newcommand{\decay}[2]{\ensuremath{#1\!\to #2}\xspace} 

\def\to                 {\ensuremath{\rightarrow}\xspace}

\def\eps   {{\ensuremath{\varepsilon}}\xspace}

\def\CP                {{\ensuremath{C\!P}}\xspace}

\newcommand{\phis}{{\ensuremath{\phi_{\squark}}}\xspace}

\def\AT#1     {\ensuremath{A_{\mathrm{T}}^{#1}}\xspace}           %

\def\C#1      {\ensuremath{\mathcal{C}_{#1}}\xspace}                       %
\def\Cp#1     {\ensuremath{\mathcal{C}_{#1}^{'}}\xspace}                    %
\def\Ceff#1   {\ensuremath{\mathcal{C}_{#1}^{\mathrm{(eff)}}}\xspace}        %
\def\Cpeff#1  {\ensuremath{\mathcal{C}_{#1}^{'\mathrm{(eff)}}}\xspace}       %
\def\Ope#1    {\ensuremath{\mathcal{O}_{#1}}\xspace}                       %
\def\Opep#1   {\ensuremath{\mathcal{O}_{#1}^{'}}\xspace}                    %

\newcommand{\ket}[1]{\ensuremath{|#1\rangle}}              %
\newcommand{\braket}[2]{\ensuremath{\langle #1|#2\rangle}} %

\newcommand{\nospaceunit}[1]{\ensuremath{\text{#1}}}       
\newcommand{\aunit}[1]{\ensuremath{\text{\,#1}}}       
\newcommand{\tev}{\aunit{Te\kern -0.1em V}\xspace}
\newcommand{\gev}{\aunit{Ge\kern -0.1em V}\xspace}
\newcommand{\mev}{\aunit{Me\kern -0.1em V}\xspace}
\newcommand{\kev}{\aunit{ke\kern -0.1em V}\xspace}
\newcommand{\ev}{\aunit{e\kern -0.1em V}\xspace}
 
\newcommand{\mevc}{\ensuremath{\aunit{Me\kern -0.1em V\!/}c}\xspace}
\newcommand{\gevc}{\ensuremath{\aunit{Ge\kern -0.1em V\!/}c}\xspace}
\newcommand{\mevcc}{\ensuremath{\aunit{Me\kern -0.1em V\!/}c^2}\xspace}
\newcommand{\gevcc}{\ensuremath{\aunit{Ge\kern -0.1em V\!/}c^2}\xspace}

\def\mum  {\ensuremath{\,\upmu\nospaceunit{m}}\xspace}

\def\fb   {\ensuremath{\aunit{fb}}\xspace}
\def\invfb   {\ensuremath{\fb^{-1}}\xspace}

\def\ps   {\ensuremath{\aunit{ps}}\xspace}
\def\fs   {\aunit{fs}}

\def\invps{\ensuremath{\ps^{-1}}\xspace}

\newcommand{\stat}{\aunit{(stat)}\xspace}
\newcommand{\syst}{\aunit{(syst)}\xspace}

\def\gsim{{~\raise.15em\hbox{$>$}\kern-.85em
          \lower.35em\hbox{$\sim$}~}\xspace}
\def\lsim{{~\raise.15em\hbox{$<$}\kern-.85em
          \lower.35em\hbox{$\sim$}~}\xspace}

\def\sPlot{\mbox{\em sPlot}\xspace}

\def\pt         {\ensuremath{p_{\mathrm{T}}}\xspace}

\def\ptot       {\ensuremath{p}\xspace}

\def\degrees{\ensuremath{^{\circ}}\xspace}

\def\evtgen     {\mbox{\textsc{EvtGen}}\xspace}

\def\geant      {\mbox{\textsc{Geant4}}\xspace}

\def\photos     {\mbox{\textsc{Photos}}\xspace}

\def\pythia     {\mbox{\textsc{Pythia}}\xspace}

\def\tell1  {TELL1\xspace}
\def\ukl1   {UKL1\xspace}

\newcommand{\eg}{\mbox{\itshape e.g.}\xspace}
\newcommand{\ie}{\mbox{\itshape i.e.}\xspace}

\newcommand{\cf}{\mbox{\itshape cf.}\xspace}

\def\control {\decay{\Bs}{\Dsm\pip\pip\pim}}

\def\signal {\decay{\Bs}{\Dsmp\Kpm\pipm\pimp}}
\def\signalbar {\decay{\Bsb}{\Dspm\Kmp\pimp\pipm}}

\def\BsDsPi {\decay{\Bs}{\Dsm\pip}}

\def\BsDsK {\decay{\Bs}{\Dsmp\Kpm}}

\usepackage{cite} %
\usepackage{mciteplus}

\newcommand{\phs}{\ensuremath{\Phi_4}}  %
\newcommand{\dphs}{\ensuremath{\mathrm{d}\phs}}  

\newcommand{\phsPoint}{\ensuremath{\mathbf{x}}}

\newcommand{\phsPointCP}{\ensuremath{\overline{\mathbf{x}}}}

\usepackage{longtable} %
\usepackage{rotating} %
\usepackage{subcaption}
\usepackage{afterpage}
\usepackage{multirow}

\usepackage{tikz}
  \usetikzlibrary{shapes}
  \usetikzlibrary{patterns,snakes,circuits.ee.IEC}

\makeatletter
\g@addto@macro\bfseries{\boldmath}
\makeatother

\begin{document}

\renewcommand{\thefootnote}{\fnsymbol{footnote}}
\setcounter{footnote}{1}

\begin{titlepage}
\pagenumbering{roman}

\vspace*{-1.5cm}
\centerline{\large EUROPEAN ORGANIZATION FOR NUCLEAR RESEARCH (CERN)}
\vspace*{1.5cm}
\noindent
\begin{tabular*}{\linewidth}{lc@{\extracolsep{\fill}}r@{\extracolsep{0pt}}}
\ifthenelse{\boolean{pdflatex}}%
{\vspace*{-1.5cm}\mbox{\!\!\!\includegraphics[width=.14\textwidth]{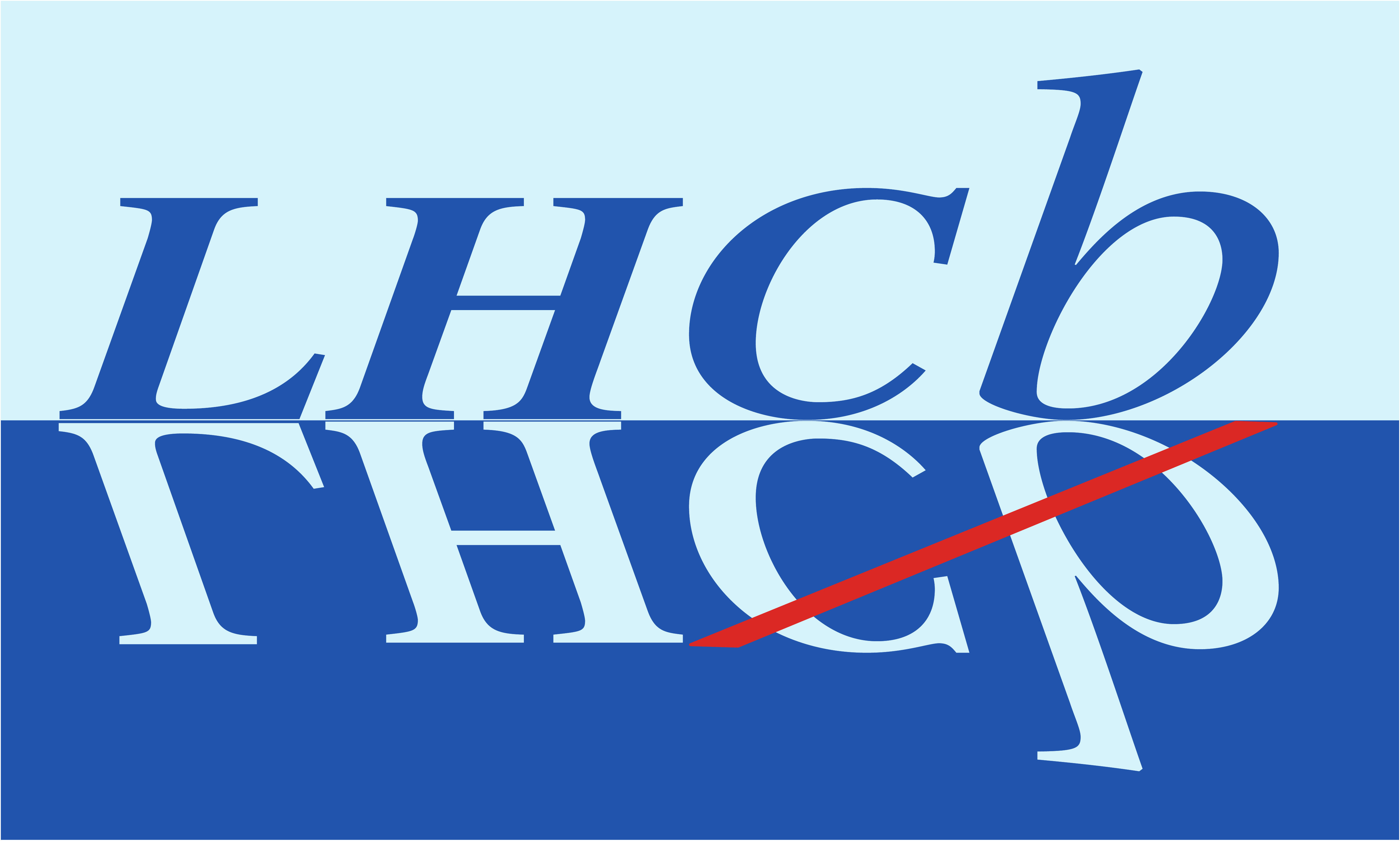}} & &}%
{\vspace*{-1.2cm}\mbox{\!\!\!\includegraphics[width=.12\textwidth]{LHCb-PAPER-2020-030-figures/lhcb-logo.eps}} & &}%
\\
 & & CERN-EP-2020-214 \\  %
 & & LHCb-PAPER-2020-030 \\  %
 & & April 12, 2021 \\ %
 & & \\
\end{tabular*}

\vspace*{1.5cm}

{\normalfont\bfseries\boldmath\huge
\begin{center}
  \papertitle
\end{center}
}

\begin{center}
\paperauthors\footnote{Authors are listed at the end of this paper.}
\end{center}

\vspace{\fill}

\begin{abstract}
  \noindent \small
The CKM angle $\gamma$ is measured for the first time
from mixing-induced \CP violation between \signal and \signalbar
decays reconstructed in proton-proton collision data corresponding to an integrated luminosity of
$9\invfb$ recorded with the LHCb detector.
A time-dependent amplitude analysis is performed to extract the
\CP-violating weak phase $\gamma-2\beta_s$ and, subsequently,
$\gamma$ by taking the $\Bs-\Bsb$ mixing phase $\beta_{s}$
as an external input.
The measurement yields $\gamma = (44 \pm 12)\degrees$ modulo $180\degrees$, where statistical and systematic uncertainties are combined.
An alternative model-independent measurement, integrating over the five-dimensional phase space of the decay, yields $\gamma = (44^{\,+\,20}_{\,-\,13})\degrees$ modulo $180\degrees$.
    Moreover, the $\Bs-\Bsb$ oscillation frequency is measured from the flavour-specific control channel \mbox{\control}
    to be
$\Delta m_s = (17.757 \pm 0.007 \stat \pm 0.008 \syst)\invps$,
consistent with and more precise than the current world-average value.

\normalsize
\end{abstract}

\vspace*{2.0cm}

\begin{center}
  Published in JHEP 03 (2021) 137
\end{center}

\vspace{\fill}

{\footnotesize
\centerline{\copyright~\papercopyright. \href{\paperlicenceurl}{\paperlicence}.}}
\vspace*{2mm}

\end{titlepage}

\newpage
\setcounter{page}{2}
\mbox{~}

\cleardoublepage

\renewcommand{\thefootnote}{\arabic{footnote}}
\setcounter{footnote}{0}

\pagestyle{plain} %
\setcounter{page}{1}
\pagenumbering{arabic}

%
%

%
%
%
%
%
%
%

%
\section{Introduction}
\label{sec:Introduction}

Within the Standard Model of particle physics, the charge-parity (\CP) symmetry between quarks and antiquarks is broken by a single complex
phase in the Cabibbo-Kobayashi-Maskawa (CKM) quark-mixing matrix~\cite{Cabibbo:1963yz,*Kobayashi:1973fv}.
The unitarity of this matrix leads to the condition $V_{ud}^{\phantom{*}}V^{*}_{ub} + V_{cd}^{\phantom{*}}V^{*}_{cb} + V_{td}^{\phantom{*}}V^{*}_{tb} = 0$, where $V_{ij}$ are the complex elements of the CKM matrix.
This equation can be visualised as a triangle in the complex plane
with angles $\alpha$, $\beta$ and~$\gamma$.
It is a key  test of the Standard Model to verify this unitarity condition by over-constraining the CKM matrix with independent measurements
sensitive to various distinct combinations of matrix elements.
In particular, measurements of
$\gamma \equiv \text{arg}[-(V_{ud}^{\phantom{*}}V_{ub}^{*})/(V_{cd}^{\phantom{*}}V_{cb}^{*})]$
in tree-level decays provide
an important benchmark of the Standard Model,
to be compared with loop-level measurements of $\gamma$ and other
CKM parameters.
The world-average value, $\gamma = \left(72.1^{\,+\,4.1}_{\,-\,4.5} \right)^\circ$~\cite{PDG20,HFLAV18}, is dominated by a combination of \lhcb measurements obtained from analyses of beauty meson decays to open-charm final states~\cite{Kenzie:2018oob}.
Improved direct measurements are needed to set
a conclusive comparison to the indirect predictions from global CKM fits,
$\gamma = \left(65.7^{\,+\,0.9}_{\,-\,2.7}\right)^\circ$~\cite{CKMfitter2015}
or
$\gamma = \left(65.8 \pm 2.2 \right)^\circ$~\cite{UTfit-UT}.

\begin{figure}[ht]
\centering
\resizebox{\linewidth}{!}{
\begin{tikzpicture}[scale=1]
\draw [->,thick] (0,0) node[left]{$s$} -- (3,0) node[right]{};
\draw [thick] (0,0) node[left]{$s$} -- (4,0) node[right]{$s$};
\draw [<-,thick] (0.5,1) node[left]{} -- (1,1) node[right]{};
\draw [<-,thick] (3,1) node[left]{} -- (4,1) node[right]{};
\draw [thick] (0,1) node[left]{$\bar{b}$} -- (4,1) node[right]{$\bar{c}$};
\draw [snake=snake] (1,1) -- (2.5,2.5);
\node at (1.5,2) {$W^+$};
\draw [<-,thick] (3.25,2.75) -- (4,3) node[right]{};
\draw [thick] (2.5,2.5) -- (4,3) node[right]{$\bar s$};
\draw [->,thick](2.5,2.5) -- (3.25,2.25) node[right]{};
\draw [thick](2.5,2.5) -- (4,2) node[right]{$u$};

\draw[decoration={brace},decorate,thick](4.5,3) -- node[right=6pt] {$K^+\pi^+\pi^-$} (4.5,2);
\draw[decoration={brace},decorate,thick] (4.5,1) -- node[right=6pt] {$D_s^-$} (4.5,0);
\draw[decoration={brace},decorate,thick] (-0.5,0) -- node[left=0.2] {\Bs} (-0.5,1);
\end{tikzpicture}
\begin{tikzpicture}[scale=1]
\draw [<-,thick] (3,0) node[left]{} -- (4,0) node[right]{};
\draw [thick] (0,0) node[left]{$\bar s$} -- (4,0) node[right]{$\bar s$};
\draw [->,thick] (0,1) node[left]{} -- (0.5,1) node[right]{};
\draw [->,thick] (0,1) node[left]{} -- (3,1) node[right]{};
\draw [thick] (0,1) node[left]{$b$} -- (4,1) node[right]{$u$};
\draw [snake=snake] (1,1) -- (2.5,2.5);
\node at (1.5,2) {$W^-$};
\draw [<-,thick] (3.25,2.75) -- (4,3) node[right]{};
\draw [thick] (2.5,2.5) -- (4,3) node[right]{$\bar c$};
\draw [->,thick](2.5,2.5) -- (3.25,2.25) node[right]{};
\draw [thick](2.5,2.5) -- (4,2) node[right]{$s$};

\draw[decoration={brace},decorate,thick](4.5,3) -- node[right=6pt] {$D_s^-$} (4.5,2);
\draw[decoration={brace},decorate,thick] (4.5,1) -- node[right=6pt] {$K^+\pi^+\pi^-$} (4.5,0);
\draw[decoration={brace},decorate,thick] (-0.5,0) -- node[left=0.2] {\Bsb} (-0.5,1);
\end{tikzpicture}
}
\caption{Leading-order Feynman diagrams for (left) \Bs and (right) \Bsb decays to the $D_s^- K^+ \pip \pim$ final state,
where the $\pip\pim$ subsystem exemplarily hadronises in conjunction with the kaon.
}
\label{fig:decay_feynman}
\end{figure}
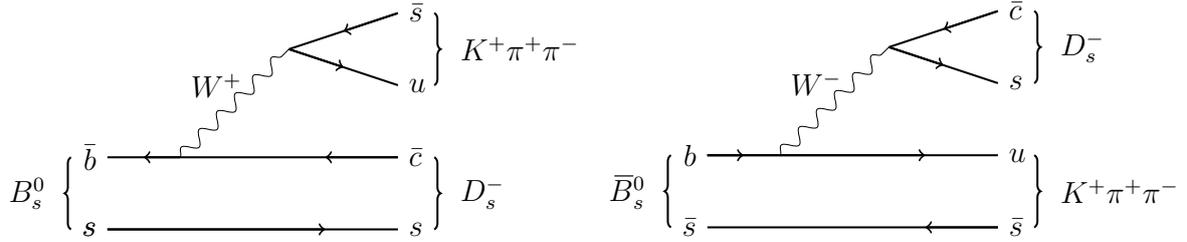

This paper presents the first measurement of the CKM angle $\gamma$ with \signal
decays.\footnote{Inclusion of charge-conjugate modes is implied throughout except where explicitly stated.}
The data set is collected with the LHCb experiment in proton-proton ($pp$) collisions at
centre-of-mass energies\footnote{Natural units with $\hbar = c = 1$ are used throughout the paper.}
of
$7, 8$ and $13\tev$, corresponding to an integrated luminosity
of $9\invfb$.
In these decays,
the interference between $\bquark\to\cquark$ and $\bquark\to\uquark$  quark-level transitions achieved through $\Bs-\Bsb$ mixing
provides sensitivity to the \CP-violating weak phase $\gamma - 2\beta_s$~\cite{Fleischer:2003yb,DeBruyn:2012jp}.
The mixing phase, $\beta_s$,
is well constrained from
$\Bs \to \jpsi K^+ K^-$~\cite{LHCb-PAPER-2014-059,LHCb-PAPER-2019-013}
and related decays~\cite{LHCb-PAPER-2014-019,LHCb-PAPER-2016-027,LHCb-PAPER-2014-051,LHCb-PAPER-2019-003} and is taken as an external input.
The leading-order Feynman diagrams
for $\Bs \to D_s^- K^+ \pip \pim$ and $\Bsb \to D_s^- K^+ \pip \pim$
decays are shown in Fig.~\ref{fig:decay_feynman}.
The amplitudes for both processes are of the same order in the Wolfenstein parameter $\lambda$\cite{PhysRevLett.51.1945}, $\mathcal O(\lambda^3)$, so that interference effects are expected to be large.
To account for the strong-phase variation across the
phase space of the decay,  a time-dependent amplitude analysis is performed.
An alternative, model-independent approach analyses the
phase-space integrated decay-time spectrum and is pursued as well;
this method is conceptually similar to the analysis of
\BsDsK  decays~\cite{LHCB-PAPER-2014-038,LHCB-PAPER-2017-047}.
However, a coherence factor needs to be introduced as an additional hadronic parameter, which dilutes the observable \CP asymmetry
since constructive and destructive interference effects cancel when integrated over the entire phase space.
The topologically similar and flavour-specific decay \control is  used to calibrate detector-induced effects. 
This mode is also employed to make a precise measurement of 
the $\Bs-\Bsb$ mixing frequency,
which can be related to one side of the unitarity triangle.
The relative branching fraction of these decay modes was measured by LHCb to be
\mbox{$\mathcal B(\signal)/\mathcal B(\control) = \left(5.2 \pm 0.5 (\text{stat}) \pm 0.3 (\text{syst})\right)\%$~\cite{LHCb-PAPER-2012-033}.}

The paper is structured as follows.
After introducing the amplitude analysis formalism and the differential decay rates in Sec.~2,
the LHCb detector, the event reconstruction and candidate selection are
described in Sec.~3.
Section 4 presents the measurement of the $\Bs$ mixing frequency, followed by the analysis of the \signal signal channel
in Sec.~5.
Experimental and model-dependent systematic uncertainties are evaluated in Sec.~6, the results are discussed in Sec.~7, and our conclusions are given in Sec.~8.
\clearpage

\section{Phenomenology of the decay}
\label{subsec:DecRates}

Assuming tree-level processes are dominant, transitions from \Bs and \Bsb flavour eigenstates to the final state $f=D_s^- K^+ \pip \pim$ are described by decay amplitudes
\begin{align}
	\braket{f}{\Bs}  &\equiv A^c(\phsPoint), & \braket{f}{\Bsb}  &\equiv r \, e^{i (\delta-\gamma)} \,A^u(\phsPoint) ,
\end{align}
with a relative magnitude $r$ and
(constant) strong- and weak-phase differences $\delta$ and $\gamma$, respectively.
A set of five independent kinematic observables (\eg invariant-mass combinations of the final state particles or helicity angles)
fully describes the phase space $\phsPoint$ of the decay.
The hadronic amplitudes $A^c(\phsPoint)$ and $A^u(\phsPoint)$,
where the superscript $c$ ($u$) refers to a $b \to c$ ($b \to u$) quark-level transition,
contain the strong-interaction dynamics and are given by
a coherent sum over intermediate-state amplitudes, $A_i(\phsPoint)$:
\begin{align}
	A^c(\phsPoint)= & \sum_i a_i^c \, A_i(\phsPoint), 	&	A^u(\phsPoint) &= \sum_i a_i^u \, A_i(\phsPoint).
	\label{eq:b2cuAmps}
\end{align}
The complex amplitude coefficients $a_i^c$ and $a_i^u$ need to be determined from data.
Since the hadronisation process is different for $\Bs \to f$ and $\Bsb \to f$ decays,
their respective amplitude coefficients are distinct ($a_i^c \ne a_i^u$).
To ensure that the parameters $r$ and $\delta$ do not depend on the convention employed for the amplitude coefficients,
the magnitude squared of the hadronic amplitudes is normalised to unity when integrated over the phase space
(with four-body phase-space element $\text d \Phi_4$)
and the overall strong-phase difference between $A^c(\phsPoint)$ and $A^u(\phsPoint)$ is set to zero, \ie
$\int \vert A^c(\phsPoint) \vert^2 \, \text d \Phi_4(\phsPoint) = \int \vert A^u(\phsPoint) \vert^2 \,\text d \Phi_4(\phsPoint) = 1$
and
$\arg \left( \int A^{c}(\phsPoint)^* \, A^u(\phsPoint) \,\text d \Phi_4(\phsPoint) \right) = 0$.
Within this convention, the decay fractions and interference fractions
for $b \to c$ ($b \to u$) transitions are defined as
\begin{align}
\label{eq:DefineFitFractions2}
	F^{c(u)}_{i} &\equiv \int \left\vert   a_{i}^{c(u)} \, A_{i}(\phsPoint) \right\vert^{2} \, \text{d}\Phi_{4},   &
    I_{ij}^{c(u)}  \equiv \int  2\,\text{Re}[a_{i}^{c(u)} \, a^{c(u)*}_{j} \, A_{i}(\phsPoint) A_{j}^{*}(\phsPoint) ] \, \text{d}\Phi_{4}.
\end{align}

\subsection{Amplitude formalism}

The isobar model is used to construct the intermediate-state amplitudes $A_{i}(\phsPoint )$\cite{isobar1,isobar,isobar2}.
Within this model, the four-body decay $\Bs \to  h_{1}\,h_{2} \, h_{3}\,h_{4}$
proceeds via two isobar states $R_{1}$ and $R_{2}$ (typically associated to intermediate resonances), which gives rise to two
distinct decay topologies;
quasi-two-body decays
$\Bs \to (R_{1} \to h_{1}\,h_{2}) \, (R_{2} \to h_{3}\,h_{4})$
or cascade decays
$\Bs \to h_{1} \, \left[R_{1} \to h_{2} \,  (R_{2} \to h_{3} \, h_{4}) \right]$.
The intermediate-state amplitude for 
a given decay channel $i$
can be parameterised as
\begin{equation}
	A_{i}(\phsPoint ) =  B_{L_{\Bs}}(\mathbf{x} ) \, [B_{L_{R_{1}}}(\phsPoint )  \, T_{R_{1}}(\phsPoint )] \, [B_{L_{R_{2}}}(\phsPoint ) \, T_{R_{2}}(\phsPoint )]  \,  S_{i}(\phsPoint )  \, ,
	\label{eq:amp4}
\end{equation}
where the form factors $B_{L}$ account for deviations from point-like interactions and the propagator $T_{R}$ describes the lineshape of resonance $R$.
The angular correlation of the final state particles subject to
total angular momentum conservation is encoded in the spin factor $S$.
The Blatt--Weisskopf penetration factors~\cite{Bl2,Blatt:1952ije} are used as form factors.
They depend on the effective interaction radius
$r_{\rm BW}$, the break-up momentum $b$
and the orbital angular momentum $L$ between the final-state particles.
The explicit expressions for $L=\{0,1,2\}$ are
\begin{align}
	B_{0}(b)  &= 1,  &
	B_{1}(b)  &= 1 / \sqrt{{1+ (b \, r_{\rm BW})^{2}}} \, \,\, \text{and}  &
	B_{2}(b)  &= 1 / \sqrt{9+3\,(b \, r_{\rm BW})^{2}+(b \, r_{\rm BW})^{4}}.
\end{align}
Resonance lineshapes are described by Breit--Wigner propagators,
\begin{equation}
	T_R(s) = (m_0^{2} - s - i\,\sqrt{s} \,\Gamma(s))^{-1},
\end{equation}
where $s$ is the square of the centre-of-mass energy.
The energy-dependent decay width, $\Gamma(s)$, is normalised to give the nominal width, $\Gamma_{0}$, when evaluated at the nominal mass, $m_{0}$.
For a decay into two stable particles $R \to AB$, the energy dependence of the decay width can be described by
\begin{equation}
	\sqrt{s} \, \Gamma_{R \to AB}(s) = m_0 \, \Gamma_{0} \, \frac{m_{0}}{\sqrt s} \, \left(\frac{b}{b_{0}}\right)^{2L+1} \, \frac{B_{L}(b)^{2}}{B_{L}(b_{0})^{2}}  \, ,
	\label{eq:gamma2}
\end{equation}
where $b_{0}$ is the value of the break-up momentum at the resonance pole \cite{BW}.
Specialised lineshape parameterisations are used for the
$f_0(500)^{0}$ (Bugg~\cite{BuggSigma}),
$K_0^*(1430)^{0}$ (LASS~\cite{Aston:1987ir,Aubert:2005ce})
and $\rho(770)^{0}$ (Gounaris--Sakurai~\cite{GS})
resonances.
The lineshapes for non-resonant states are set to a constant.

The energy-dependent width for a three-body decay $R \to ABC$
is computed by integrating the squared transition amplitude over the phase space,
\begin{equation}
	\Gamma_{R \to ABC}(s) =  \frac{1}{2 \, \sqrt s} \, \int \vert A_{R \to ABC} \vert^{2} \, \text{d}\Phi_{3}   ,
	\label{eq:gamma3}
\end{equation}
as described in Ref.~\cite{dArgent:2017gzv}.
For the $K_1(1270)^{+} \to \rho(770)^0 K^+$ cascade decay chain,  mixing
between the $\rho(770)^0$ and $\omega(782)^0$ states is included~\cite{Akhmetshin:2001ig},
with relative magnitude and phase fixed
to the values determined in Ref.~\cite{LHCb-PAPER-2018-041}.
More details are given in Appendix~\ref{a:lineshapes}.
The spin factors are constructed in the covariant Zemach (Rarita--Schwinger) tensor formalism
\cite{Zemach,Rarita,helicity3,Zou, Filippini}.
The explicit expressions for the decay topologies relevant for this analysis are taken from Refs.~\cite{dArgent:2017gzv,jdalseno_2019_2585535}.

\subsection{Decay rates}

Since \Bs mesons can convert into \Bsb and vice versa, the
flavour eigenstates are an admixture of the physical mass eigenstates $B_L$ and $B_H$,
\begin{align}
	\ket{B_L} &= p \ket{\Bs} + q \ket{\Bsb},	 &
       \ket{B_H}  &= p \ket{\Bs} - q \ket{\Bsb}   ,
       \label{eq:massStates}
\end{align}
where the complex coefficients are normalised such that $\vert p \vert^2 + \vert q \vert^2  = 1$.
The light, $\ket{B_L}$,  and heavy, $\ket{B_H}$, mass eigenstates have distinct masses, $m_L$ and $m_H$,
and decay widths, $\Gamma_L$ and $\Gamma_H$.
Their arithmetic means (differences) are denoted as
$m_s$ and $\Gamma_s$ ($\Delta m_s = m_H - m_L$ and $\Delta\Gamma_s = \Gamma_L - \Gamma_H$).
The time evolution of the flavour and mass eigenstates can be described
by an effective Schr\"odinger equation~\cite{PhysRev.106.340,Dunietz:2000cr}
resulting in the following
differential decay rates for initially produced $\Bs$ or $\Bsb$ mesons \cite{Artuso:2015swg,Dunietz:2000cr,PhysRevD.37.3186}:
\begin{equation}
\footnotesize
\label{eq:decayRateBs}
\begin{split}
	\frac{\text d \Gamma(\Bs \to f)}{\text dt \, \text d \Phi_4(\phsPoint) }
	= \vert \braket{f}{\Bs(t)} \vert^2 \propto e^{-\Gamma_s \, t}
	&\Biggl[	 \left( \vert A^c(\phsPoint) \vert^2 + r^2 \, \vert A^u(\phsPoint) \vert^2 \right) \, \text{cosh} \left( \frac{\Delta \Gamma_s \, t}{2}\right)   \\
	 &+  \left( \vert A^c(\phsPoint) \vert^2 - r^2 \, \vert A^u(\phsPoint)  \vert^2 \right) \, \text{cos} \left(\Delta m_s \, t \right)  \\
	 & -2 \, r  \, \text{Re}\left( A^c(\phsPoint)^{*}  \, A^u(\phsPoint)  \, e^{i \left(\delta - (\gamma - 2\beta_s) \right)}  \right) \, \text{sinh} \left( \frac{\Delta \Gamma_s \, t}{2}\right)  \\
	 & -2 \, r \,  \text{Im}\left( A^c(\phsPoint)^{*} \, A^u(\phsPoint)  \, e^{i \left(\delta - (\gamma - 2\beta_s) \right)}   \right)\, \text{sin} \left(\Delta m_s \, t \right)
	 \Biggr],    \\
	\frac{\text d \Gamma(\Bsb \to f)}{\text dt \, \text d \Phi_4(\phsPoint) }
	= \vert \braket{f}{\Bsb(t)} \vert^2 \propto e^{-\Gamma_s \, t}
	&\Biggl[	 \left( \vert A^c(\phsPoint) \vert^2 + r^2 \, \vert A^u(\phsPoint) \vert^2 \right) \, \text{cosh} \left( \frac{\Delta \Gamma_s \, t}{2}\right)   \\
	 &-  \left( \vert A^c(\phsPoint) \vert^2 - r^2 \, \vert A^u(\phsPoint)  \vert^2 \right) \, \text{cos} \left(\Delta m_s \, t \right)  \\
	 & -2 \, r  \, \text{Re}\left( A^c(\phsPoint)^{*}  \, A^u(\phsPoint)  \, e^{i \left(\delta - (\gamma - 2\beta_s) \right)}  \right) \, \text{sinh} \left( \frac{\Delta \Gamma_s \, t}{2}\right)  \\
	 & +2 \, r \,  \text{Im}\left( A^c(\phsPoint)^{*} \, A^u(\phsPoint)  \, e^{i \left(\delta - (\gamma - 2\beta_s) \right)}   \right)\, \text{sin} \left(\Delta m_s \, t \right)
	 \Biggr].    \\	
\end{split}
\end{equation}
\noindent Here,
the magnitude of $q/p$ is set to unity (\ie no \CP violation in mixing is assumed~\cite{HFLAV18,LHCb-PAPER-2013-033,LHCb-PAPER-2016-013})
and the phase between $q$ and $p$ can be related to
the mixing phase $\beta_s$, \mbox{$\arg(q/p) \approx -2\beta_s$~\cite{Artuso:2015swg,PhysRevD.37.3186}.}
The decay rates to the \CP-conjugate final state $\bar f=D_s^+ K^- \pim \pip$ (with phase-space point $\phsPointCP \equiv CP \, \phsPoint$),
$\text d \Gamma(\Bsb \to \bar f)$  and $\text d \Gamma(\Bs \to \bar f)$,
follow from  the expressions for
$\text d \Gamma(\Bs \to f)$  and $ \text d \Gamma(\Bsb \to f)$
in Eq.~(\ref{eq:decayRateBs})
by replacing $A^c(\phsPoint) \to A^c(\phsPointCP)$, $A^u(\phsPoint) \to A^u(\phsPointCP)$ and $-(\gamma - 2\beta_s) \to + (\gamma - 2\beta_s)$.
This assumes no \CP violation in the hadronic decay amplitudes (\ie, $a_i^c = \bar a_i^c$ and $a_i^u = \bar a_i^u$), as expected for tree-level-dominated decays.

It is also instructive to examine the decay rates as functions of the decay time only, by marginalising the phase space
\begin{equation}
\label{eq:timePdf}
\small
\begin{split}
    \frac{\text d \Gamma(\Bs \to f)}{\text dt}
	 &\propto
	\Biggl[ \text{cosh} \left( \frac{\Delta \Gamma_s \, t}{2}\right)
	 +  \, C_f \, \text{cos} \left(\Delta m_s \, t \right)  \\
	 & \phantom{\propto [} +  A^{\Delta\Gamma}_f \, \text{sinh} \left( \frac{\Delta \Gamma_s \, t}{2}\right)
	  -  \, \, S_f \, \text{sin} \left(\Delta m_s \, t \right)
	 \Biggr] e^{-\Gamma_s \, t},    \\
	 \frac{\text d \Gamma(\Bsb \to f)}{\text dt}
	 &\propto
	\Biggl[ \text{cosh} \left( \frac{\Delta \Gamma_s \, t}{2}\right)
	 -  \, C_f \, \text{cos} \left(\Delta m_s \, t \right)  \\
	 & \phantom{\propto [} +  A^{\Delta\Gamma}_f \, \text{sinh} \left( \frac{\Delta \Gamma_s \, t}{2}\right)
	 + \, \, S_f \, \text{sin} \left(\Delta m_s \, t \right)
	 \Biggr] e^{-\Gamma_s \, t}.    \\
	 \end{split}
\end{equation}
Analogous expressions
for the \CP-conjugate processes can be written by replacing $A^{\Delta\Gamma}_f$ with $A^{\Delta\Gamma}_{\bar f}$
and $S_f$ with $-S_{\bar f}$,
where the \CP coefficients
are defined as

\small
\begin{flalign}
\label{eq:CPcoeff}
	C_f =  \frac{1-r^2}{1+r^2},  & \phantom{\,\,\,}  \nonumber \\
	A^{\Delta\Gamma}_f =   -\frac{2 \, r \, \kappa \, \text{cos}\left(\delta- \, (\gamma-2\beta_s)\right)}{1+r^2},  & \phantom{\quad}
		A^{\Delta\Gamma}_{\bar f} =   -\frac{2 \, r \, \kappa \, \text{cos}\left(\delta+ \, (\gamma-2\beta_s)\right)}{1+r^2},   \nonumber \\
	S_f^{\phantom{\Delta\Gamma}} =  \,  + \, \frac{2 \, r \, \kappa \, \text{sin}\left(\delta- \, (\gamma-2\beta_s)\right)}{1+r^2},  & \phantom{\quad}
  S_{\bar f}^{\phantom{\Delta\Gamma}} =  \, - \, \frac{2 \, r \, \kappa \, \text{sin}\left(\delta+ \, (\gamma-2\beta_s)\right)}{1+r^2}.
\end{flalign}
\normalsize
The coherence factor, $\kappa$, results from the integration over the interfering amplitudes across the phase space,
$	\kappa \equiv  \int  A^c(\phsPoint)^{*} \,  A^u(\phsPoint)  \, \dphs$.
It is bounded between zero and unity, and dilutes the sensitivity to the weak phase.
For the two-body decay $\BsDsK$ the coherence factor is $\kappa =1$
\cite{LHCB-PAPER-2017-047}.
Measured values of $A^{\Delta\Gamma}_f \ne A^{\Delta\Gamma}_{\bar f}$ or $S_f \ne - S_{\bar f}$
signify time-dependent \CP violation and lead to
different mixing asymmetries
for decays into the $f$ or $\bar f$ final states.
These mixing asymmetries are defined as~\cite{Fleischer:2003yb,DeBruyn:2012jp}
\begin{align}
	A_{\text{mix}}^f(t) &= \frac{N_f(t) - \bar N_f(t)}{N_f(t) + \bar N_f(t)} = \frac{C_f \cos\left(\Delta m_s \, t \right) - S_f \sin\left(\Delta m_s \, t \right) }{\cosh \left( \frac{\Delta \Gamma_s \, t}{2}\right)  + A^{\Delta\Gamma}_f \sinh\left( \frac{\Delta \Gamma_s \, t}{2}\right) },  \nonumber \\
		A_{\text{mix}}^{\bar f}(t) &= \frac{\bar N_{\bar f}(t) - N_{\bar f}(t)}{\bar N_{\bar f}(t) +  N_{\bar f}(t)} = \frac{C_f \cos\left(\Delta m_s \, t \right) +  S_{\bar f} \sin\left(\Delta m_s \, t \right) }{\cosh \left( \frac{\Delta \Gamma_s \, t}{2}\right)  + A^{\Delta\Gamma}_{\bar f} \sinh\left( \frac{\Delta \Gamma_s \, t}{2}\right) },
		\label{eq:mixAsym}
\end{align}
where $N_f(t)$ ($\bar N_f(t)$) and $N_{\bar f}(t)$ ($\bar N_{\bar f}(t)$) denote the number of initially produced $\Bs$ ($\Bsb$) mesons
decaying at  proper time $t$ to the final states $f$ and $\bar f$, respectively.
Flavour-specific decay modes such as \control have $r=0$
and consequently $C_f=1$ as well as $A^{\Delta\Gamma}_f=A^{\Delta\Gamma}_{\bar f}=S_f=S_{\bar f} = 0$.

\section{Event reconstruction}
\label{sec:Detector}

The \lhcb detector~\cite{Alves:2008zz,LHCb-DP-2014-002} is a single-arm forward
spectrometer covering the \mbox{pseudorapidity} range $2<\eta <5$,
designed for the study of particles containing \bquark or \cquark
quarks. The detector includes a high-precision tracking system
consisting of a silicon-strip vertex detector (VELO) surrounding the $pp$
interaction region~\cite{LHCb-DP-2014-001}, a large-area silicon-strip detector located
upstream of a dipole magnet with a bending power of about
$4{\mathrm{\,Tm}}$, and three stations of silicon-strip detectors and straw
drift tubes~\cite{LHCb-DP-2013-003,LHCb-DP-2017-001} placed downstream of the magnet.
The polarity of the dipole magnet is reversed periodically throughout the data-taking process to control systematic asymmetries.
The tracking system provides a measurement of the momentum, \ptot, of charged particles with
a relative uncertainty that varies from 0.5\% at low momentum to 1.0\% at 200\gev.
The minimum distance of a track to a primary vertex (PV), the impact parameter (IP),
is measured with a resolution of $(15+29/\pt)\mum$,
where \pt is the component of the momentum transverse to the beam, in\,\gev.
Different types of charged hadrons are distinguished using information
from two ring-imaging Cherenkov detectors~\cite{LHCb-DP-2012-003}.
The online event selection is performed by a trigger~\cite{LHCb-DP-2012-004},
which consists of a hardware stage, based on information from the calorimeter and muon
systems, followed by a software stage, which applies a full event
reconstruction.
At the hardware trigger stage, events are required to have a muon with high \pt or a
hadron, photon or electron with high transverse energy in the calorimeters. For hadrons,
the transverse energy threshold is 3.5\gev.
The software trigger requires a two-, three- or four-track
secondary vertex with a significant displacement from any primary
$pp$ interaction vertex. At least one charged particle
must have a transverse momentum $\pt > 1.6\gev$ and be
inconsistent with originating from a PV.
A multivariate algorithm~\cite{BBDT} is used for
the identification of secondary vertices consistent with the decay
of a \bquark hadron.
The data-taking period from 2011 to 2012 with centre-of-mass energies of 7 and 8 \tev (2015 to 2018 with centre-of-mass energy of 13 \tev) is referred to as Run 1 (Run 2) throughout the paper.

Simulated events are used to study the detector acceptance and
specific background contributions.
In the simulation, $pp$ collisions are generated using
\pythia~\cite{Sjostrand:2006za,*Sjostrand:2007gs} with a specific \lhcb
configuration~\cite{LHCb-PROC-2010-056}.  Decays of hadrons
are described by \evtgen~\cite{Lange:2001uf}, in which final-state
radiation is generated using \photos~\cite{Golonka:2005pn}.
The simulated signal decays are generated according to a simplified
amplitude model
with an additional pure phase-space component.
The interaction of the generated particles with the detector, and its response,
are implemented using the \geant
toolkit~\cite{Allison:2006ve, *Agostinelli:2002hh} as described in
Ref.~\cite{LHCb-PROC-2011-006}.

\subsection{Candidate selection}
\label{sec:Selection}

The selection of \signal and \control candidates is performed by first reconstructing $\Dsm\to\Km\Kp\pim$, $\Dsm\to\Km\pim\pip$ and $\Dsm\to\pim\pip\pim$ candidates from charged particle tracks with high momentum and transverse momentum originating from a common displaced vertex.
Particle identification (PID) information is used to assign a kaon or pion hypothesis to the tracks.
Candidate \Dsm mesons with a reconstructed invariant mass within $25\mev$ of the known \Dsm mass~\cite{PDG20} are
combined with three additional charged tracks to form a \Bs vertex, which must be displaced from any PV.
The PV with respect to which the \Bs candidate has the smallest
impact parameter significance is considered as the production vertex.
The reconstructed invariant mass of the \Bs candidate is required to be between $5200\mev$ and $5700\mev$.
The mass resolution is improved by performing a kinematic fit~\cite{Hulsbergen:2005pu}
where the \Bs candidate is constrained to
originate from the PV
and the reconstructed $\Dsm$ mass is constrained to the world-average $\Dsm$ mass~\cite{PDG20}.
When deriving the decay time, $t$, of the \Bs candidate and the
phase-space observables, $\phsPoint$,
the reconstructed \Bs mass is constrained to its known value~\cite{PDG20}.
The \Bs proper time is required to be larger than $0.4\ps$ to suppress most of the prompt combinatorial background.
The considered phase-space region is limited to $m(K^+\pi^+\pi^-) < 1950 \mev$,
$m(K^+\pi^-) < 1200 \mev$ and $m(\pi^+\pi^-) < 1200 \mev$
since the decay proceeds predominantly through the low-mass axial-vector states $K_{1}(1270)^+$ and $K_{1}(1400)^+$~\cite{LHCb-PAPER-2012-033}, while the combinatorial background is concentrated at high $K^+\pi^+\pi^-$, $K^+\pi^-$ and $\pi^+\pi^-$ invariant masses.
A combination of PID information and kinematic requirements is used
to veto
charmed meson or baryon decays reconstructed as \Dsm candidates
due to the misidentification of protons or pions as kaons.

A boosted decision tree (BDT)~\cite{Breiman,AdaBoost} with gradient boosting is used to suppress background from random combinations of charged particles.
The multivariate classifier is trained using a background-subtracted \control data sample as signal proxy, while \signal candidates with invariant mass greater than $5500 \mev$ are used as background proxy.
The features used in the BDT are topological variables
related to the vertex separation,
such as the impact parameters of the \Bs candidate and final-state particles,
the flight distance of the \Dsm candidate with respect to the secondary vertex, as well as several criteria on the track quality and vertex reconstruction
and estimators of the isolation of the \Bs candidate from other tracks
in the event.
The working point of the BDT classifier is chosen to optimise the significance of the \signal signal.

\subsection{Data sample composition}
\label{sec:Massfit}

Irreducible background contributions to the selected \control and \mbox{\signal} data samples
are disentangled from signal decays on a statistical basis
by means of an extended maximum likelihood fit to the reconstructed $m(D_s^\mp h^\pm \pi^\pm \pi^\mp)$ invariant mass,
where $h$ is either a pion or a kaon.
A Johnson's $S_\text{U}$ function~\cite{10.2307/2332539} is used as
probability density function (PDF) for the signal component.
The shape parameters are initially determined from simulation.
To account for small differences between simulation and data, scale factors for the mean and standard deviation of the signal PDF are introduced.
These are determined from a fit to the \control calibration sample and thereafter fixed
when fitting \signal candidates.
Background decays of $\Bd$ mesons are described by the same PDF shifted by the known mass difference between $\Bs$ and $\Bd$ mesons~\cite{PDG20}.
The combinatorial background is modelled with a second-order polynomial function.
The shapes for partially reconstructed $\Bs\to \Dssm \pip\pip\pim$,
$\Bs\to \Dssm \Kp \pip\pim$ and $\Bd\to \Dssm \Kp\pip\pim$ decays,
where the $\Dssm$ meson decays to $\Dsm\gamma$ or $\Dsm\piz$,
are derived from simulated decays.
The same applies to the shape for misidentified
\control and $\Bs \to \Dssm \pip \pip \pim$ decays contributing to the \signal sample.
The expected yields of these cross-feed background contributions
are estimated by determining the probability of a pion to pass the PID requirement imposed on the kaon candidate from a control sample of
$D^{*+} \to \left( D^0 \to K^- \pip \right) \pip$ decays~\cite{LHCb-PUB-2016-021}.
All other yields are determined from the fit.

\begin{figure}[ht]
\centering
\includegraphics[height=!,width=0.49\textwidth]{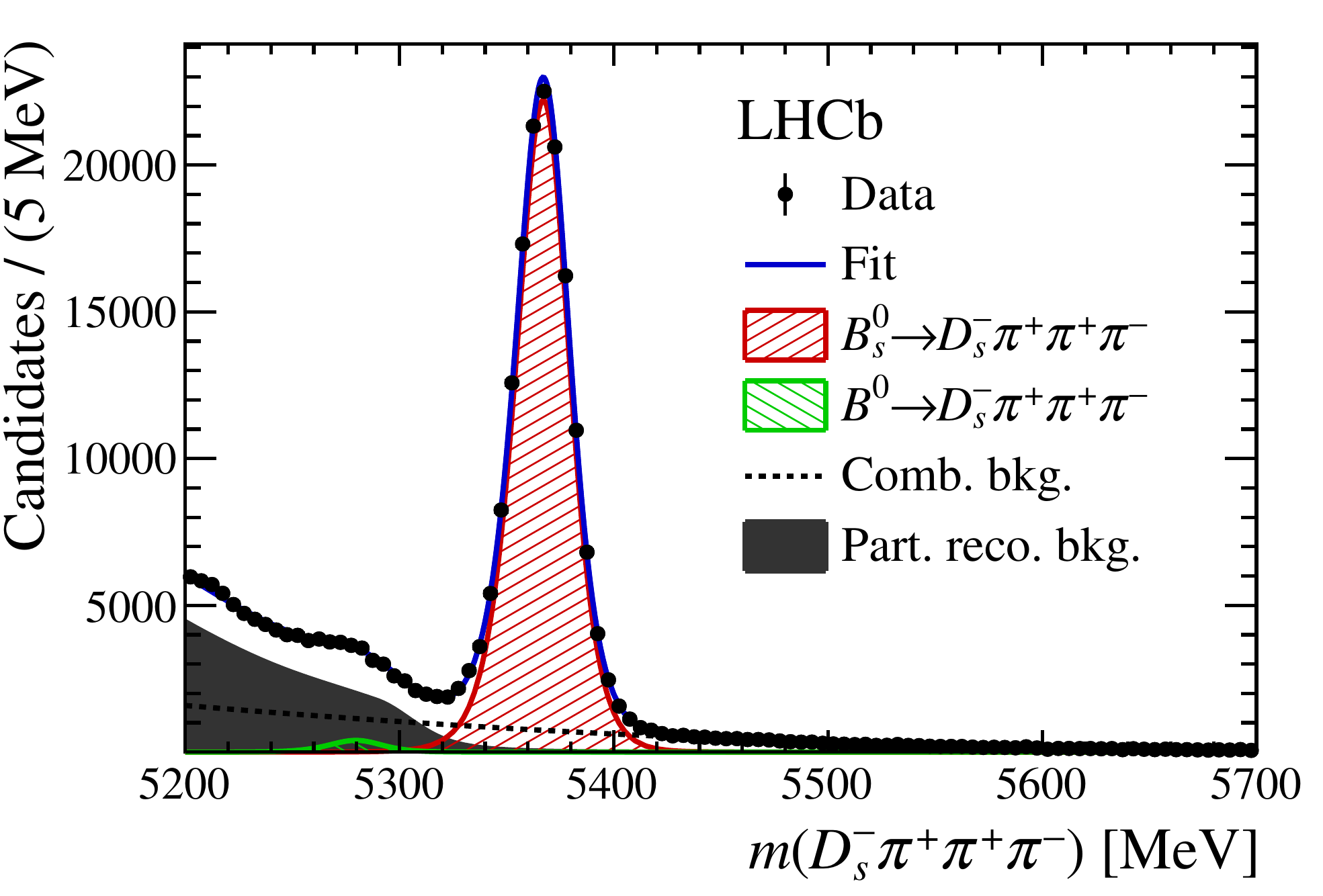}
\includegraphics[height=!,width=0.49\textwidth]{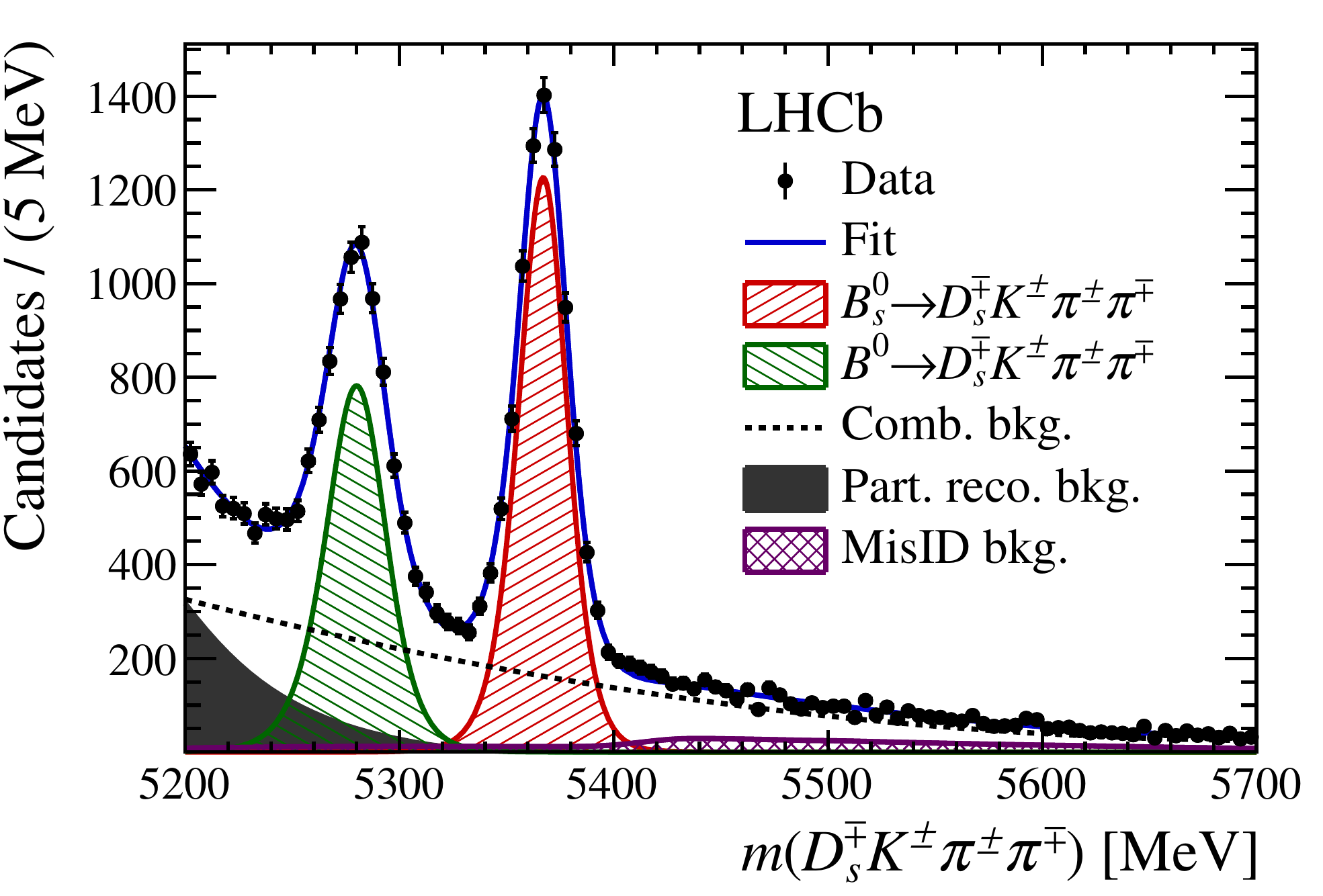}
\caption{Invariant mass distribution of selected (left) \control  and  (right) \mbox{\signal} candidates with fit projections overlaid.}
\label{fig:massFit}
\end{figure}

Figure~\ref{fig:massFit} displays 
the invariant mass distributions of \control and  \mbox{\signal} candidates with fit projections overlaid.
A signal yield of
$148 \, 000 \pm 400$ ($7 500 \pm 100$)
is obtained
for \control (\signal) decays.
The results are used to assign weights to the candidates
to statistically  subtract the background
with the \sPlot technique~\cite{Pivk:2004ty}.
Here, the $m(D_s^\mp h^\pm \pi^\pm \pi^\mp)$ invariant mass is used as discriminating variable
when performing fits to the decay-time and phase-space distributions~\cite{Xie:2009rka}.

\section{Measurement of the \Bs mixing frequency}

A likelihood fit to the background-subtracted decay-time spectrum of the
\control control channel is performed in order
to calibrate detector-induced effects and flavour-tagging algorithms
as well as to measure the mixing frequency $\Delta m_s$.

\subsection{Decay-time resolution}
\label{sec:Resolution}

Excellent decay-time resolution is essential in order to resolve the fast $\Bs-\Bsb$ mixing.
The global kinematic fit to the decay topology
provides an estimate of the decay-time resolution
for each candidate.
The per-candidate decay-time uncertainty, $\delta_t$,
is calibrated by
reconstructing  $\Bs$ candidates
from particles originating directly from the PV.
These prompt $\Bs$ candidates have a known true decay time of zero,
such that the
width of the decay-time distribution is
a measure of the true resolution, $\sigma_t$.
It is determined in
equally populated slices of $\delta_t$.
A linear calibration function
is used
to map the per-candidate decay-time uncertainty to the actual resolution.
On average, the resolution
amounts to $\langle \sigma_t \rangle = 36.6 \pm 0.5 \fs$, where
the uncertainty is statistical only.
A decay-time bias, $\mu_t$, of approximately $-2\fs$ is observed which results from the applied selection requirements and the precision of the detector alignment.
The implementation of the decay-time bias and decay-time resolution in the fit are discussed in Sec.~\ref{sec:DecayTimeFitDs3pi}.

\subsection{Flavour tagging}
\label{sec:Tagging}

Two complementary methods are used to determine the flavour of the $\Bs$
candidates at production.
The opposite-side (OS) tagger~\cite{LHCb-PAPER-2011-027, LHCb-PAPER-2015-027} exploits the fact
that \bquark quarks are predominantly produced in quark-antiquark pairs,
which leads to a second \bquark hadron alongside the signal $\Bs$ meson.
The flavour of the non-signal \bquark hadron is determined using the
charge of the lepton ($\mu,e$) produced in semileptonic  decays,
the charge of a reconstructed secondary charm hadron,
the charge of the kaon from the $b \to c \to s$ decay chain,
and/or the charge of the inclusive secondary vertex reconstructed from the \bquark hadron decay products.
The same-side (SS) tagger~\cite{LHCB-PAPER-2015-056} determines the flavour of the signal
candidate by identifying the charge of the kaon produced together with the $\Bs$ meson in the fragmentation process.

Each tagging algorithm provides a flavour-tagging decision, $d$,
which takes the value $d=+1$ ($d=-1$) for a candidate tagged as a \Bs (\Bsb) meson
and $d=0$ if no decision can be made (untagged).
The tagging efficiency $\epsilon_{\text{tag}}$ is defined
as the fraction of selected candidates with non-zero tag decision.
The tagging algorithms also provide
an estimate, $\eta$, of the probability that the decision is incorrect. %
The tagging decision and mistag estimate are obtained
using flavour specific, self-tagging, decays. 
Multivariate classifiers are employed combining
various inputs such as kinematic variables of tagging particles and of the signal candidate.
These are trained on simulated samples of \BsDsPi decays for the SS tagger and on data samples
of $B^+ \to \jpsi \Kp$ decays for the $\mathrm{OS}$ tagger.
The mistag estimate
does not necessarily represent the true mistag probability, $\omega$,
since the algorithms might perform differently on the signal decay
than on the decay modes used for the training.
Therefore, it is calibrated for each tagger
using the flavour-specific decay \control.

\subsection{Decay-time fit}
\label{sec:DecayTimeFitDs3pi}

The signal PDF describing the \control proper-time spectrum
is based on the theoretical decay rate in Eq.~(\ref{eq:timePdf})
taking several experimental effects into account~\cite{LHCB-PAPER-2014-038,LHCB-PAPER-2017-047,LHCB-PAPER-2013-006,LHCB-PAPER-2018-009}

  \begin{minipage}{\linewidth}
  \begin{flalign}
	&\mathcal P(t,  d_{\mathrm{OS}}, d_{\mathrm{SS}}, q_f \vert \delta_t,  \eta_{\mathrm{OS}}, \eta_{\mathrm{SS}}) \propto \nonumber \\ &\left[ (1- q_f A_D) \, p(t^\prime,  d_{\mathrm{OS}}, d_{\mathrm{SS}}, q_f \vert  \delta_t,  \eta_{\mathrm{OS}}, \eta_{\mathrm{SS}}) \, e^{-\Gamma_s \, t^\prime}
	\otimes R(t-t^\prime \vert \mu_t, \sigma_t(\delta_t)) \right] \, \epsilon(t),   \nonumber \\ \nonumber
\end{flalign}
  \end{minipage}
where

\noindent
  \begin{flalign}
    & p(t^\prime,  d_{\mathrm{OS}}, d_{\mathrm{SS}}, q_f \vert  \delta_t,  \eta_{\mathrm{OS}}, \eta_{\mathrm{SS}}) \equiv \nonumber \\ 
    &\Biggl[
	(1-A_P)  h(d_{\mathrm{OS}}, \eta_{\mathrm{OS}})  h(d_{\mathrm{SS}}, \eta_{\mathrm{SS}})
	\biggl( \text{cosh} \left( \frac{\Delta \Gamma_s \, t^\prime}{2}\right)
	 + \,q_f \, \text{cos} \left(\Delta m_s \, t^\prime \right)   \biggr)\Biggr.       \nonumber   \\
	& \Biggl. + (1+A_P)  \bar h(q_{\mathrm{OS}}, \eta_{\mathrm{OS}})  \bar h(d_{\mathrm{SS}}, \eta_{\mathrm{SS}})
	\biggl( \text{cosh} \left( \frac{\Delta \Gamma_s \, t^\prime}{2}\right)
	 - \,q_f \, \text{cos} \left(\Delta m_s \, t^\prime \right)  \biggr) \Biggr],    
	 \nonumber   \\
	 & h(d, \eta) \equiv \vert d \vert \Biggl[1+ d \, \biggl(1-2 \, w(\eta)\biggr) \Biggr]   \, \epsilon_{\text{tag}} + 2 \, (1 - \vert d \vert) \, (1-\epsilon_{\text{tag}}),    \nonumber   \\
	 & \bar h(d, \eta) \equiv \vert d \vert  \Biggl[1- d \, \biggl(1-2 \, \bar w(\eta)\biggr) \Biggr]  \, \bar \epsilon_{\text{tag}} + 2 \, (1 - \vert d \vert) \, (1-\bar \epsilon_{\text{tag}}). 
\end{flalign}\label{eq:timePdfMod}
This PDF depends on the tagging decisions of the $\mathrm{OS}$ and $\mathrm{SS}$ taggers,
$d_{\mathrm{OS}}$ and $d_{\mathrm{SS}}$;
on the final state, $q_f = +1 (-1)$
for $f = D_s^-\pip\pip\pim (\bar f = D_s^+\pim\pim\pip$);
and is conditional on the per-event observables~\cite{Punzi:2004wh} $\delta_t,  \eta_{\mathrm{OS}}$ and $\eta_{\mathrm{SS}}$, describing the estimated decay-time error and the estimated mistag rates of the OS and SS taggers, respectively.
The parameters of the Gaussian resolution model,
$R(t-t^\prime \vert \mu_t, \sigma_t(\delta_t))$,
are fixed to the values determined from the prompt candidate data sample.
The decay-time-dependent efficiency, $\epsilon(t)$, of
reconstructing and selecting signal decays
is modelled by a B-spline curve~\cite{2001:GMS:500858,10.2307/41133848},
whose cubic polynomials are uniquely defined by a set of knots.
These are placed across the considered decay-time range
to account for local variations~\cite{Karbach:2014qba}.
Six knots are chosen such that there is an approximately equal amount of data in-between two consecutive knots.
By fixing the decay width $\Gamma_s=(0.6624\pm0.0018)\textrm{ ps}^{-1}$ and decay width difference $\Delta\Gamma_s=(0.090\pm0.005)\textrm{ ps}^{-1}$ to their world-average values~\cite{HFLAV18},
the spline coefficients can be directly determined in the fit to the data.
The correlation $\rho(\Gamma_s,\,\Delta\Gamma_s)=-0.080$ 
is taken into account when evaluating the systematic uncertainties.

Both taggers are simultaneously calibrated during the fit as described by the functions
$h(d, \eta)$ and $\bar h(d, \eta)$.
These also take into account
a small dependence of the tagging performance on the
initial $\Bs$ flavour
by introducing linear calibration functions
for initial \Bs and \Bsb  mesons
denoted as $\omega(\eta)$ and $\bar \omega(\eta)$, respectively.
Similarly,  $\epsilon_{\text{tag}}$ ($\bar \epsilon_{\text{tag}}$)
denotes the tagging efficiency for an initial \Bs (\Bsb) meson.
The calibrated responses of the OS and SS taggers are then explicitly combined in the PDF.
Since the tagging algorithms have been retuned for the Run 2 data-taking period~\cite{Fazzini:2018dyq}
to account for the changed conditions,
separate calibrations for the two data-taking periods are performed.

The production asymmetries between \Bs and \Bsb mesons at centre-of-mass energies of $7 \tev$ and $8 \tev$
are taken from an LHCb measurement
using \BsDsPi decays~\cite{LHCb-PAPER-2016-062}.
After correcting for kinematic differences between \BsDsPi and \control decays,
the effective production asymmetry, $A_P = (N(\Bsb) - N(\Bs))/(N(\Bsb) + N(\Bs))$, for Run 1 data amounts to
$A_P = (-0.1 \pm 1.0 ) \%$.
The production asymmetry at a centre-of-mass energy of $13 \tev$ is determined in the fit.
A detection asymmetry between the final states is caused by the different interaction cross-sections of positively and negatively charged kaons with the detector material.
The detection asymmetry is defined as $A_D = (\eps(\bar f) - \eps(f))/(\eps(\bar f) + \eps(f)) $,
where $\eps(\bar f)$ $(\eps(f))$ denotes the detection efficiency of
final state $\bar f$ $(f)$.
It is computed by comparing the charge asymmetries in
$D^\pm \to K^\mp \pi^\pm \pi^\pm$ and $D^\pm \to \KS \pi^\pm$
calibration samples~\cite{Davis:2310213}, weighted to match the kinematics of the signal kaon.
Only the decay mode $\Dsm \to \Km\pim\pip$ is a possible source of detection asymmetry for \control decays resulting in an average detection asymmetry of $A_{D} = (-0.07 \pm 0.15) \%$ for Run 1
and $A_{D} = (-0.08 \pm 0.21) \%$ for Run 2 data. 
A sufficiently large subsample of the Run 2 data set is used to reconstruct the calibration modes for this study.

\begin{figure}[b]
	\centering
		\includegraphics[width=0.49\textwidth, height = !]{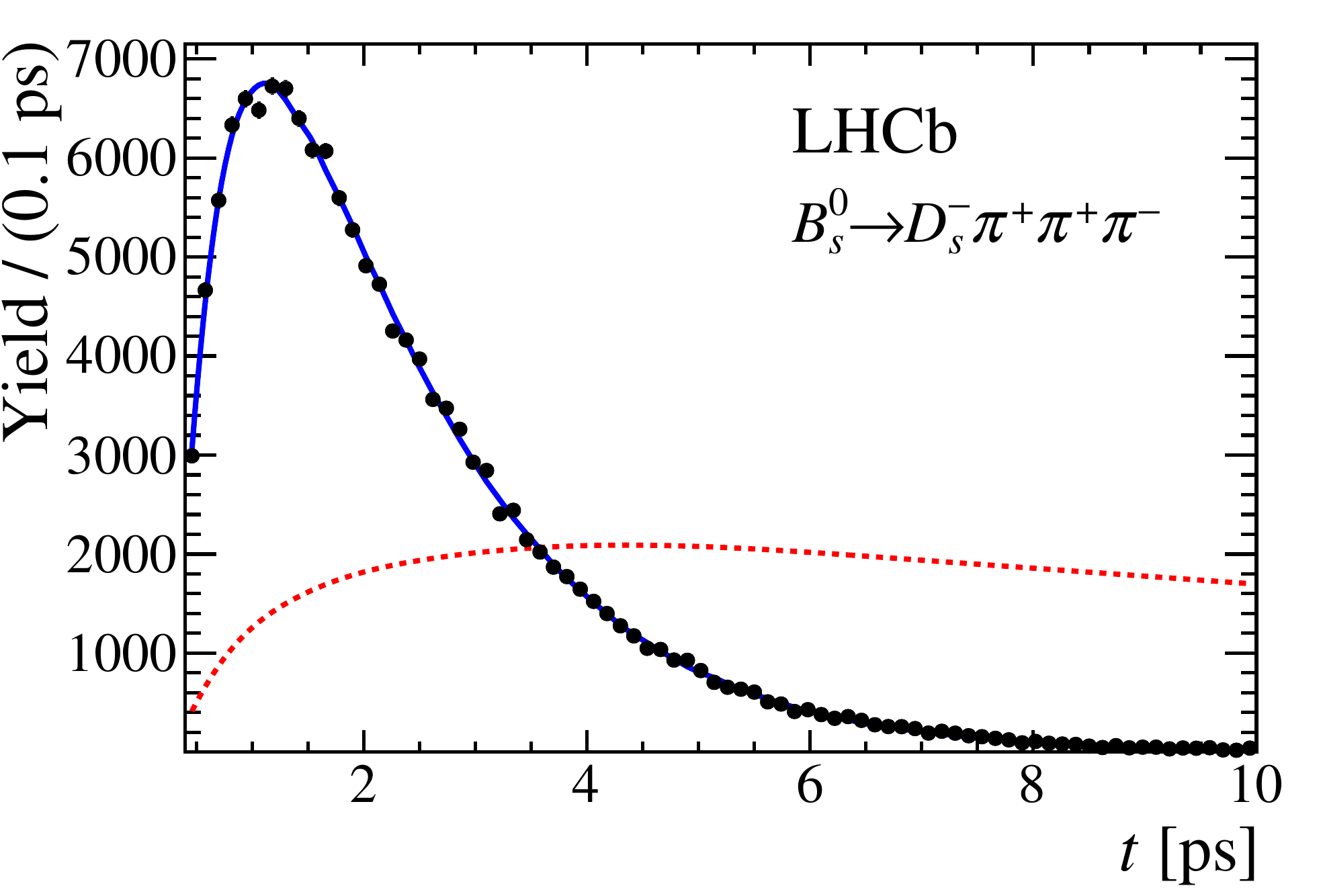}
		
		\includegraphics[width=0.49\textwidth, height = !]
		{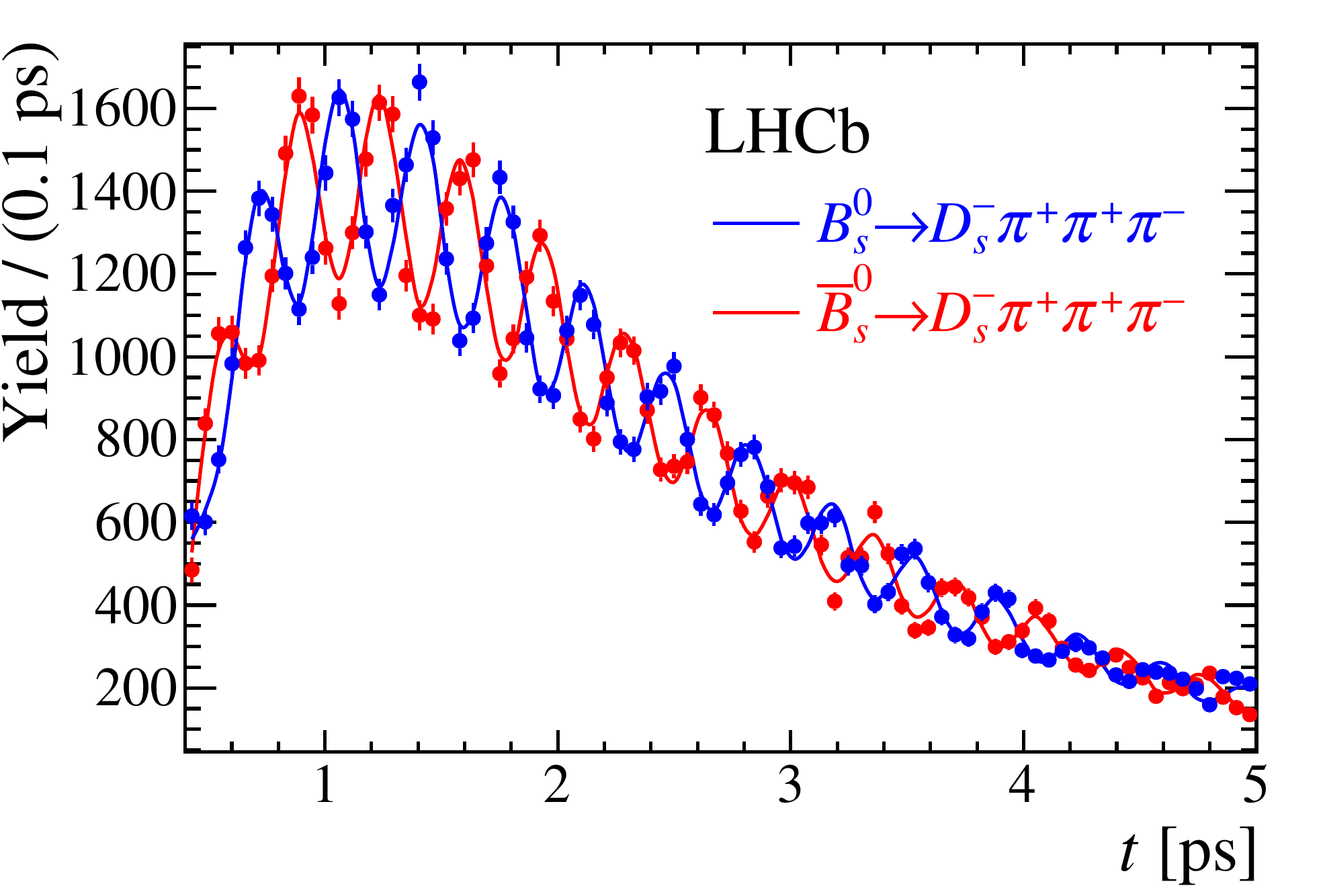}
		\includegraphics[width=0.49\textwidth, height = !]{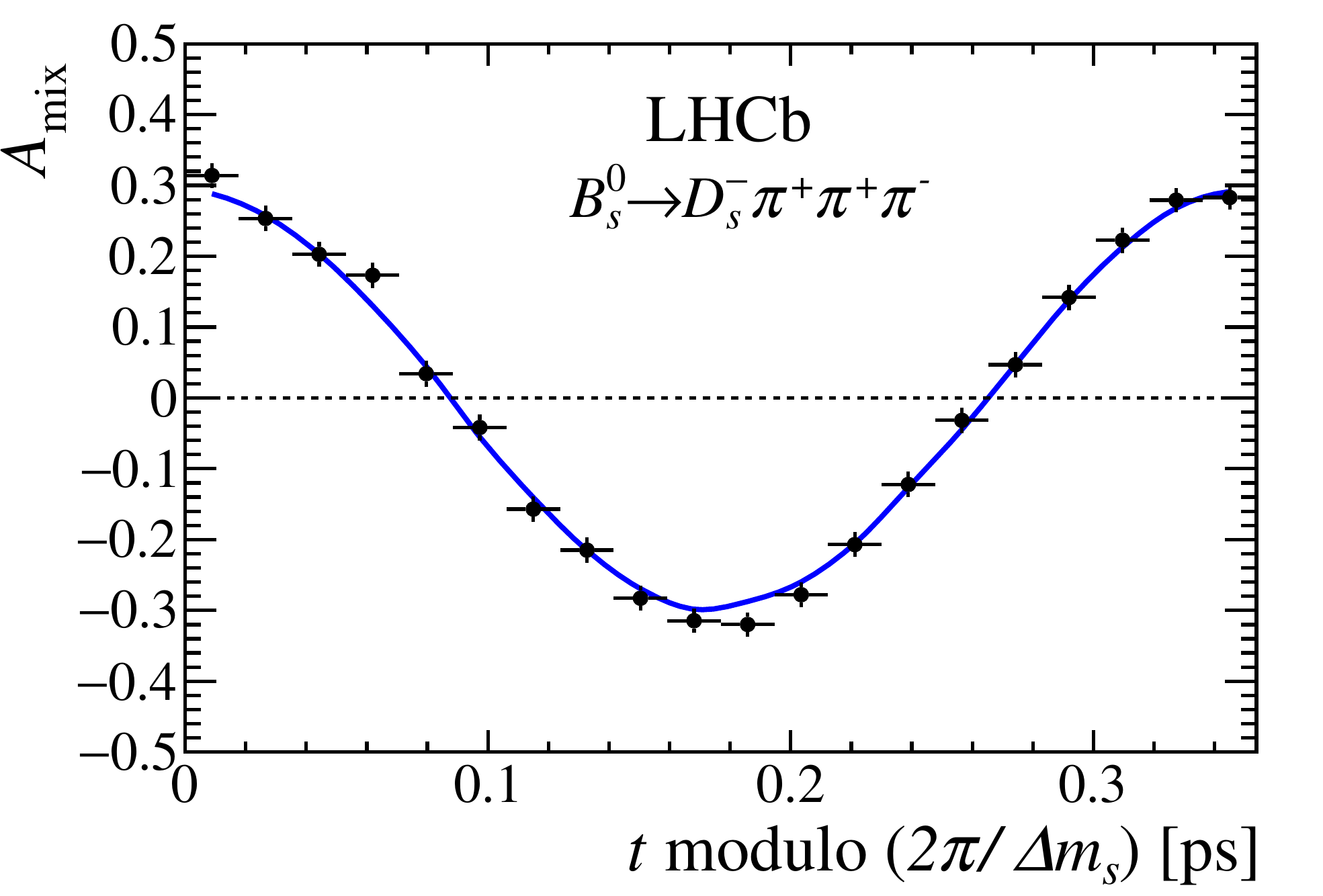} 	
		\caption{
		  Background-subtracted decay-time distribution of (top) all and (bottom left) tagged \control candidates as well as (bottom right) the dilution-weighted mixing asymmetry folded into one oscillation period along with the fit projections (solid lines). The decay-time acceptance (top) is overlaid in an arbitrary scale (dashed line).
		 } 		
		\label{fig:tFitNorm}
\end{figure}

Figure~\ref{fig:tFitNorm} displays the decay-time distribution
and the mixing asymmetry
for \mbox{\control} signal candidates.
The latter is weighted by the per-candidate time resolution and flavour-tagging dilution to enhance the visible asymmetry.
All features are well reproduced by the fit projections which are overlaid.
The $\Bs-\Bsb$
mixing frequency is determined to be
\begin{equation}
\Delta m_s = (17.757 \pm 0.007 \pm 0.008)\invps, \nonumber
\end{equation}
where the first uncertainty is statistical and the second systematic.
The systematic studies are discussed in Sec.~\ref{sec:Systematics}.
Within uncertainties, the measured production asymmetry for Run 2 data,
$A_P = (0.1 \pm 0.5 \pm 0.1)\%$, is consistent with zero.
The calibrated per-candidate mistag probabilities,  $\omega_i$,
are used to compute the effective tagging power as
$\epsilon_{\text{eff}} = \frac{1}{N} \sum_i ( 1 - 2 \omega_i )^2$,
where $N$ is the total number of signal candidates and a value of $\omega_i = 0.5$ is assigned to untagged candidates.
Table~\ref{tab:tagPerfRun1} reports the observed tagging performance for Run 1 and Run 2 data
considering three mutually exclusive categories: tagged by the $\mathrm{OS}$ combination algorithm only, tagged by the $\mathrm{SS}$ kaon algorithm only and tagged by both $\mathrm{OS}$ and $\mathrm{SS}$ algorithms.
While the flavour taggers suffer from the higher track multiplicity during the Run 2 data-taking period,
they profit from the harder momentum spectrum of the produced $\bquark \bar \bquark$ quark pair.
Combined, this results in a net 
relative improvement of $14\%$
in effective tagging power.
\begin{table}[t]
\centering
\small
\caption{The flavour-tagging performance for only OS-tagged, only SS-tagged and both OS- and SS-tagged \control signal candidates.}
\begin{subtable}[ph]{\textwidth}
\centering
\caption{Run 1 data.}
\begin{tabular}{c c c c}
\hline
\hline
 & $\epsilon_{\text{tag}} [\%]$ & $\langle \omega \rangle [\%] $ & $\epsilon_{\text{eff}} [\%]$ \\
\hline
Only OS & 14.74 $\pm$ 0.11 & 39.09 $\pm$ 0.80 & 1.25 $\pm$ 0.16\\
Only SS & 35.38 $\pm$ 0.18 & 44.26 $\pm$ 0.62 & 1.05 $\pm$ 0.18\\
Both OS-SS & 33.04 $\pm$ 0.30 & 37.33 $\pm$ 0.73 & 3.41 $\pm$ 0.33\\
\hline
Combined & 83.16 $\pm$ 0.37 & 40.59 $\pm$ 0.70 & 5.71 $\pm$ 0.40\\
\hline
\hline
\end{tabular}
 \end{subtable}
\begin{subtable}[h]{\textwidth}
\caption{Run 2 data.}
\centering
\begin{tabular}{c c c c}
\hline
\hline
 & $\epsilon_{\text{tag}} [\%]$ & $\langle \omega \rangle [\%] $ & $\epsilon_{\text{eff}} [\%]$ \\
\hline
Only OS & 11.91 $\pm$ 0.04 & 37.33 $\pm$ 0.41 & 1.11 $\pm$ 0.05\\
Only SS & 40.95 $\pm$ 0.08 & 42.41 $\pm$ 0.29 & 1.81 $\pm$ 0.10\\
Both OS-SS & 28.96 $\pm$ 0.12 & 35.51 $\pm$ 0.32 & 3.61 $\pm$ 0.13\\
\hline
Combined & 81.82 $\pm$ 0.15 & 39.23 $\pm$ 0.32 & 6.52 $\pm$ 0.17\\
\hline
\hline
\end{tabular}
 \end{subtable}

\label{tab:tagPerfRun1}

\end{table}

\section{Measurement of the CKM angle $\gamma$}
\label{sec:timeFit}

This section first describes the phase-space-integrated decay-time analysis of the  signal channel \signal that allows the determination of the CKM angle $\gamma$ in a model-independent way.
Afterwards, the resonance spectrum in \signal decays is studied and a full time-dependent amplitude analysis is performed for a model-dependent determination of the CKM angle $\gamma$.

\subsection{Model-independent analysis}
\label{ssec:modelIndependent}

\begin{figure}[t]
	\centering
		\includegraphics[width=0.49\textwidth, height = !]{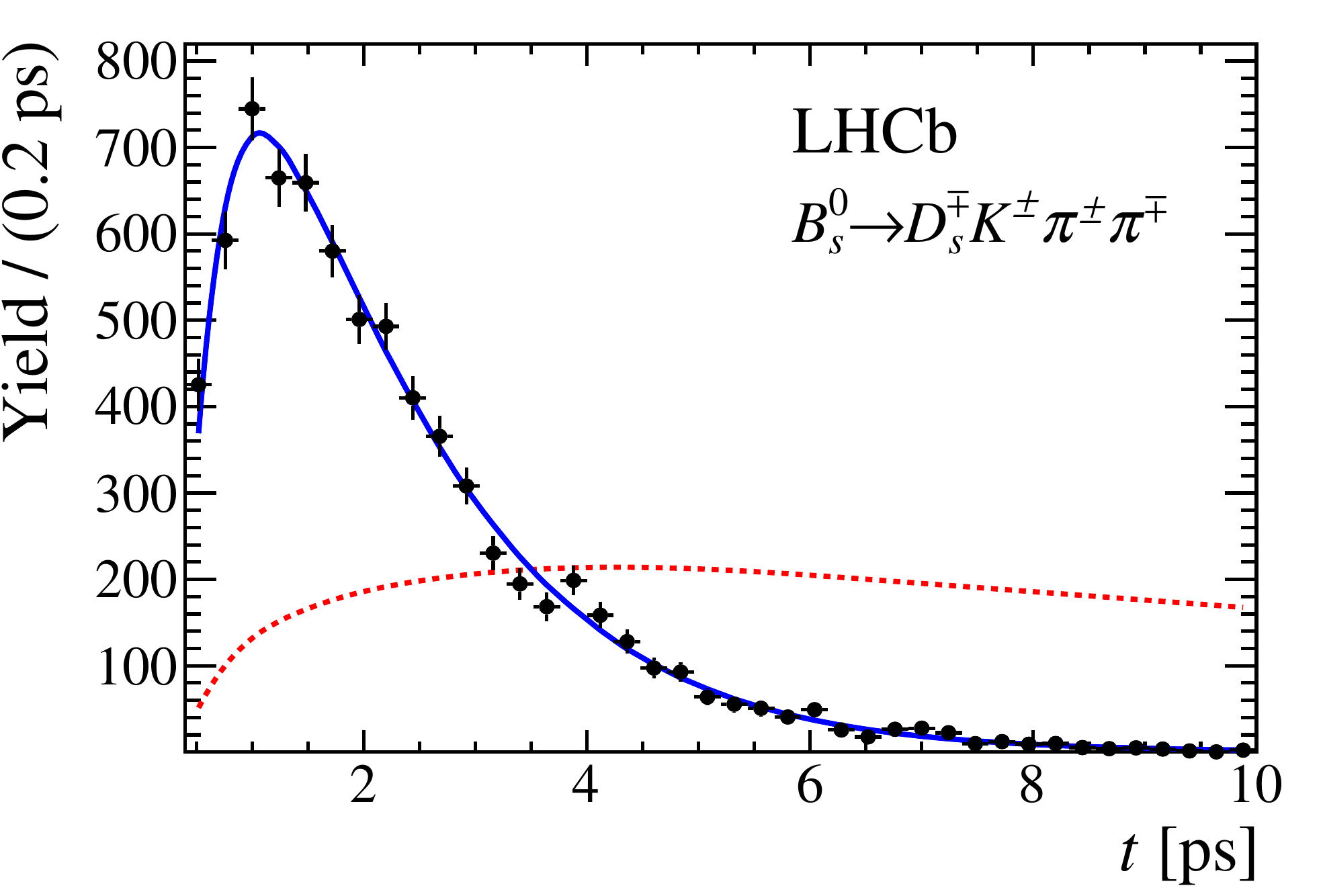} \includegraphics[width=0.49\textwidth, height = !]{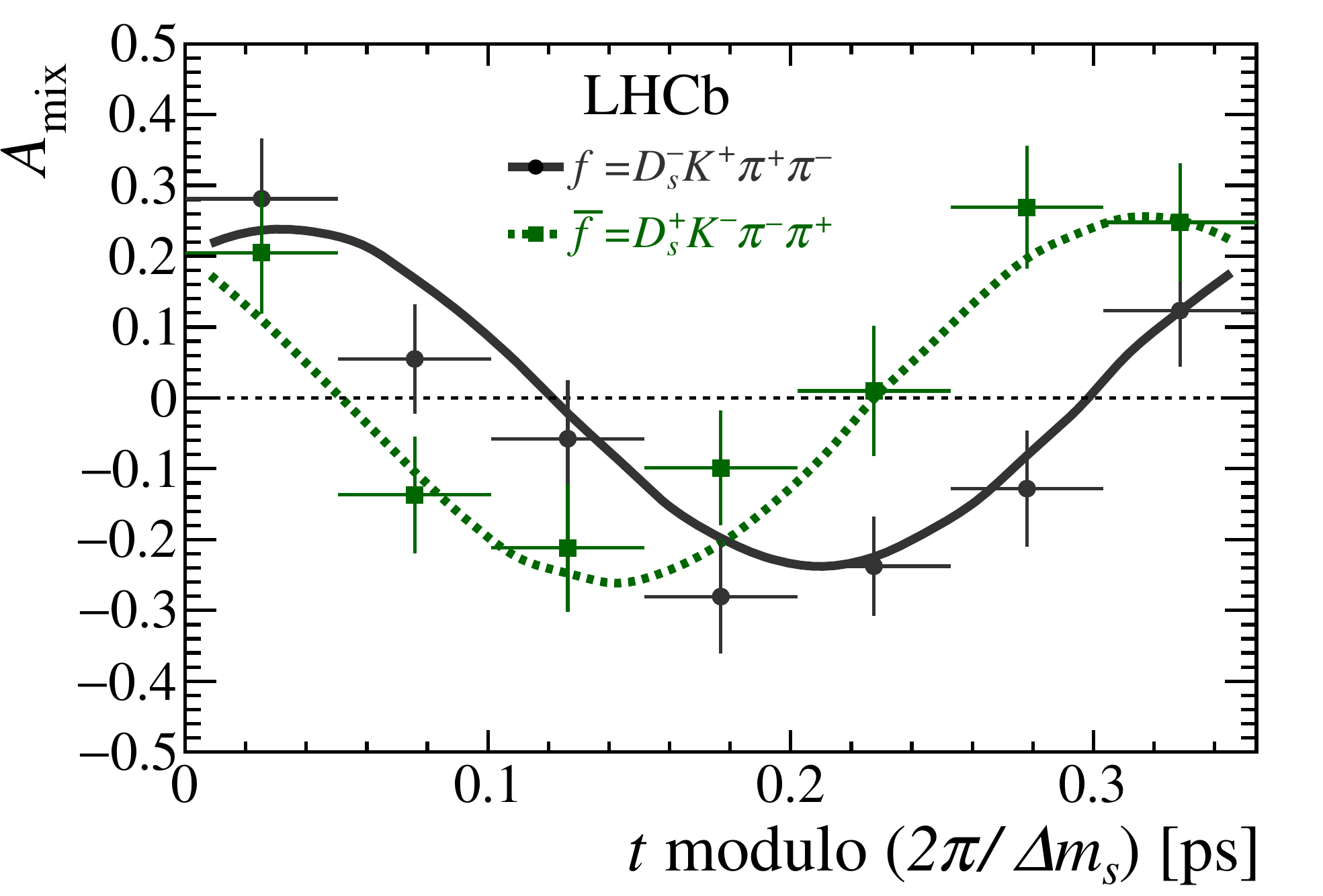} 		
		\caption{
		  Decay-time distribution of (left) background-subtracted \signal candidates and (right) dilution-weighted mixing asymmetry along with the model-independent fit projections
(lines). The decay-time acceptance (left) is overlaid in an arbitrary scale (dashed line).
		 }
		\label{fig:tFitSig}
\end{figure}

The decay-time fit to the \signal candidates uses a signal PDF based on Eq.~(\ref{eq:timePdf})
with modifications accounting for the experimental effects described in Sec.~\ref{eq:timePdfMod}.
The $\Bs$ production asymmetry for Run 2 data
and the $\Bs-\Bsb$ mixing frequency are fixed to the values
obtained from the \control data sample,
whereas the tagging calibration parameters
are allowed to vary within Gaussian constraints
taking into account their correlation. 
The kaon detection asymmetry for \signal decays is
determined in a similar manner as for \control decays and amounts to
$A_{D} = (-1.02 \pm 0.15) \%$ for Run 1
and $A_{D} = (-0.91 \pm 0.22) \%$ for Run 2 data.
The decay-time acceptance is also fixed to the \control
result, corrected by a decay-time dependent correction factor
derived from simulation to account for small differences in the selection and decay kinematics between the decay modes.
Otherwise, the fit strategy is identical to that discussed in the previous section.
Figure~\ref{fig:tFitSig} shows the decay-time distribution
and mixing asymmetries together with the fit projections.
The mixing asymmetries for $D_s^- K^+ \pip \pim$
and $D_s^+ K^- \pim \pip$ final states are shifted with respect to each other indicating mixing-induced \CP violation, \cf Eq.~(\ref{eq:mixAsym}).
The \CP coefficients $C_f,A^{\Delta\Gamma}_f,A^{\Delta\Gamma}_{\bar f}, S_f$ and $S_{\bar f}$  determined from
the fit are reported in Table~\ref{tab:sigFitResults}.
They are converted to the parameters of interest $r,\kappa,\delta$ and $\gamma-2\beta_s$ in Sec.~\ref{ssec:results}.

\begin{table}[t]
\centering
\caption{\CP coefficients determined from the phase-space fit to the \signal decay-time distribution. The uncertainties are statistical and systematic (discussed in Sec.~\ref{sec:Systematics}).}
	\renewcommand{\arraystretch}{1.25}
\begin{tabular}{c r }
\hline
\hline
Fit parameter & \multicolumn{1}{c}{Value}  \\
\hline
$C_f$ & 0.631 $\pm$ 0.096 $\pm$ 0.032 \\
$A^{\Delta\Gamma}_f$ & $-0.334$ $\pm$ 0.232 $\pm$ 0.097 \\
$A^{\Delta\Gamma}_{\bar{f}}$ & $-0.695$ $\pm$ 0.215 $\pm$ 0.081 \\
$S_f$ & $-0.424$ $\pm$ 0.135 $\pm$ 0.033 \\
$S_{\bar{f}}$ & $-0.463$ $\pm$ 0.134 $\pm$ 0.031 \\
\hline
\hline
\end{tabular}
 \label{tab:sigFitResults}
\end{table}

\subsection{Time-dependent amplitude analysis}
\label{ssec:ampAna}

To perform the time-dependent amplitude fit, a
signal PDF is employed which replaces the phase-space integrated decay rate with the full decay rate given in Eq.~(\ref{eq:decayRateBs}),
but is otherwise identical to the PDF used in Sec.~\ref{ssec:modelIndependent}.
Variations of the selection efficiency over the phase space are incorporated by evaluating the likelihood normalisation integrals
with the Monte Carlo (MC) integration technique using fully simulated decays~\cite{dArgent:2017gzv,MarkIIIK3pi,FOCUS4pi,KKpipi}.

The light meson spectrum comprises
a large number of resonances potentially contributing to the \signal decay
in various decay topologies and angular momentum configurations.
The full list of considered intermediate-state amplitudes can be found in Table~\ref{tab:ampsDsKpipiLASSO}.
A significant complication arises from the fact that two (quasi-independent) amplitude models need to be developed simultaneously:
one amplitude describes decays via $b \to c$ ($A^c(\phsPoint)$), the other decays via $b \to u$ ($A^u(\phsPoint)$) quark-level transitions.
A model building procedure is applied
to obtain a good description of the observed phase-space distribution while keeping the number of included amplitudes as small as possible.
This is accomplished in two stages.
The first stage identifies the set of intermediate-state amplitudes contributing at a significant level to either decays via $b \to c$ or $b \to u$ quark-level transitions or to both.
To that end, the time-integrated and flavour-averaged phase-space distribution
is examined.
A single total amplitude,
$A^{\text{eff}}(\phsPoint) = \sum_i a^{\text{eff}}_i  A_i(\phsPoint),$
is sufficient in this case, which effectively describes the incoherent superposition of the $b\to c$ and $b\to u$ amplitudes,
$ \vert A^{\text{eff}}(\phsPoint)  \vert^{2} =   \vert A^c(\phsPoint)  \vert^{2} +   r^2 \, \vert A^u(\phsPoint)   \vert^{2}$.
This significantly simplified fitting procedure allows the initial inclusion of the whole pool of considered intermediate-state amplitudes, limiting the model complexity with the  LASSO technique\cite{Tibshirani94regressionshrinkage,Guegan:2015mea,dArgent:2017gzv}.
This method adds a penalty term to the likelihood function,
\begin{equation}
	-2 \, \log \mathcal L \to -2 \, \log \mathcal L + \lambda \, \sum_{i} \sqrt{ \int \vert a^{\text{eff}}_i  A_i(\phsPoint) \vert^{2} \, \text{d}\Phi_{4}  },
\end{equation}
which shrinks the amplitude coefficients towards zero.
The optimal value for the LASSO parameter $\lambda$,
which controls the model complexity,
is found by minimising the Bayesian information criterion
$\text{BIC}(\lambda) = - 2 \, \log \mathcal L + k  \, \log N_{\rm Sig}$~\cite{BIC},
where $N_{\rm Sig}$ is the signal yield and $k$ is the number of amplitudes with a decay fraction above the threshold of $0.5 \%$.
The amplitudes with a decay fraction above the threshold at the minimum $\text{BIC}(\lambda)$ value are selected.
The second stage of the model selection performs
a full time-dependent amplitude fit.
The components selected by the first stage are included for both $b\to c$ and $b\to u$ transitions and
a LASSO penalty term for each is added to the likelihood function.
As the strong interaction is \CP symmetric, the subdecay modes of three-body resonances
and their conjugates are constrained to be the same.
The final set of $b\to c$ and $b\to u$ amplitudes
is henceforth referred to as the baseline model.
The LASSO penalty term is only used to select
the model and discarded in the final fit to avoid biasing the parameter uncertainties.

\begin{figure}[h]
	\centering		
		\includegraphics[width=0.49\textwidth, height = !]{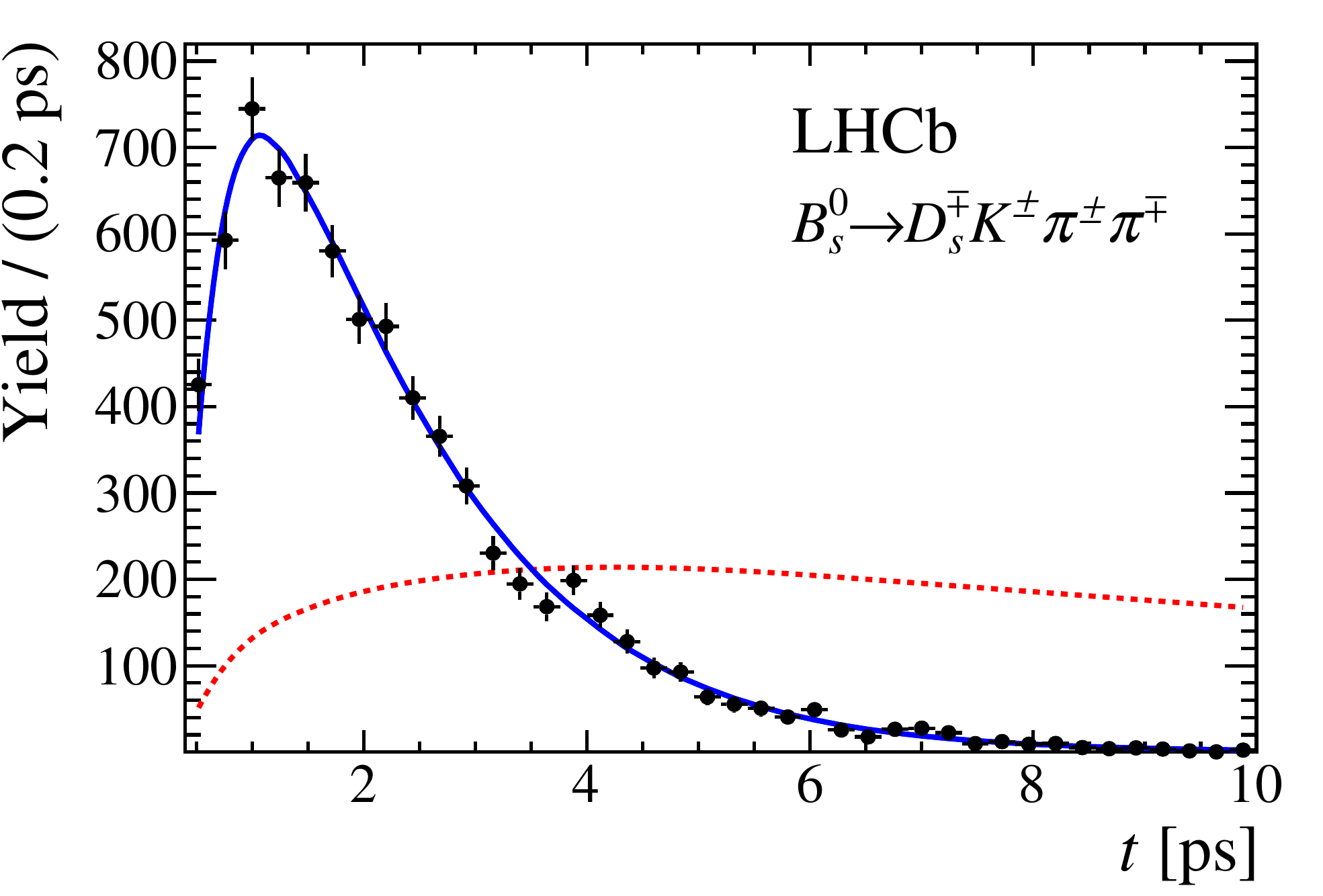}
		\includegraphics[width=0.49\textwidth, height = !]{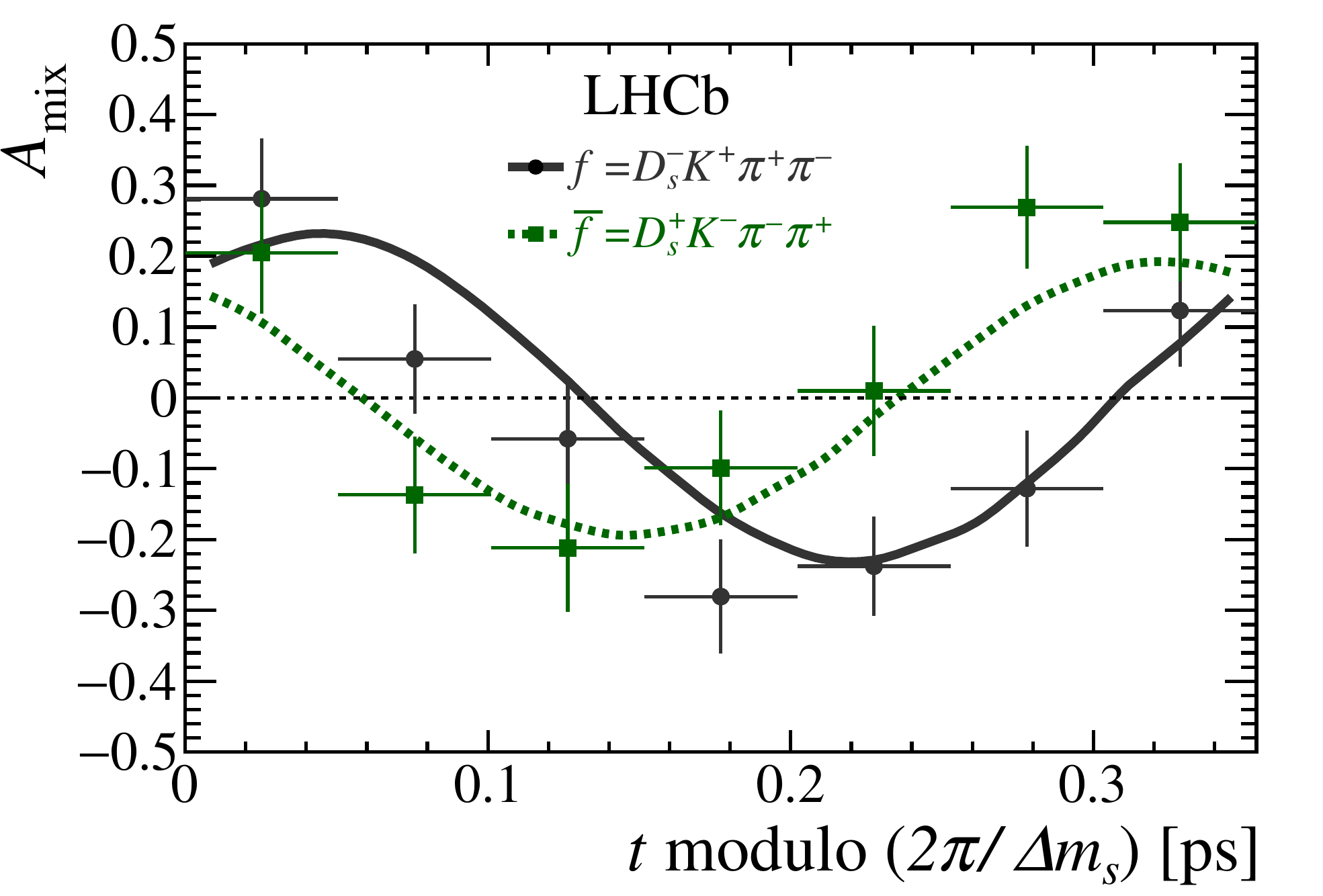} 		
		\caption{
 Decay-time distribution of (left) background-subtracted \signal candidates and (right) dilution-weighted mixing asymmetry along with the model-dependent fit projections
(lines). The decay-time acceptance (left) is overlaid in an arbitrary scale (dashed line).
		 } 		
		\label{fig:fullFitTime}		
\end{figure}	

\begin{figure}[h]
\centering
		\includegraphics[width=0.4\textwidth, height = !]{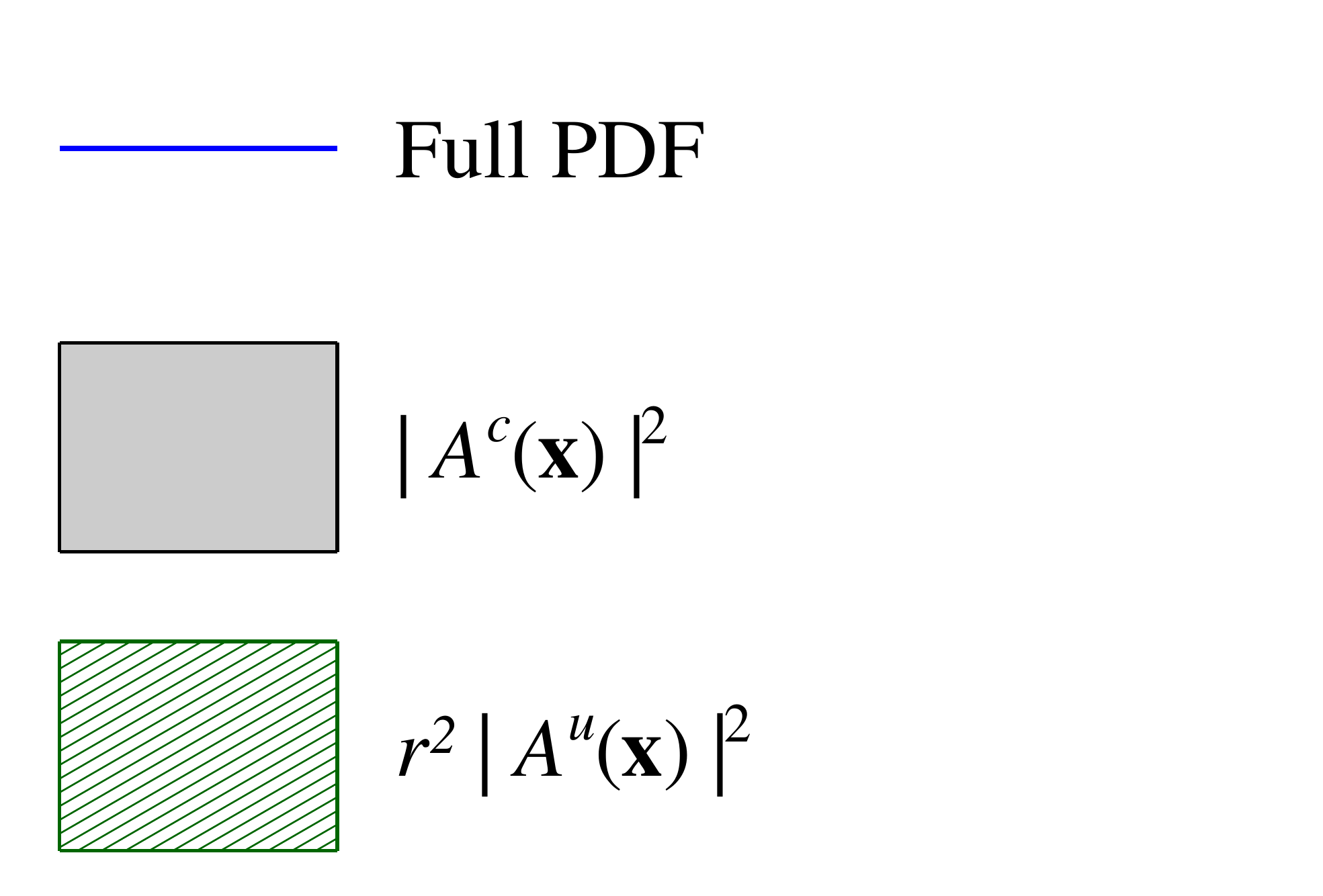} 		
		\includegraphics[width=0.4\textwidth, height = !]{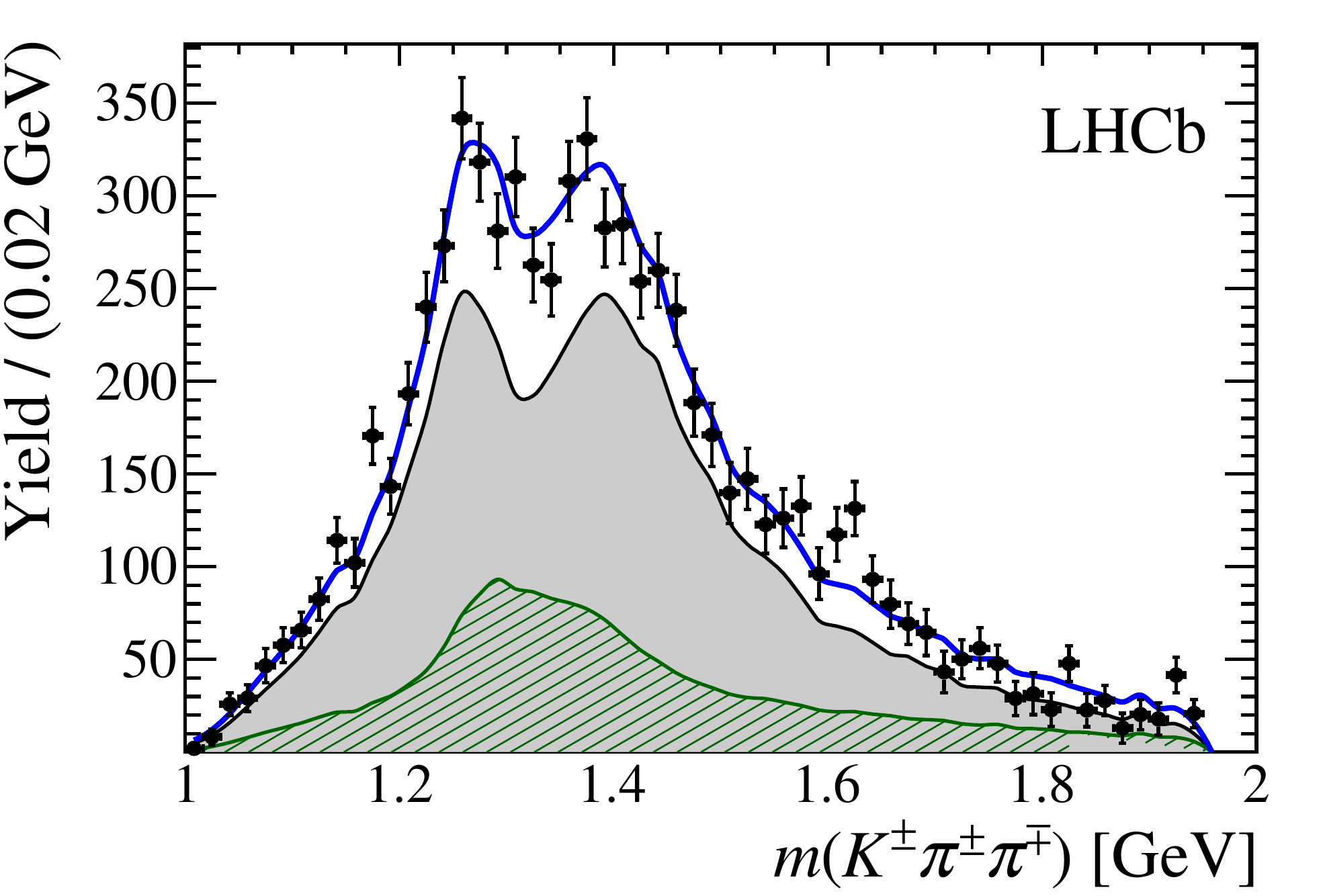} 	
			
		\includegraphics[width=0.4\textwidth, height = !]{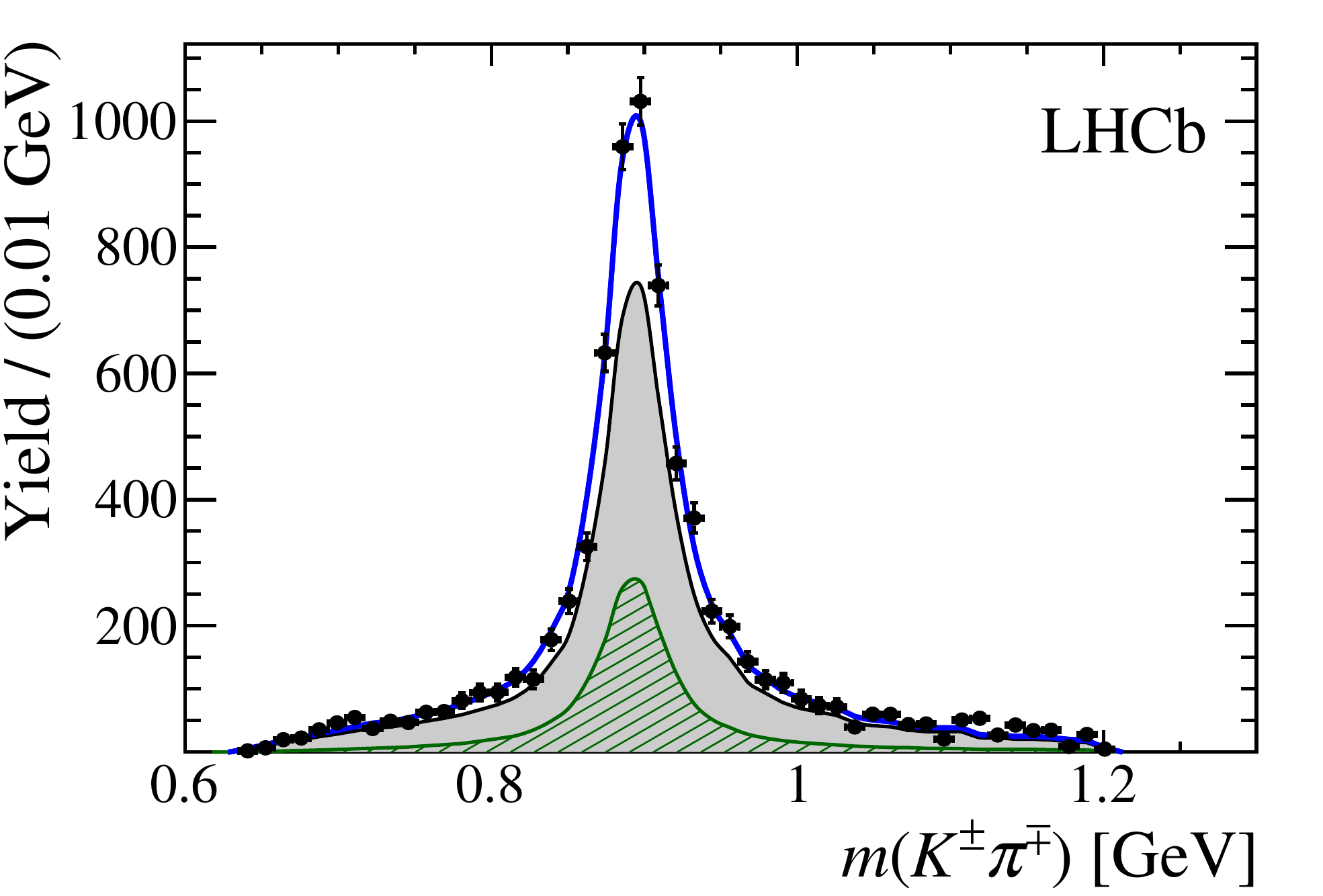}
		\includegraphics[width=0.4\textwidth, height = !]{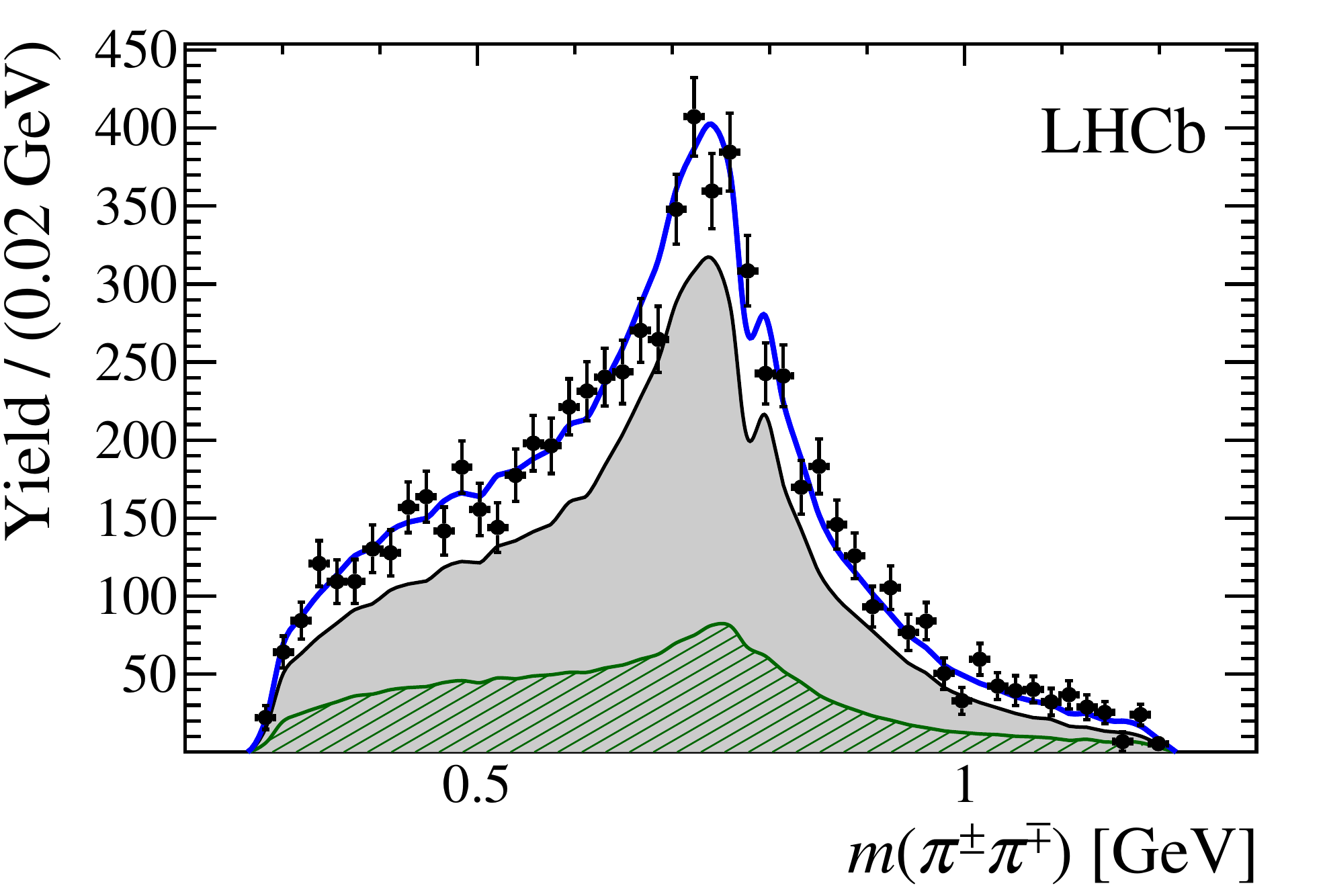}
		
		\includegraphics[width=0.4\textwidth, height = !]{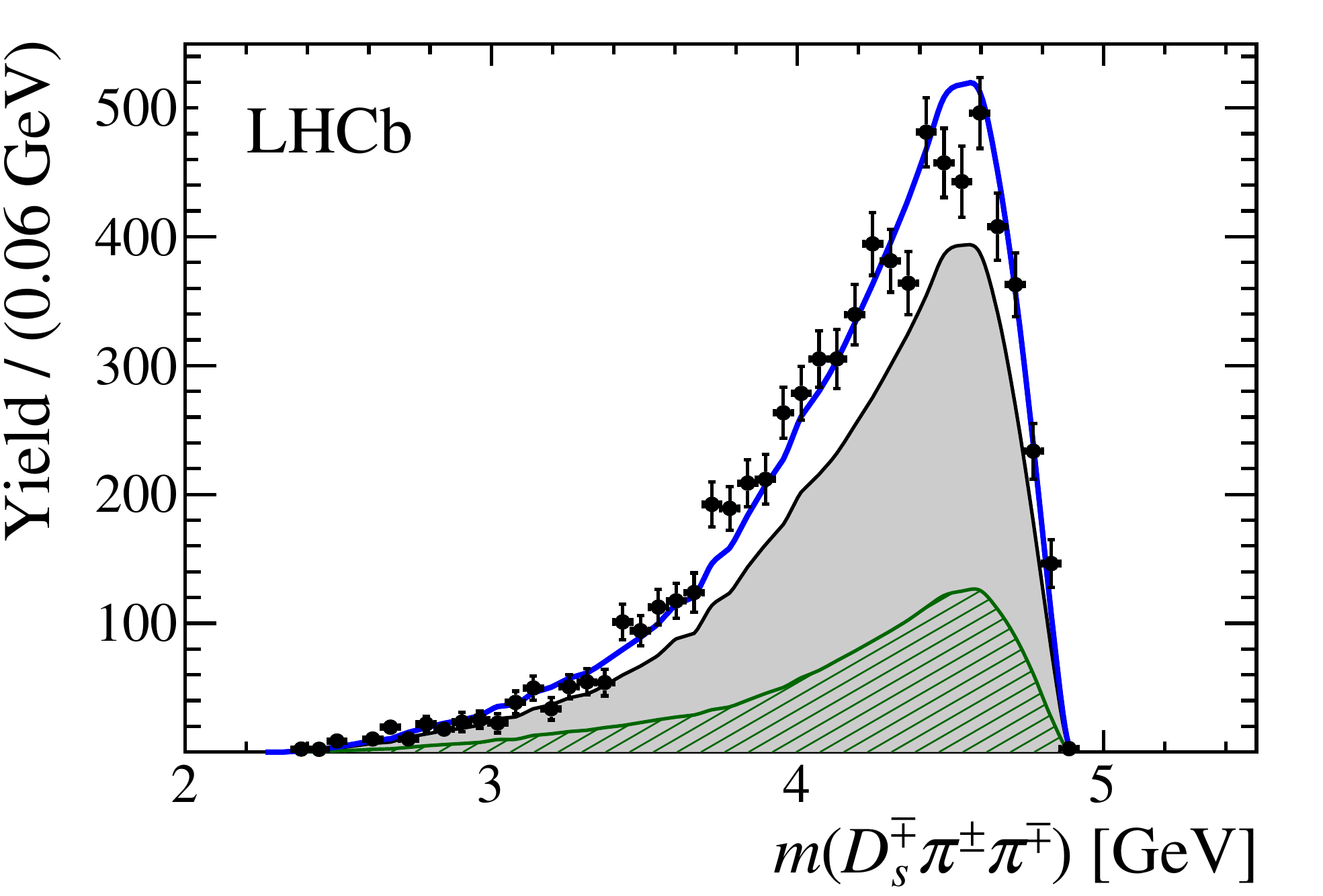}
		\includegraphics[width=0.4\textwidth, height = !]{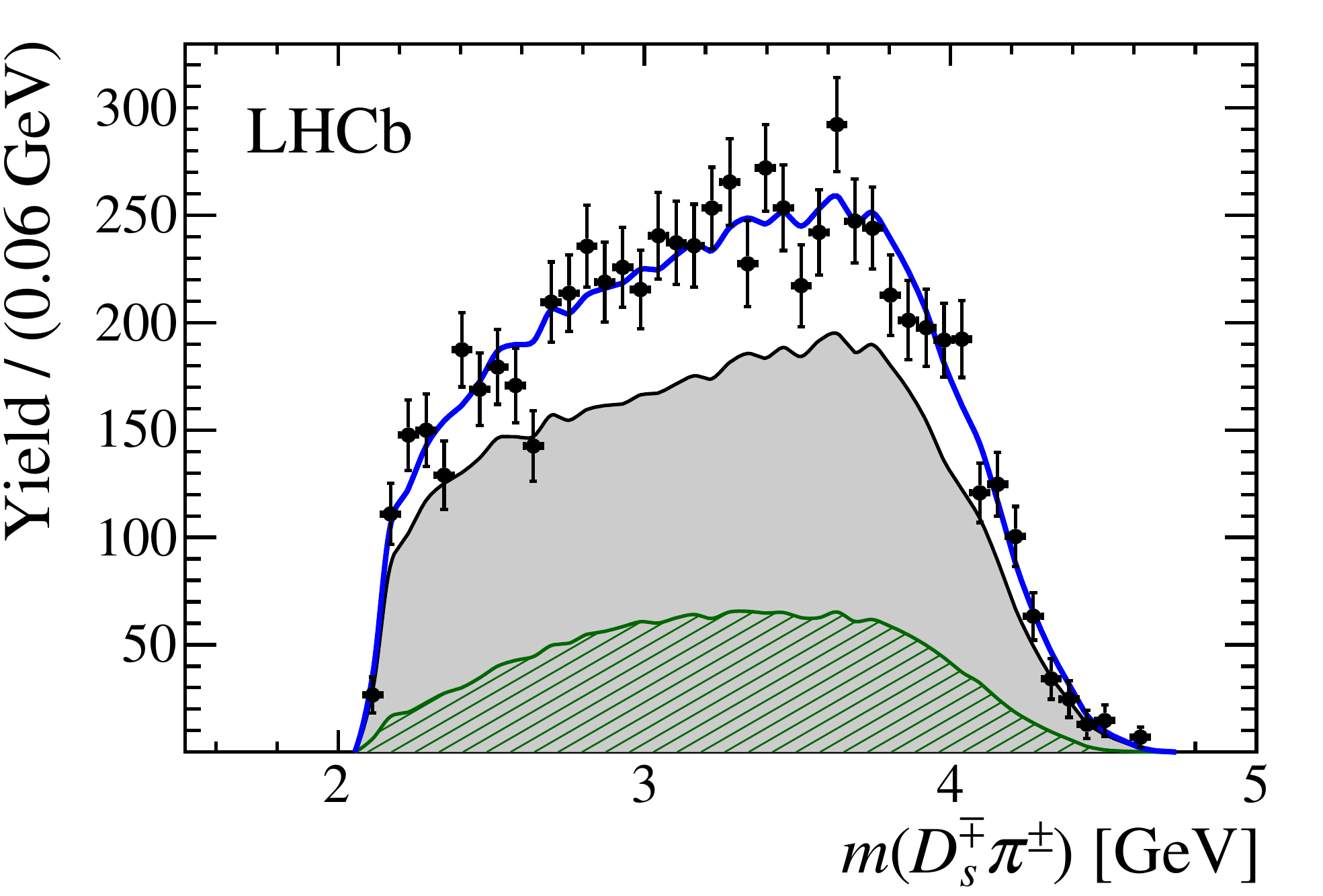}
		
		\caption{Invariant-mass distribution of background-subtracted \signal candidates (data points) and fit projections (blue solid line). Contributions from $b\to c$ and $b\to u$ decay amplitudes are overlaid.
		  } 		
		\label{fig:fullFit}
\end{figure}

\begin{figure}[h]
\centering
   		\includegraphics[width=0.4\textwidth, height = !]{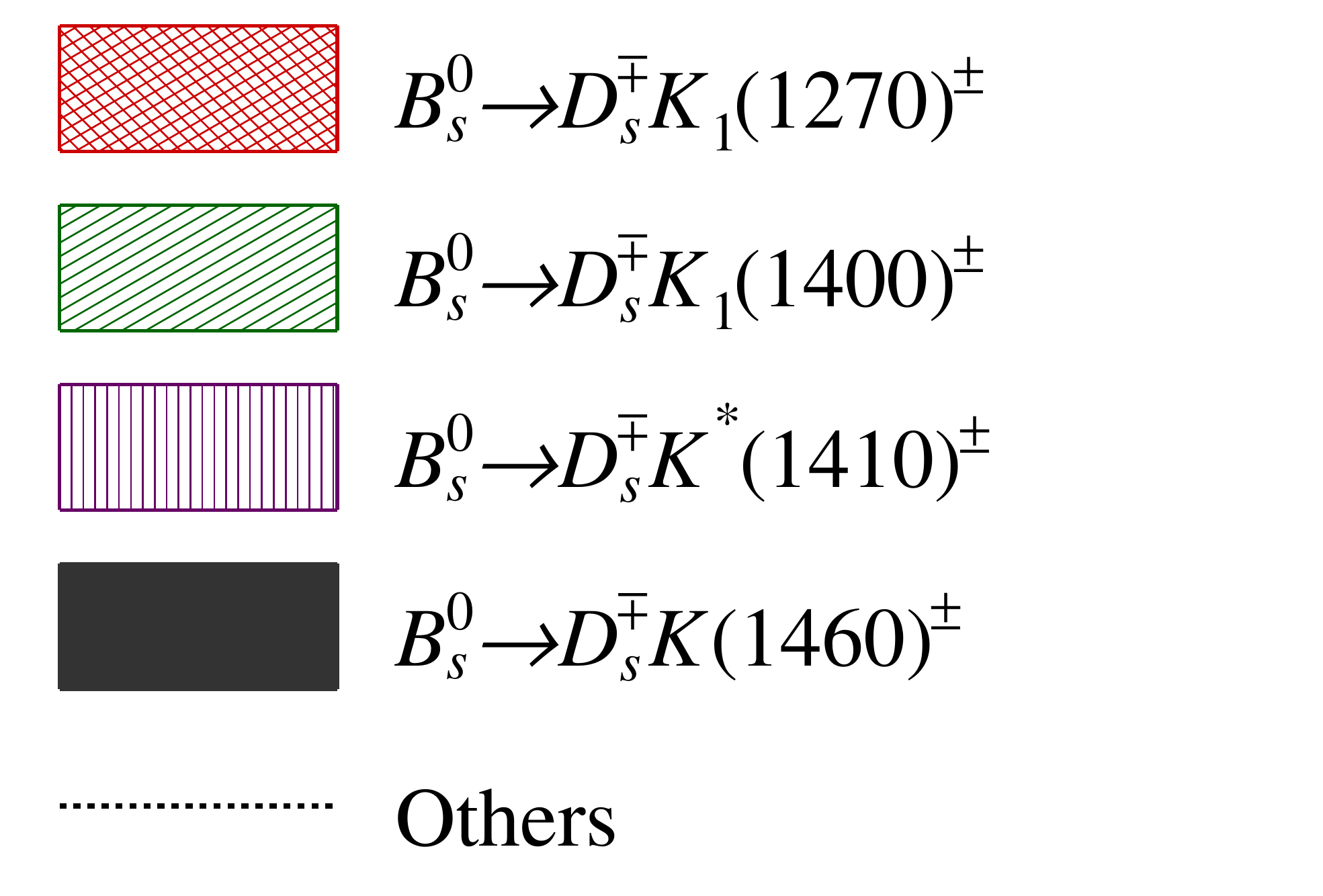} 		
		\includegraphics[width=0.4\textwidth, height = !]{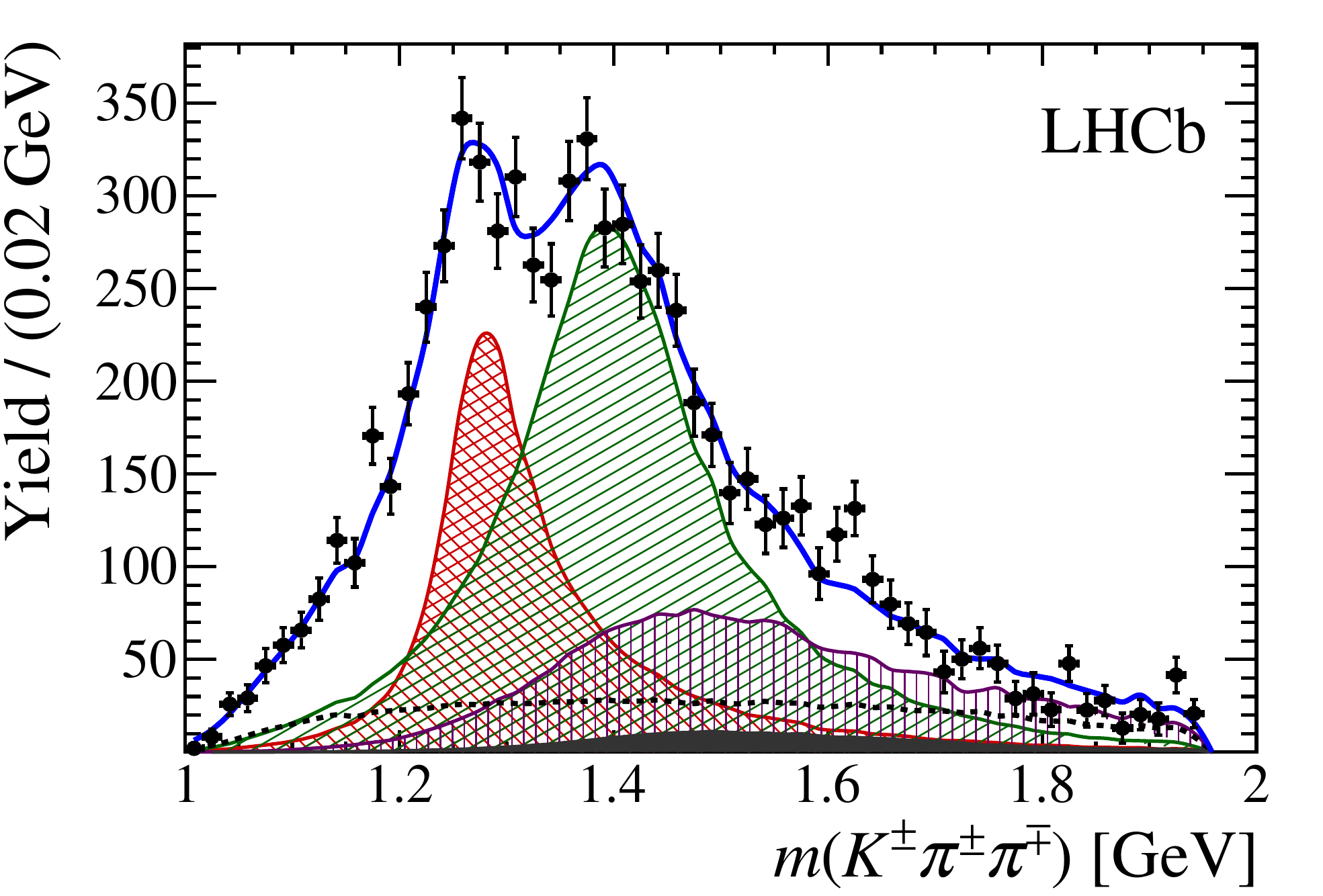} 	
			
		\includegraphics[width=0.4\textwidth, height = !]{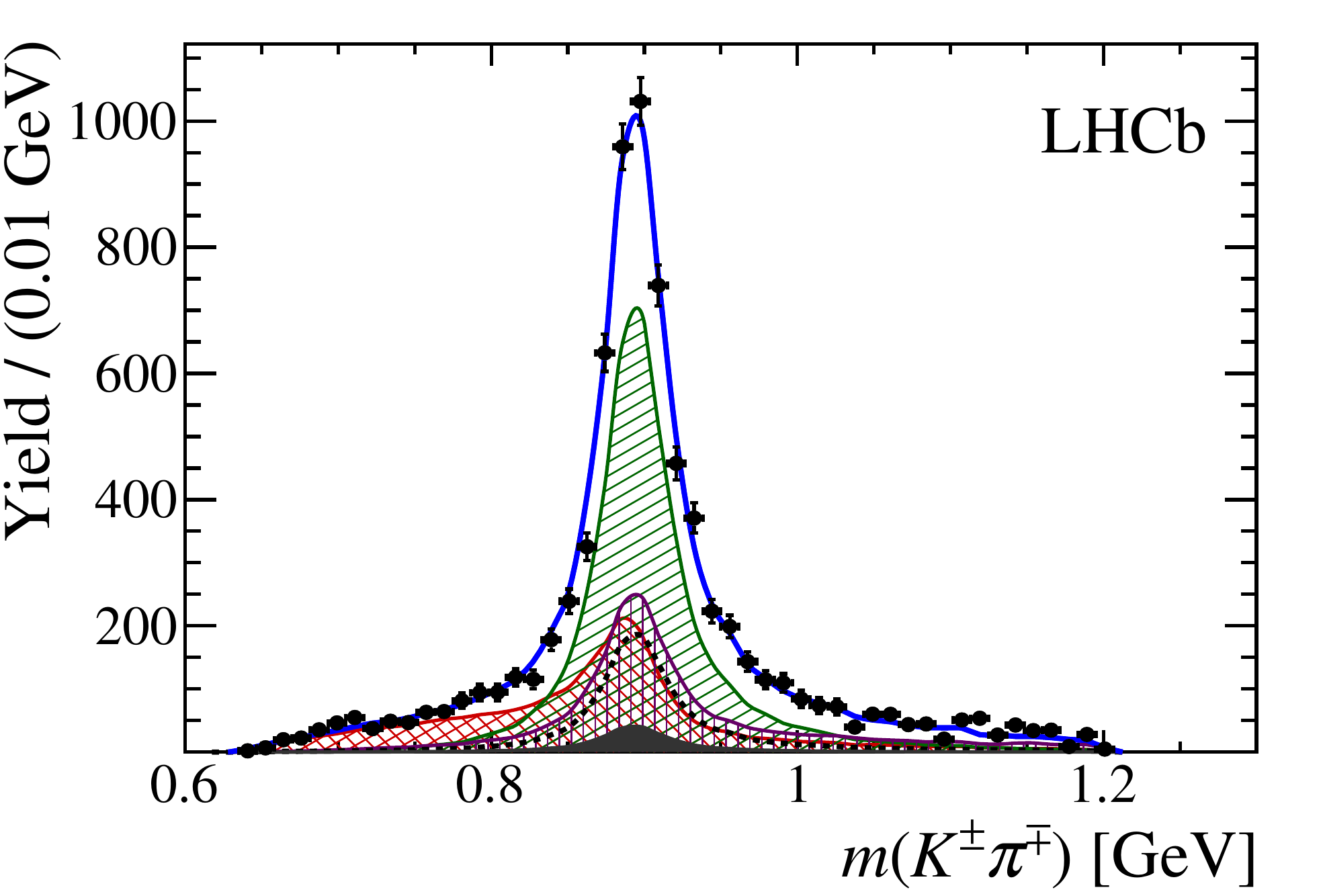}
		\includegraphics[width=0.4\textwidth, height = !]{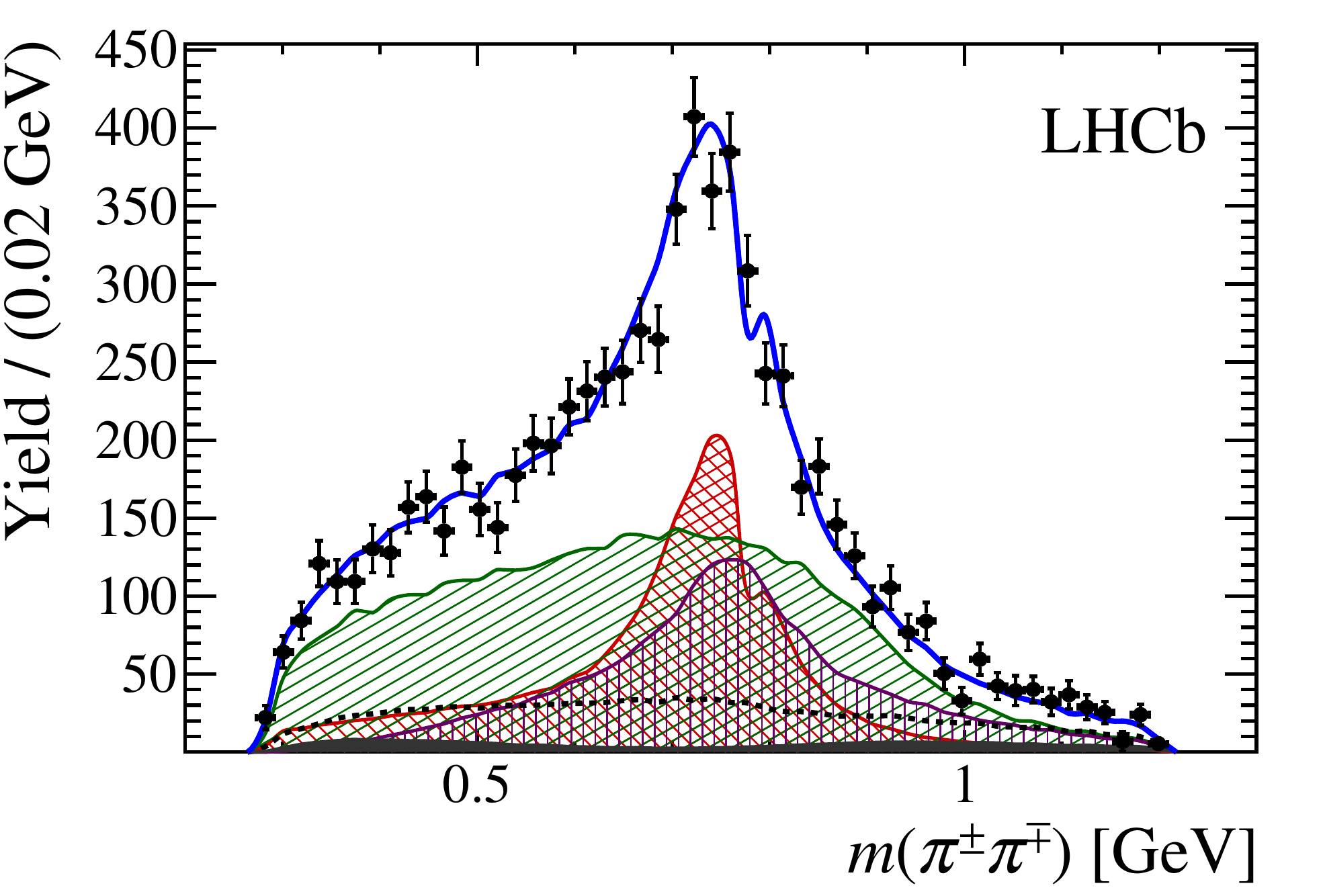}
		
		\includegraphics[width=0.4\textwidth, height = !]{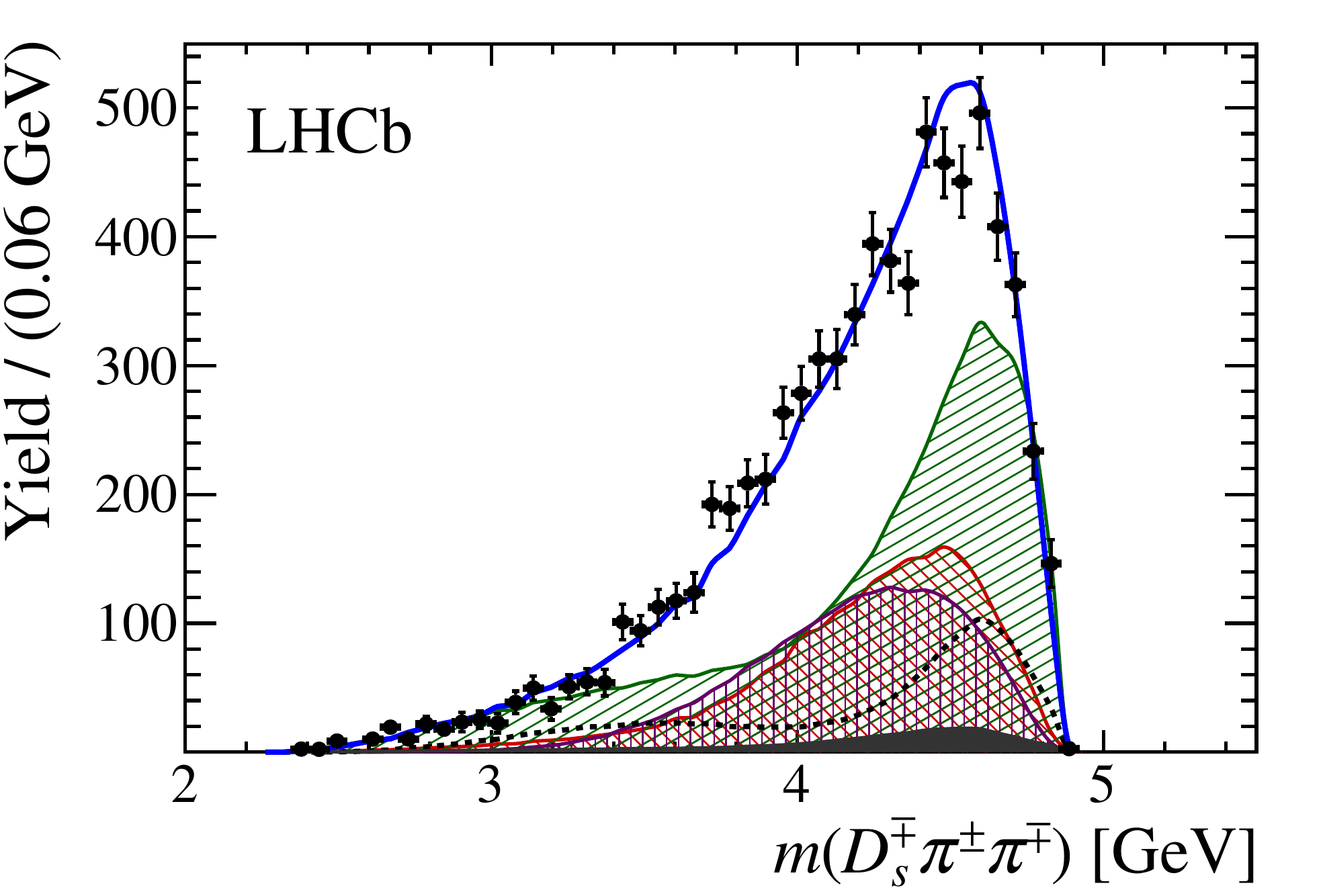}
		\includegraphics[width=0.4\textwidth, height = !]{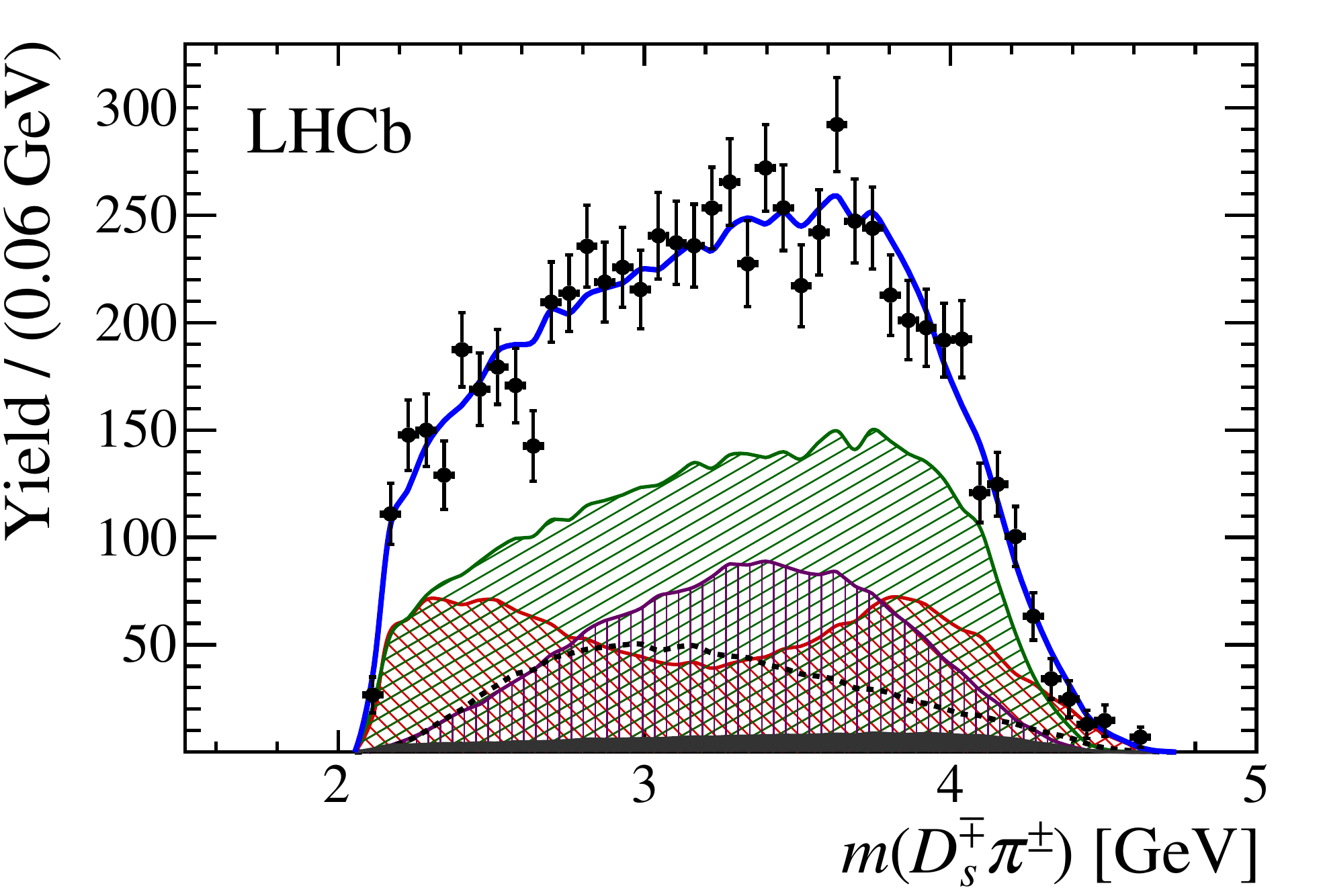}
		
		\caption{Invariant-mass distribution of background-subtracted \signal candidates (data points) and fit projections (blue solid line).  Incoherent contributions from intermediate-state components are overlaid.
		  } 		
		\label{fig:fullFit2}
\end{figure}

Table~\ref{tab:fullResult} and~\ref{tab:fullResultCascade} list the moduli and phases of the complex amplitude coefficients
obtained by fitting the baseline model to the background-subtracted \signal signal candidates.
The corresponding decay fractions for the $b\to c$ and $b\to u$ amplitudes
are given in Table~\ref{tab:fullFractions}.
The decay-time projection and mixing asymmetries shown in Fig.~\ref{fig:fullFitTime} are  consistent with those of the phase-space integrated fit in Fig.~\ref{fig:tFitSig}.
Invariant-mass projections are shown in Figs.~\ref{fig:fullFit} and~\ref{fig:fullFit2}
indicating that the model provides a reasonable description of the data.
Decays via $b \to c$ quark-level transitions are found to be dominated by the axial-vector states $K_1(1270)^+$ and $K_1(1400)^+$.
These resonances are produced by the external weak current (see Fig.~\ref{fig:decay_feynman}).
The sub-leading contribution comes from the vector resonance $K^*(1410)^+$.
In $b \to u$ quark-level transitions, the excited kaon states are produced by the spectator-quark interaction.
Here, no clear hierarchy is observed. There are  sizeable contributions from the axial-vector resonances but also from the pseudoscalar state $K(1460)^+$ and from the quasi-two-body
process $\Bs \to ( \Dsmp \, \pipm)_{P} \, \, K^{*}(892)^0$,
where $( \Dsmp \, \pipm)_{P}$ denotes a non-resonant two-particle system in a
$P$-wave ($L=1$) configuration.
Interference fractions of the $b\to c$ and $b\to u$ intermediate-state amplitudes are given in
Tables~\ref{tab:intFracDsKpipi}
and~\ref{tab:intFracDsKpipi2}.
Sizeable interference effects between the decay modes
$\Bs \to \Dsmp \, ( K_1(1270)^\pm \to K^{*}(892)^0 \, \pipm )$,
\mbox{$\Bs \to \Dsmp \, ( K_1(1400)^\pm \to K^{*}(892)^0 \, \pipm )$}
and $\Bs \to ( \Dsmp \, \pipm)_{P} \, \, K^{*}(892)^0$
are observed since
the overlap of their phase-space distributions is significant.
A net constructive (destructive) interference effect of all amplitude components 
of around $+26\%$ ($-12\%$) remains for $b\to c$ ($b\to u$) quark-level transitions when integrated over the phase space.

\begin{table}[t]
\centering
\caption{
Decay fractions of the intermediate-state amplitudes contributing to decays via $b \to c$ and $b \to u$ quark-level transitions.
The uncertainties are statistical, systematic
and due to alternative amplitude models considered.
}
\resizebox{\linewidth}{!}{
	\renewcommand{\arraystretch}{1.1}
\begin{tabular}{l
r@{$\,\pm\,$}c@{$\,\pm\,$}c@{$\,\pm$\,}l
r@{$\,\pm\,$}c@{$\,\pm\,$}c@{$\,\pm$\,}l
}
\hline
\hline
\multicolumn{1}{c}{Decay channel} & \multicolumn{4}{c}{$F^{c}_i [\%]$} & \multicolumn{4}{c}{$F^{u}_i [\%]$}  \\
\hline
$\Bs \to \Dsmp \, ( K_1(1270)^\pm \to K^{*}(892)^0 \, \pipm )$ & 13.0 & 2.4 & 2.7 & 3.4 & 4.1 & 2.2 & 2.9 & 2.6 \\
$\Bs \to \Dsmp \, ( K_1(1270)^\pm \to \Kpm \, \rho(770)^0 )$ & 16.0 & 1.4 & 1.8 & 2.1 & 5.1 & 2.2 & 3.5 & 2.0 \\
$\Bs \to \Dsmp \, ( K_1(1270)^\pm \to K^{*}_{0}(1430)^0 \, \pipm )$ & 3.4 & 0.5 & 1.0 & 0.4 & 1.1 & 0.5 & 0.6 & 0.5 \\
$\Bs \to \Dsmp \, ( K_1(1400)^\pm \to K^{*}(892)^0 \, \pipm )$ & 63.9 & 5.1 & 7.4 & 13.5 & 19.3 & 5.2 & 8.3 & 7.8 \\
$\Bs \to \Dsmp \, ( K^{*}(1410)^\pm \to K^{*}(892)^0 \, \pipm )$ & 12.8 & 0.8 & 1.5 & 3.2 & 12.6 & 2.0 & 2.6 & 4.1 \\
$\Bs \to \Dsmp \, ( K^{*}(1410)^\pm \to \Kpm \, \rho(770)^0 )$ & 5.6 & 0.4 & 0.6 & 0.7 & 5.6 & 1.0 & 1.2 & 1.8 \\
$\Bs \to \Dsmp \, ( K(1460)^\pm \to K^{*}(892)^0 \, \pipm )$ & \multicolumn{4}{c}{} & 11.9 & 2.5 & 2.9 & 3.1 \\
$\Bs \to ( \Dsmp \, \pipm)_{P} \, \, K^{*}(892)^0$ & 10.2 & 1.6 & 1.8 & 4.5 & 28.4 & 5.6 & 6.4 & 15.3 \\
$\Bs \to ( \Dsmp \, \Kpm)_{P} \, \, \rho(770)^0$ & 0.9 & 0.4 & 0.5 & 1.0 &  \multicolumn{4}{c}{}  \\
\hline
$\text{Sum}$ & 125.7 & 6.4 & 6.9 & 19.9 & 88.1 & 7.0 & 10.0 & 20.9 \\
\hline
\hline
\end{tabular}
 }
\label{tab:fullFractions}
\end{table}

The mass and width of the $K_1(1400)^+$ and $K^*(1410)^+$ resonances
are determined from the fit to be
\begin{align}
	m_{K_1(1400)} &= (1406 \pm 7 \pm 6 \pm 11) \mev, &
	\Gamma_{K_1(1400)} &=  (195 \pm 11 \pm 12 \pm 16) \mev, \nonumber \\
	m_{K^{*}(1410)} &= (1433 \pm 10 \pm 23 \pm 8) \mev, &
	\Gamma_{K^{*}(1410)} &= (402 \pm 24 \pm 47 \pm 22) \mev, \nonumber
\end{align}
in good agreement with the PDG average values~\cite{PDG20}.
The uncertainties are statistical,  systematic and due to alternative models considered as detailed in Sec.~\ref{sec:Systematics}.
The ratio of the
$B_s^0 \to D_s^- K^+ \pip \pim$ and $\bar B_s^0 \to D_s^- K^+ \pip \pim$
decay amplitudes as well as their strong- and weak-phase difference are
measured to be
\begin{align}
r &=0.56 \pm 0.05  \pm 0.04  \pm 0.07 ,  \nonumber \\ 
\delta &= (-14 \pm 10  \pm 4  \pm 5)  \degrees, \nonumber \\
\gamma - 2 \beta_{s} &= (\phantom{-}42 \pm 10  \pm 4 \pm 5)  \degrees,  \nonumber
\end{align}
where the angles are given modulo $180 \degrees$.
The coherence factor is computed
by numerically integrating over the phase space using the baseline model
resulting in
\begin{align}
\kappa = 0.72 \pm 0.04 \stat \pm 0.06 \syst \pm 0.04 \,(\text{model}). \nonumber
\end{align}

\clearpage

\section{Systematic uncertainties}
\label{sec:Systematics}

The systematic uncertainties on the measured observables are summarised
in Table~\ref{tab:normSys} for the decay-time fits to \control
and \mbox{\signal} decays
and in Tables~\ref{tab:sigSys} and~\ref{tab:sigSys3} for the time-dependent amplitude fit to \mbox{\signal} decays.
The various sources of systematic uncertainties are described in the following.

The overall fit procedure is tested by generating pseudoexperiments
from the default fit model using the measured values and subsequently fitting them with the same model.
For each pseudoexperiment and fit parameter, a pull is calculated by dividing the difference between the fitted
and generated values by the statistical uncertainty.
The means of the pull distributions are assigned as systematic uncertainties due to an intrinsic fit bias.
A closure test using a large sample of fully simulated signal candidates
shows a
non-significant bias
for the determination of $\Delta m_s$,
which is assigned as a systematic uncertainty.

The statistical subtraction of the residual background relies on the correct description of the reconstructed $m(D_s^\mp h^\pm \pi^\pm \pi^\mp)$ invariant mass distribution.
Alternative parameterisations are tested for the signal and each background component. The yields of the cross-feed contributions to the \signal candidates estimated from a combination of simulated data and control modes are fixed to zero or doubled.
The standard deviations of the obtained differences to the default fit values are assigned as a systematic uncertainty due to the background subtraction.
The background subtraction technique relies on the assumption of  independence between the reconstructed $m(D_s^\mp h^\pm \pi^\pm \pi^\mp)$ invariant mass and the observables
that are used in the final fit.
The impact of ignoring the small correlation between reconstructed mass and decay time
observed for the combinatorial background
is determined with pseudoexperiments
in which the correlation is included at generation and neglected
in the fit.

\begin{table}[t]
\centering
\caption{Systematic uncertainties on the $\Bs$ mixing frequency determined from the fit to \control signal candidates and
on the fit parameters of the phase-space integrated fit to \signal signal candidates in units of the statistical standard deviations.}
	\renewcommand{\arraystretch}{1.1}

\begin{tabular}{l | c |  c  c  c  c  c   c }
\hline
\hline
Systematic  & $\Delta m_{s}$ & $C_{f}$ & $A^{\Delta\Gamma}_{f}$ & $A^{\Delta\Gamma}_{\bar{f}}$ & $S_{f}$ & $S_{\bar{f}}$ \\
\hline
Fit bias ($+$ closure test) & 0.39 & 0.09 & 0.00 & 0.01 & 0.02 & 0.11 \\
Background subtraction & 0.18 & 0.17 & 0.19 & 0.18 & 0.11 & 0.06 \\
Correlations & 0.09 & 0.11 & 0.32 & 0.24 & 0.04 & 0.06 \\
Acceptance & 0.02 & 0.03 & 0.19 & 0.22 & 0.02 & 0.02 \\
Resolution & 0.16 & 0.24 & 0.01 & 0.03 & 0.12 & 0.10 \\
Decay-time bias & 1.00 & 0.06 & 0.00 & 0.00 & 0.10 & 0.08 \\
Nuisance asymmetries & 0.02 & 0.06 & 0.06 & 0.06 & 0.06 & 0.05 \\
VELO $z$ scale & 0.26 & & & & & \\
VELO alignment & 0.44 & & & & & \\
$\Delta m_{s}$ & & 0.02 & 0.01 & 0.01 & 0.12 & 0.12 \\
\hline
Total & 1.21 & 0.34 & 0.42 & 0.38 & 0.24 & 0.23 \\
\hline
\hline
\end{tabular}
 \label{tab:normSys}
\end{table}

The systematic uncertainties related to the decay-time acceptance as well as due to the limited experimental knowledge of $\Gamma_s$ and $\Delta\Gamma_s$ are studied simultaneously. Pseudoexperiments are generated assuming the default configuration
and subsequently fitted under both this default and an alternative configuration in which the acceptance parameters together with $\Gamma_s$ and $\Delta\Gamma_s$ are randomised within their uncertainties (taking their correlation into account).
The bias of the mean of the resulting pull distribution is added in quadrature to the pull width in order to arrive at the final systematic uncertainty.

Systematic effects originating from the calibration of the decay-time uncertainty estimate are studied with two alternative parameterisations which either slightly overestimate or underestimate the decay-time resolution.
Due to the high correlation between the decay-time resolution and the calibration of the flavour taggers, their systematic uncertainty is studied simultaneously.
As a first step, the decay-time fit to the \control candidates is repeated using the alternative decay-time error calibration functions.
New tagging calibration parameters are obtained
which are then used (together with the respective decay-time error calibration function) for the fits to the \signal candidates.
The largest deviations of the central values from their default values are assigned as a systematic uncertainty for each fit parameter.
A systematic uncertainty due to the limited knowledge of the decay-time bias which is fixed in the fit is evaluated by
randomising the value within its uncertainty.

The systematic uncertainty from the production and detection asymmetries and $\Delta m_s$ (in case of \signal decays) which are fixed in the fit are evaluated by means of pseudoexperiments, analogously to the procedure performed for the decay-time acceptance.

The precision with which the $\Bs$ flight distance
can be determined is limited by the knowledge of
the overall length of the VELO detector (VELO $z$ scale) and the position of the individual VELO modules (VELO alignment).
This VELO-reconstruction uncertainty 
translates into
a relative uncertainty on $\Delta m_s$ of $0.02\%$~\cite{LHCB-PAPER-2013-006} with other parameters being unaffected.
In the fit to the \signal candidates, the
VELO-reconstruction uncertainty is then implicitly included in the systematic
error due to the $\Delta m_s$ uncertainty described above.

\begin{table}[t]
\centering
\caption{Systematic uncertainties on the physical observables and resonance parameters determined from
the full time-dependent amplitude fit to \signal data in units of the statistical standard deviations. The systematic uncertainties for the amplitude coefficients are given in Table~\ref{tab:sigSys3}.
}
\resizebox{\linewidth}{!}{
	\renewcommand{\arraystretch}{1.2}

\begin{tabular}{l |  c  c  c  c | c  c  c   c }
\hline
\hline
Systematic  & $m_{K_1(1400)} $ & $\Gamma_{K_1(1400)}$ & $m_{K^{*}(1410)}$ & $\Gamma_{K^{*}(1410)}$ & $r$ & $\delta$ & $\gamma - 2 \beta_{s}$ \\
\hline
Fit bias & 0.00 & 0.14 & 0.14 & 0.42 & 0.06 & 0.13 & 0.13 \\
Background subtraction & 0.15 & 0.28 & 0.15 & 0.41 & 0.15 & 0.15 & 0.22 \\
Correlations & 0.23 & 0.27 & 0.18 & 0.07 & 0.13 & 0.05 & 0.08 \\
Time acceptance & 0.01 & 0.01 & 0.00 & 0.00 & 0.06 & 0.04 & 0.07 \\
Resolution & 0.11 & 0.33 & 0.09 & 0.03 & 0.30 & 0.27 & 0.26 \\
Decay-time bias & 0.00 & 0.02 & 0.01 & 0.00 & 0.02 & 0.12 & 0.01 \\
Nuisance asymmetries & 0.00 & 0.01 & 0.00 & 0.00 & 0.01 & 0.07 & 0.03 \\
$\Delta m_{s}$ & 0.00 & 0.01 & 0.00 & 0.00 & 0.03 & 0.06 & 0.03 \\
Phase-space acceptance & 0.17 & 0.34 & 0.32 & 0.21 & 0.35 & 0.08 & 0.06 \\
Acceptance factorisation & 0.25 & 0.41 & 0.76 & 0.36 & 0.49 & 0.08 & 0.05 \\
Lineshapes & 0.52 & 0.53 & 0.43 & 0.29 & 0.34 & 0.14 & 0.13 \\
Resonances $m,\Gamma$ & 0.07 & 0.06 & 0.02 & 0.02 & 0.02 & 0.01 & 0.01 \\
Form factors & 0.51 & 0.60 & 2.03 & 1.74 & 0.12 & 0.13 & 0.07 \\
Amplitude model & 1.59 & 1.50 & 0.77 & 0.88 & 1.60 & 0.51 & 0.47 \\ \hline
Total & 1.80 & 1.86 & 2.38 & 2.10 & 1.79 & 0.67 & 0.63 \\
\hline
\hline
\end{tabular}
 }
\label{tab:sigSys}
\end{table}

The treatment of the phase-space acceptance relies on simulated data.
The integration error due to the limited size of the simulated sample
used to normalise the signal PDF is below $0.2\%$ and thus negligibly small.
To assess the uncertainty due to possible data-simulation differences,
alternative phase-space acceptances are derived by varying the selection requirements (for the simulated sample only) on quantities that
are expected not to be well described by the simulation.
It is assumed that the phase-space acceptance is independent of the decay time. This assumption is tested by using only simulated candidates within specific decay-time intervals to calculate the MC normalisation integrals. Four approximately equally populated decay-time intervals are chosen and the sample variance of the fitted values is assigned as a systematic uncertainty.

The lineshape parameterisations for the $\rho(770)^0$ and $K_0^*(1430)^0$ resonances are replaced by a relativistic Breit--Wigner propagator given by Eq.~(\ref{eq:gamma2}) as part of the systematic studies.
Moreover, energy-dependent decay widths of  three-body resonances are recomputed taking only the dominant $\Kpm\pipm\pimp$ decay mode into account.
For each  alteration, a  time-dependent amplitude fit is performed and
the standard deviation of the obtained fit results is assigned as a systematic uncertainty.
Systematic uncertainties due to fixed resonance masses and widths
are computed with the same procedure used for the other fixed parameters
mentioned above.
Similarly, pseudoexperiments are performed in which
the Blatt--Weisskopf radial parameter $r_{\mathrm{BW}}$
(set to $1.5 \gev^{-1}$ by default)
is varied uniformly within the interval $[0,3]\gev^{-1}$
to assign a systematic uncertainty due to the form-factor modelling.

Several modifications to the baseline model
are tested to assign an additional uncertainty
due to the choice of amplitude components:
	all amplitudes selected by the first stage of the model selection are included for both $b\to c$ and $b\to u$ transitions,
	the amplitudes with the smallest decay fraction are removed,
	additional sub-decay modes of selected three-body resonances are considered,
	higher orbital angular momentum states are included where applicable,
	the orbital angular momentum state of non-resonant two-body states is set to other allowed values,
additional cascade and quasi-two-body amplitudes (which were removed by the first stage of the model selection) are considered.
In total, twelve amplitude models with similar fit quality as the baseline model are identified.
The fit results for those are summarised in Tables~\ref{tab:altModelsDsKpipi} and~\ref{tab:altModelsDsKpipi2}.
The largest deviations from the baseline values for $r, \delta $ and $\gamma-2\beta$
are $0.19, 10\degrees$ and $12\degrees$
observed for alternative models 8, 1 and 11, respectively.
The standard deviation of the twelve fit results is taken as the model uncertainty.
No model uncertainty is assigned to the amplitude coefficients since they are, by definition, parameters of a given model.

No statistically significant effect on the results is observed when repeating the analysis on subsets of the data, 
splitting by data-taking period or tagging category (OS-tagged, only SS-tagged or both OS- and SS-tagged). 

\section{Results}
\label{ssec:results}

To interpret the parameters determined in the model-independent fit, \mbox{$P_{\text{obs}} \equiv (C_f,A^{\Delta\Gamma}_f,A^{\Delta\Gamma}_{\bar f},S_f, S_{\bar f})$},
in terms of the physical observables $\Lambda \equiv (r,\kappa,\delta,\gamma-2\beta_s)$,
the equations for the \CP coefficients in terms of these physical variables reported in  Eq.~(\ref{eq:CPcoeff}), $P(\Lambda)$,
need to be inverted.
This is accomplished by minimising the likelihood function~\cite{GammaCombo, *LHCb-PAPER-2016-032}
\begin{equation}
	-2 \mathcal L (\Lambda) = -2 \, \exp\left( -\frac{1}{2} (P(\Lambda) - P_{\text{obs}})^T \, V^{-1} \, (P(\Lambda) - P_{\text{obs}}) \right).
\end{equation}
Here, $V$ denotes the experimental (statistical and systematic) covariance matrix
of the measured observables,
see Appendix~\ref{a:CPcoeff}.
Figure~\ref{fig:FitCL} displays the
confidence levels (CL)
for the physical parameters $\Lambda$ obtained from the model-independent method,
where the physical boundary of the coherence factor
is enforced.
The $1-\text{CL} = 68.3\%$ ($1\sigma$) confidence intervals are given in Table~\ref{tab:ResultSummary} together with the results of the full time-dependent amplitude analysis.
Considering the difference in statistical sensitivity of the two methods and that the model-dependent uncertainty only affects the
full time-dependent amplitude fit, a good agreement between the measurements is observed.
As a cross-check, pseudoexperiments are performed to study
the distribution of the test statistic
\mbox{$Q^2 = \sum_i (\Lambda_i^{\text{MI}} - \Lambda_i^{\text{MD}})^2/ (\sigma_{\text{stat}}(\Lambda_i^{\text{MI}})^2 - \sigma_{\text{stat}}(\Lambda_i^{\text{MD}})^2)$},
where $\Lambda_i^{\text{MI}}$ ($\Lambda_i^{\text{MD}}$) and $\sigma_{\text{stat}}(\Lambda_i^{\text{MI}})$ ($\sigma_{\text{stat}}(\Lambda_i^{\text{MD}})$)
denote the measured value of the physical observable $\Lambda_i$ and its statistical uncertainty
obtained with the model-independent (model-dependent) method.
It is found that $p=33\%$ of the pseudoexperiments have a larger $Q^2$ value than observed on data,
considering only the statistical uncertainty.
The p-value increases to $p=49\%$ when the uncertainty due to the amplitude modelling is included.

The measured ratio of the $b\to u$ and $b\to c$ decay amplitudes is qualitatively consistent with the naive expectation based on the involved CKM elements ($r \approx 0.4$). Note that the parameters $r,\kappa$ and $\delta$ are determined in a limited phase-space region
(\cf Sec.~\ref{sec:Selection}) and might differ when the full phase space is considered.

\begin{figure}[h]
	\centering
		\includegraphics[width=0.425\textwidth, height = !]{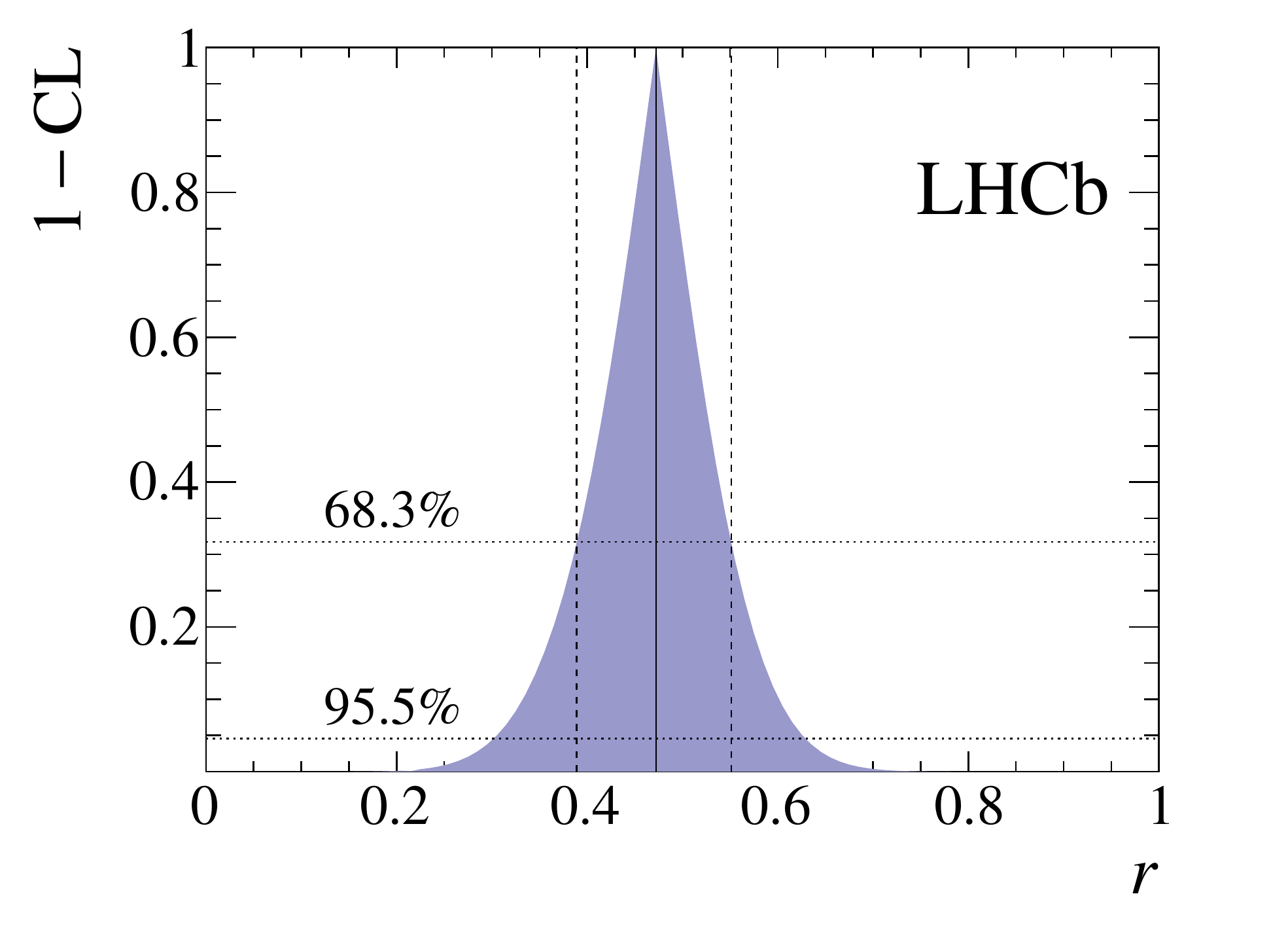}
		\includegraphics[width=0.425\textwidth, height = !]{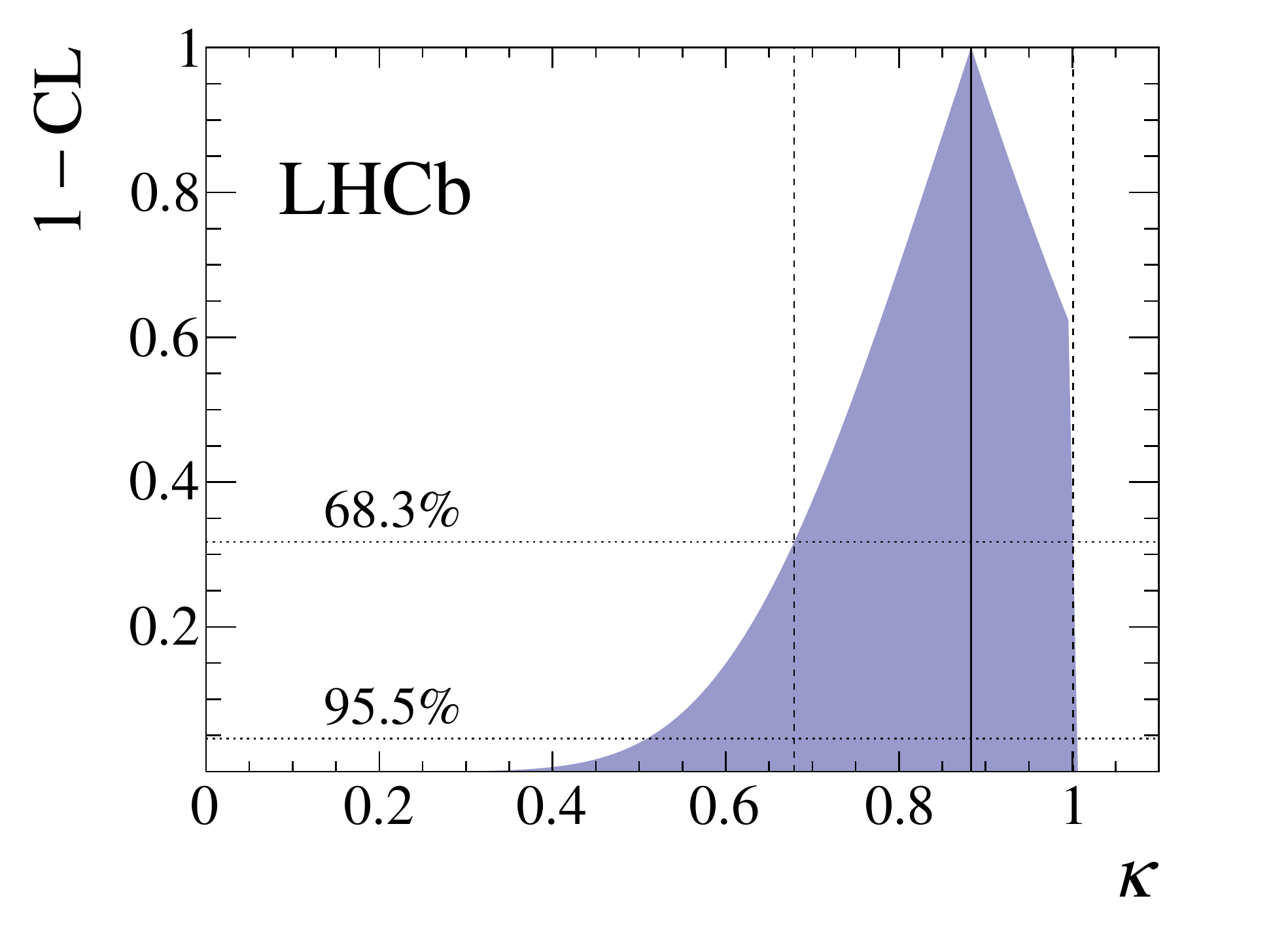}
		
		\includegraphics[width=0.425\textwidth, height = !]{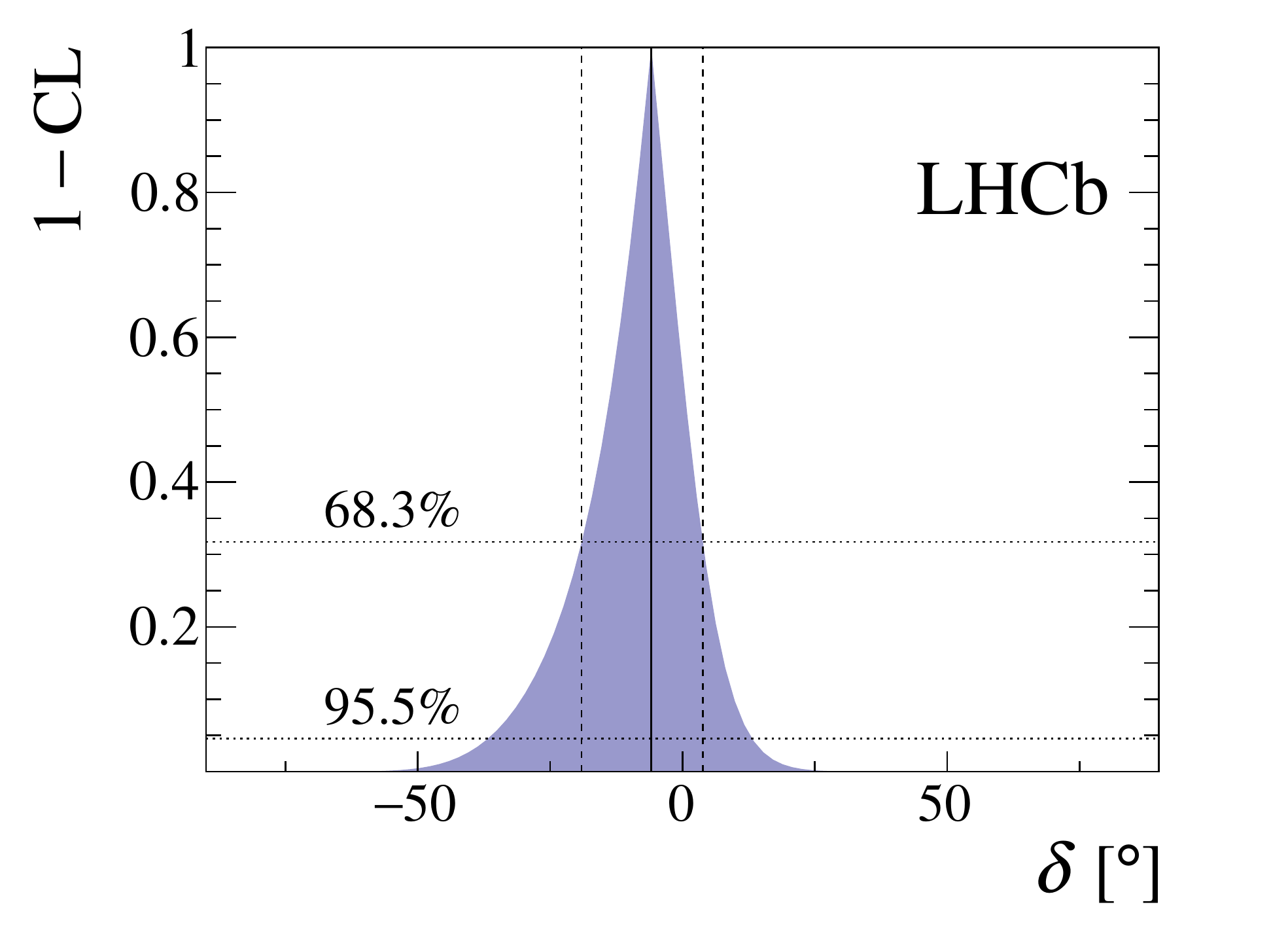}
		\includegraphics[width=0.425\textwidth, height = !]{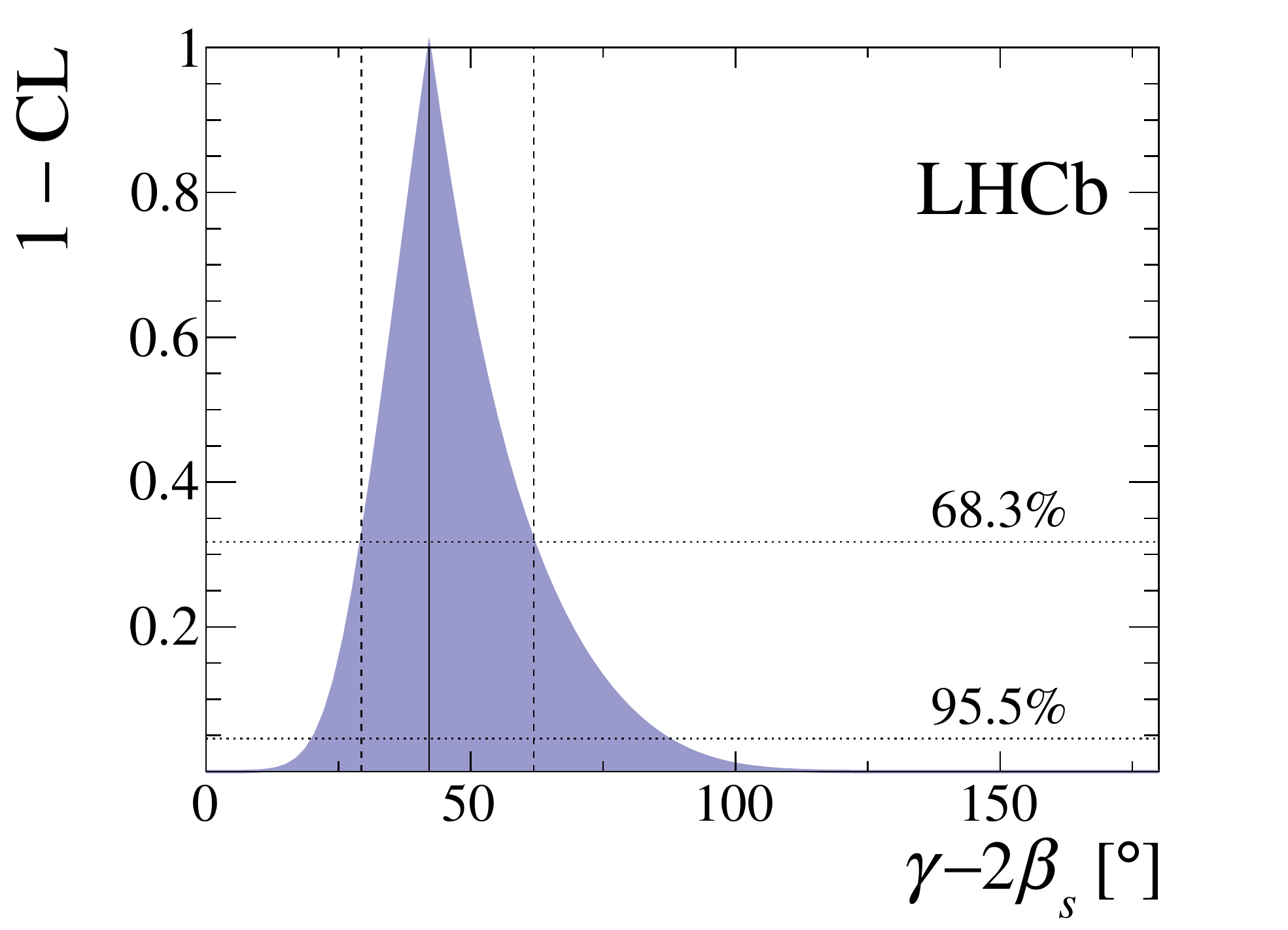}
		\caption{The 1$-$CL contours for the physical observables $r,\kappa,\delta$ and $\gamma-2\beta_s$ obtained with the model-independent fit. }
		\label{fig:FitCL}	
\end{figure}
\begin{table}[h]
\centering
\caption{Parameters determined from the
model-independent
and model-dependent
fits to the \signal signal candidates.
The uncertainties are statistical, systematic
and (if applicable) due to alternative amplitude models considered.
The angles are given modulo $180 \degrees$.
 }
	\renewcommand{\arraystretch}{1.2}
\begin{tabular}{l
r@{}c@{}l
r@{$\,\pm\,$}c@{$\,\pm\,$}c@{$\,\pm\,$}l  }
\hline
\hline
\multicolumn{1}{c}{Parameter} & \multicolumn{3}{c}{ Model-independent} & \multicolumn{4}{c}{Model-dependent}  \\
\hline
$r$ & $0.47$ &${}^{\,+\,0.08}_{\,-\,0.08}$ &${}^{\,+\,0.02}_{\,-\,0.03}$  & 0.56 & 0.05 & 0.04 & 0.07 \\
$\kappa$ & $0.88$ &${}^{\,+\,0.12}_{\,-\,0.19}$ &${}^{\,+\,0.04}_{\,-\,0.07}$  & 0.72 & 0.04 & 0.06 & 0.04  \\
$\delta$ [$\degrees$] &  $-6$ &${}^{\,+\,10}_{\,-\,12}$ &${}^{\,+\,2}_{\,-\,4}$ & $-14$ & 10 & 4 & 5 \\
$\gamma - 2 \beta_s$ [$\degrees$] & $42$ & ${}^{\,+\,19}_{\,-\,13} $ & ${}^{\,+\,6}_{\,-\,2} $ & 42 & 10 & 4 & 5 \\
\hline
\hline
\end{tabular}
\label{tab:ResultSummary}
\end{table}

\section{Conclusion}

Mixing-induced \CP violation in \signal decays is studied for the first time using $9\invfb$ of proton-proton collision data recorded by the LHCb detector.
A time-dependent amplitude analysis is performed, disentangling
the various intermediate-state components contributing
via $b \to c$ or $b \to u$ quark-level transitions.
The \CP-violating weak phase $\gamma-2\beta_s$ is measured
and used to determine the CKM angle $\gamma$
by taking the mixing phase,
$-2\beta_s= \phi_s = (-2.35 \pm 1.43)^\circ$~\cite{LHCb-PAPER-2019-013},
as an external input.
This results in a measured value of $\gamma = (44\pm 12)^\circ$ modulo 180\degrees.
A model-independent fit to the phase-space integrated
decay-time spectrum yields a compatible, but statistically less precise, value of
$\gamma = (44^{\,+\,20}_{\,-\,13})^\circ$ modulo 180\degrees.
These results correspond to $4.4\sigma$ and $4.6\sigma$
evidence for mixing-induced \CP violation
and agree with the world-average value of the CKM angle $\gamma$~\cite{PDG20,HFLAV18} within
$2.2\sigma$
and $1.4\sigma$
for the model-dependent and  model-independent methods, respectively.
The $\Bs-\Bsb$ oscillation frequency is measured from flavour-specific \control decays to be
$\Delta m_s = (17.757 \pm 0.007 \pm 0.008)\invps$.
This is the most precise measurement of this quantity and
consistent with the world-average value~\cite{PDG20}
and theoretical predictions~\cite{King:2019lal}.
Both the measurement of the CKM angle $\gamma$ and of the $\Bs-\Bsb$ mixing frequency are vital inputs for global fits of the CKM matrix.

\section*{Acknowledgements}
\noindent We express our gratitude to our colleagues in the CERN
accelerator departments for the excellent performance of the LHC. We
thank the technical and administrative staff at the LHCb
institutes.
We acknowledge support from CERN and from the national agencies:
CAPES, CNPq, FAPERJ and FINEP (Brazil); 
MOST and NSFC (China); 
CNRS/IN2P3 (France); 
BMBF, DFG and MPG (Germany); 
INFN (Italy); 
NWO (Netherlands); 
MNiSW and NCN (Poland); 
MEN/IFA (Romania); 
MSHE (Russia); 
MICINN (Spain); 
SNSF and SER (Switzerland); 
NASU (Ukraine); 
STFC (United Kingdom); 
DOE NP and NSF (USA).
We acknowledge the computing resources that are provided by CERN, IN2P3
(France), KIT and DESY (Germany), INFN (Italy), SURF (Netherlands),
PIC (Spain), GridPP (United Kingdom), RRCKI and Yandex
LLC (Russia), CSCS (Switzerland), IFIN-HH (Romania), CBPF (Brazil),
PL-GRID (Poland) and OSC (USA).
We are indebted to the communities behind the multiple open-source
software packages on which we depend.
Individual groups or members have received support from
AvH Foundation (Germany);
EPLANET, Marie Sk\l{}odowska-Curie Actions and ERC (European Union);
A*MIDEX, ANR, Labex P2IO and OCEVU, and R\'{e}gion Auvergne-Rh\^{o}ne-Alpes (France);
Key Research Program of Frontier Sciences of CAS, CAS PIFI,
Thousand Talents Program, and Sci. \& Tech. Program of Guangzhou (China);
RFBR, RSF and Yandex LLC (Russia);
GVA, XuntaGal and GENCAT (Spain);
the Royal Society
and the Leverhulme Trust (United Kingdom).

\clearpage
\section*{Appendices}

\appendix

\section{Lineshapes}
\label{a:lineshapes}

\setcounter{table}{0}
\setcounter{equation}{0}

\renewcommand{\thetable}{A.\arabic{table}}
\renewcommand{\theequation}{A.\arabic{equation}}

\begin{figure}[h]
\centering
\includegraphics[height=!,width=0.49\textwidth]{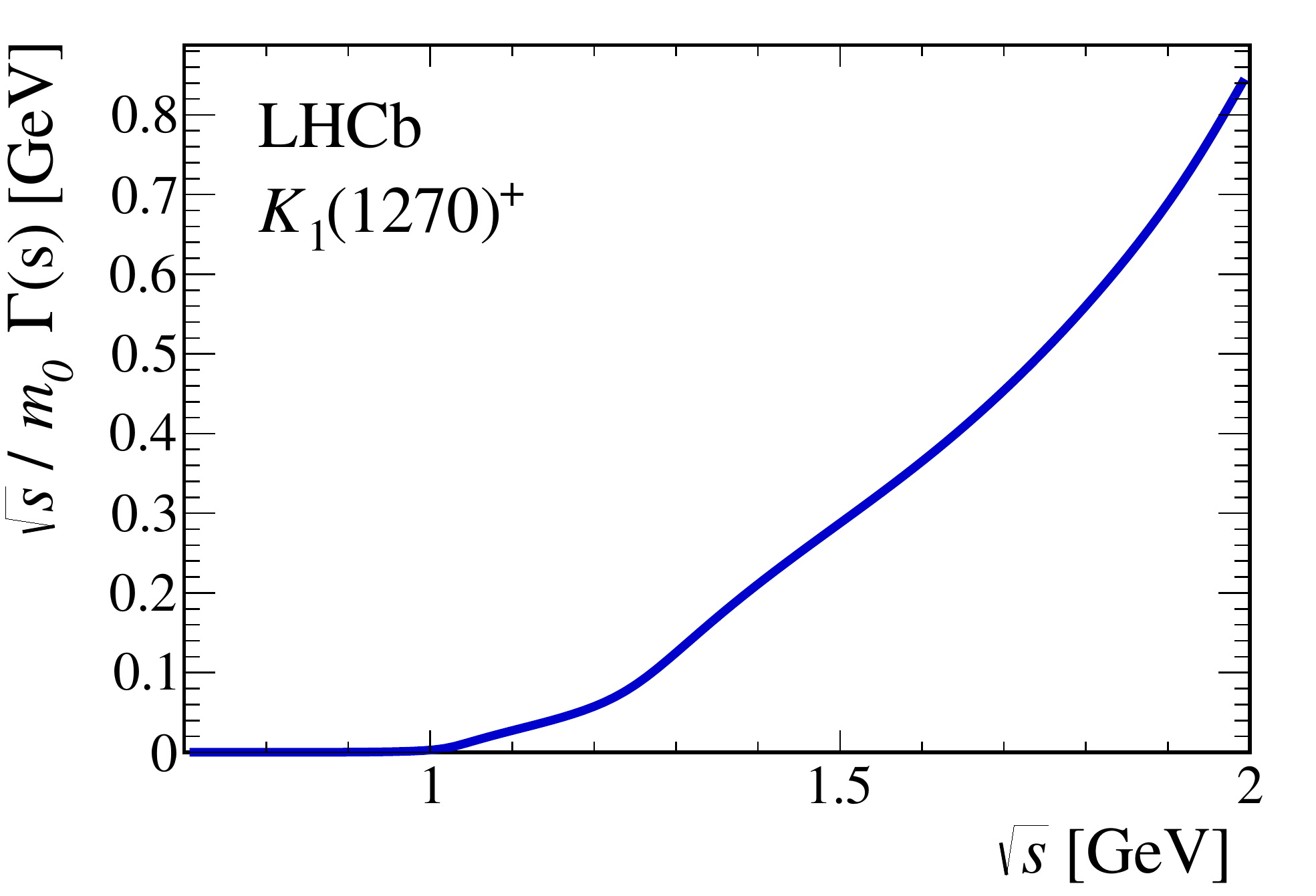}
\includegraphics[height=!,width=0.49\textwidth]{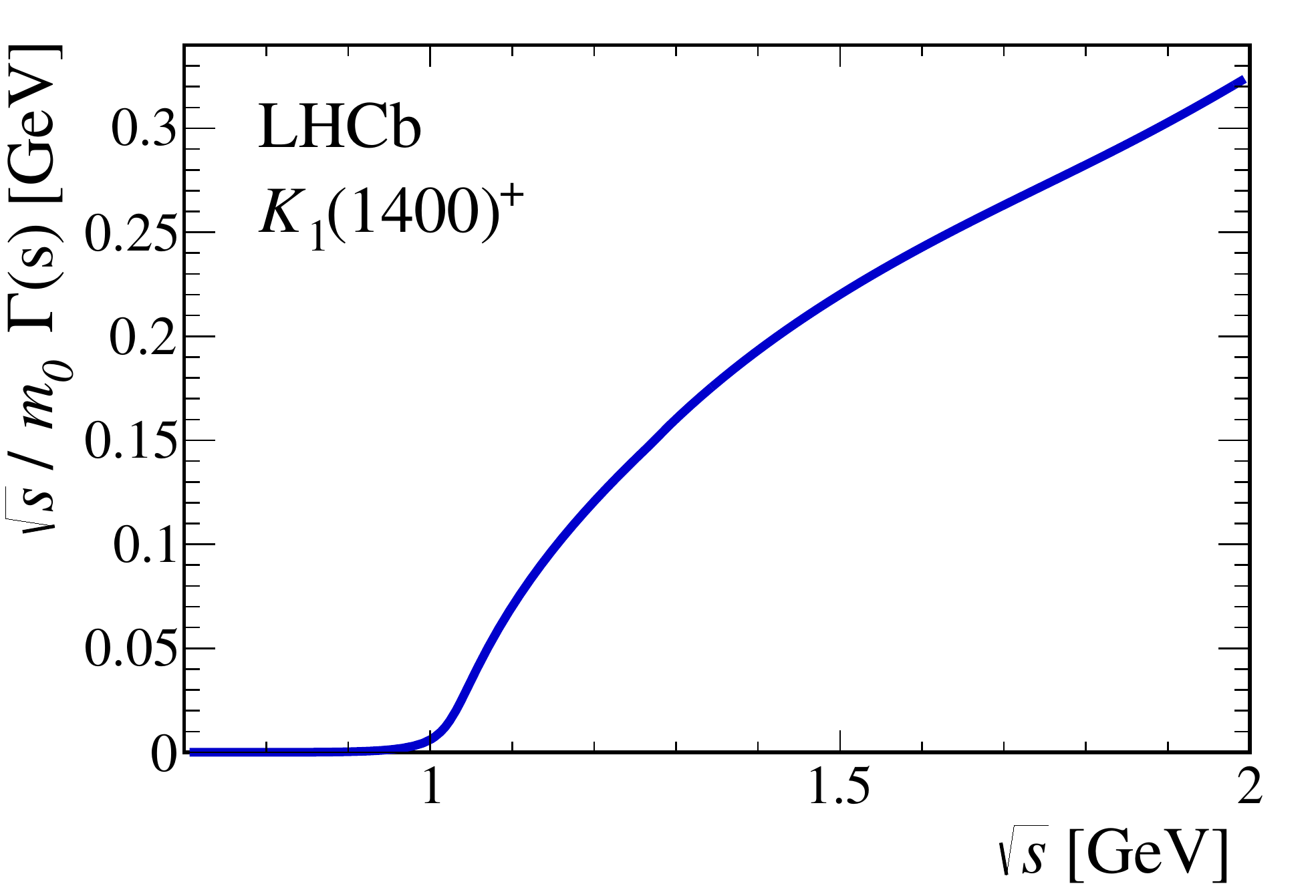}

\includegraphics[height=!,width=0.49\textwidth]{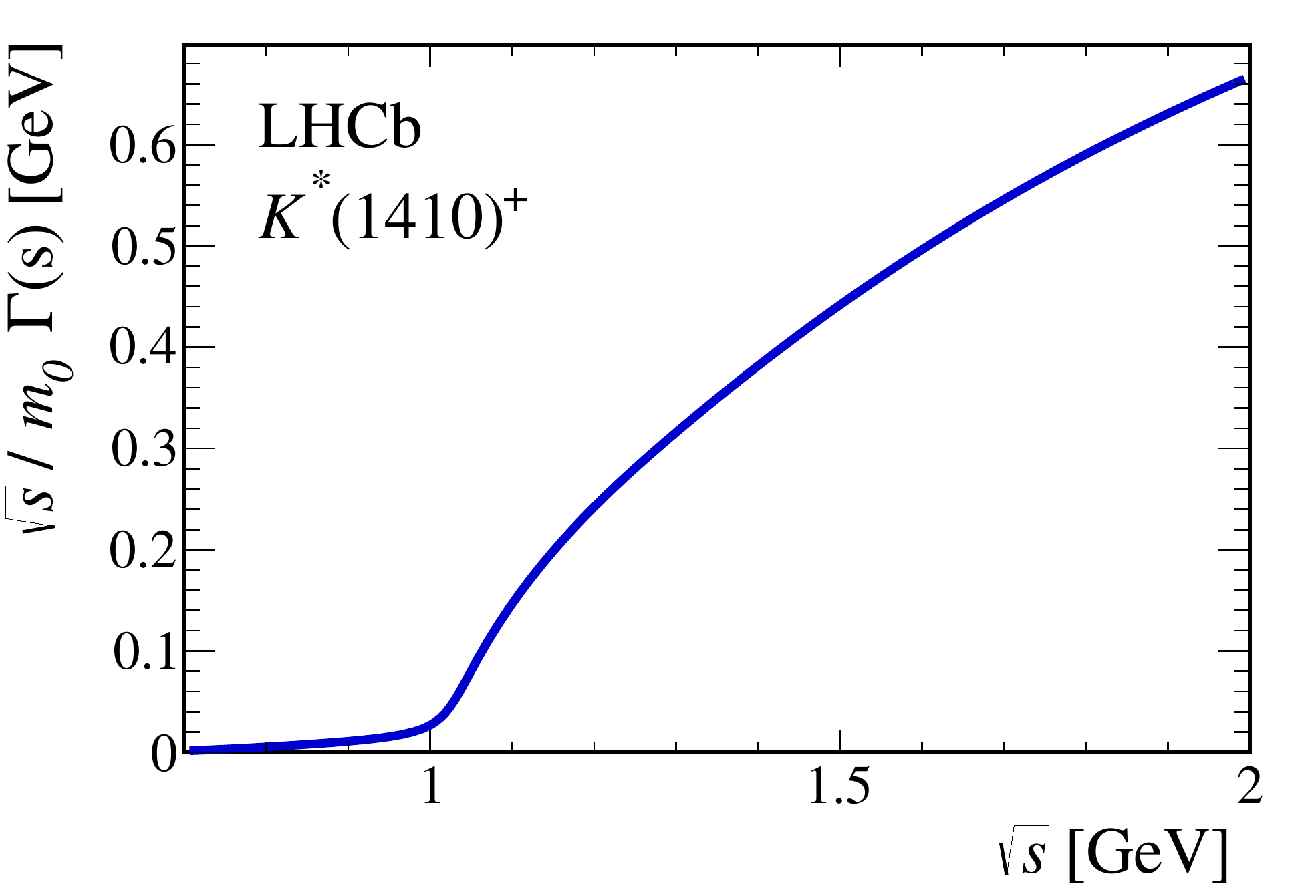}
\includegraphics[height=!,width=0.49\textwidth]{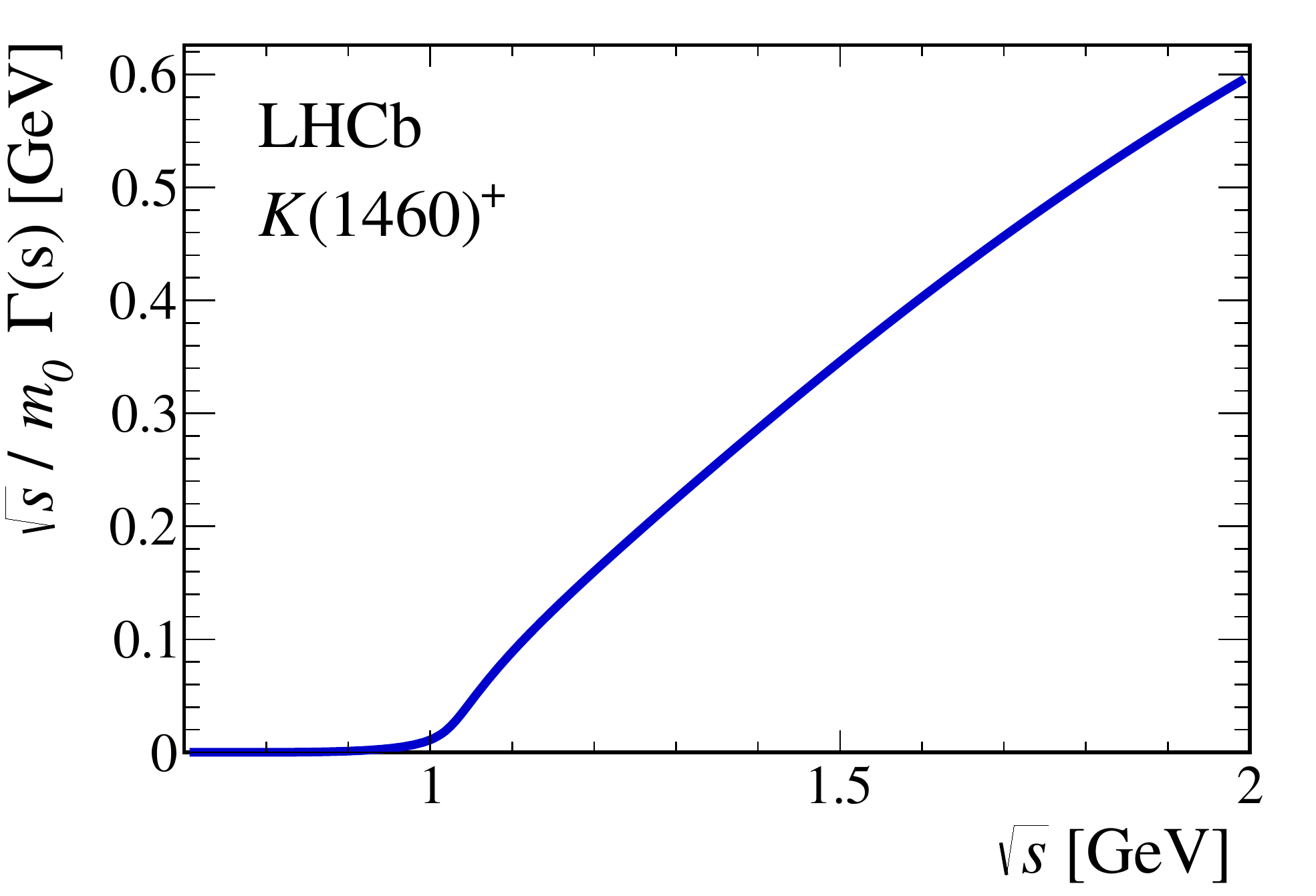}
\caption{Running width distributions of the three-body resonances included in the baseline model for \signal decays:
(top left) $K_1(1270)^+$, (top right) $K_1(1400)^+$,
(bottom left) $K^*(1410)^+$
and
(bottom right) $K(1460)^+$.
}
\label{fig:rw}
\end{figure}

Lineshape parameters for resonances contributing to \signal decays are fixed to the values given in Table~\ref{tab:resoParam}.
The running-width distributions of the three-body resonances are shown in Figure~\ref{fig:rw}.

\begin{table}[h]
\centering
\caption{Parameters of the resonances included in the \signal baseline model.}
\begin{tabular}{l r@{$\,\pm\,$}l  r@{$\,\pm\,$}l   c c}
\hline
\hline
\multicolumn{1}{c}{Resonance} &  \multicolumn{2}{c}{$m_0$ $[\mev]$} &  \multicolumn{2}{c}{$\Gamma_0$ $[\mev]$} & $J^P$ & Source \\
\hline
$\rho(770)^0$ & $775.26 $ & $  0.25$ & $149.1 $ & $ 0.8$ & $1^-$ & ~\cite{PDG20}\\
$K^*(892)^0$ & $895.55 $ & $ 0.20$ & $47.3 $ & $ 0.5$ & $1^-$ & ~\cite{PDG20}\\
$K_{1}(1270)^+$ & $1289.81 $ & $ 1.75$ & $116.11 $ & $ 3.4$ & $1^+$ &  ~\cite{LHCb-PAPER-2017-040}\\
$K(1460)^+$ & $1482.4 $ & $ 15.6 $ & $335.6  $ & $ 10.6$ & $0^-$ & ~\cite{LHCb-PAPER-2017-040}\\
$K_{0}^{*}(1430)^0$ & $1425 $ & $ 50$ & $270 $ & $ 80$ & $0^+$ & ~\cite{PDG20}\\
\hline
\hline
\end{tabular}
		\label{tab:resoParam}
\end{table}

\section{Amplitude model}
\label{a:altModelsGamma}
\setcounter{table}{0}
\setcounter{equation}{0}

\renewcommand{\thetable}{B.\arabic{table}}
\renewcommand{\theequation}{B.\arabic{equation}}

\begin{table}[b]
    \caption{Intermediate-state components considered for the \signal  LASSO model building procedure. The letters in square brackets and subscripts refer to the relative orbital angular momentum of
the decay products in spectroscopic notation.
   If no angular momentum is specified, the lowest angular momentum state compatible with angular momentum conservation and, where appropriate, parity conservation, is used.}
 \label{tab:ampsDsKpipiLASSO}
\footnotesize
	\renewcommand{\arraystretch}{1.2}
    \begin{subtable}{.4\linewidth}
      \centering
        \caption{\small Cascade decays}
        \begin{tabular}{ll}
        \hline
              $\Bs \to D_s^\mp \, [K_{1}(1270)^\pm[S,D]\to \pi^\pm \, K^{*}(892)^{0}]$ \\
              $\Bs \to D_s^\mp \, [K_{1}(1270)^\pm\to \pi^\pm \, K_0^{*}(1430)^{0}]$ \\
              $\Bs \to D_s^\mp \, [K_{1}(1270)^\pm\to K^\pm \, f_0(500)^{0}]$ \\
             $\Bs \to D_s^\mp \, [K_{1}(1270)^\pm\to K^\pm \, f_0(980)^{0}]$ \\
              $\Bs \to D_s^\mp \, [K_{1}(1270)^\pm[S,D]\to K^\pm \, \rho(770)^{0}]$ \\
              $\Bs \to D_s^\mp \, [K_{1}(1270)^\pm[S,D]\to K^\pm \, \rho(1450)^{0}]$ \\
              $\Bs \to D_s^\mp \, [K_{1}(1400)^\pm[S,D]\to \pi^\pm \, K^{*}(892)^{0}]$ \\
              $\Bs \to D_s^\mp \, [K_{1}(1400)^\pm[S,D]\to K^\pm \, \rho(770)^{0}]$ \\
              $\Bs \to D_s^\mp \, [K(1460)^\pm\to K^\pm \, f_0(500)^0]$ \\
              $\Bs \to D_s^\mp \, [K(1460)^\pm\to K^\pm \, \rho(770)^{0}]$ \\
              $\Bs \to D_s^\mp \, [K(1460)^\pm\to \pi^\pm \, K^{*}(892)^{0}]$ \\
              $\Bs \to D_s^\mp \, [K(1460)^\pm\to \pi^\pm \, K^{*}(1430)^{0}]$ \\
              $\Bs \to D_s^\mp \, [K^{*}(1410)^\pm\to \pi^\pm\, K^{*}(892)^{0}]$ \\
              $\Bs \to D_s^\mp \, [K^{*}(1410)^\pm\to K^\pm \, \rho(770)^{0}]$ \\
              $\Bs \to D_s^\mp \, [K_{2}^*(1430)^\pm\to \pi^\pm \, K^{*}(892)^{0}]$ \\
              $\Bs \to D_s^\mp \, [K_{2}^*(1430)^\pm\to K^\pm \, \rho(770)^{0}]$ \\
              $\Bs \to D_s^\mp \, [K_{1}(1650)^\pm \to \pi^\pm \, K^{*}(892)^{0}]$ \\
              $\Bs \to D_s^\mp \, [K_{1}(1650)^\pm \to K^\pm \, \rho(770)^{0}]$ \\
              $\Bs \to D_s^\mp \, [K^*(1680)^\pm\to \pi^\pm \, K^{*}(892)^{0}]$ \\
              $\Bs \to D_s^\mp \, [K^*(1680)^\pm\to K^\pm \, \rho(770)^{0}]$ \\
               $\Bs \to D_s^\mp \, [K_{2}(1770)^\pm\to \pi^\pm \, K^{*}(892)^{0}]$ \\
              $\Bs \to D_s^\mp \, [K_{2}(1770)^\pm\to K^\pm \, \rho(770)^{0}]$ \\
               $\Bs \to D_s^\mp \, [K_{2}(1770)^\pm\to \pi^\pm \, K_2^{*}(1430)^{0}]$ \\
              $\Bs \to D_s^\mp \, [K_{2}(1770)^\pm\to K^\pm \, f_2(1270)^0]$ \\
               $\Bs \to D_s^\mp \, [K_{2}^*(1980)^\pm\to \pi^\pm \, K^{*}(892)^{0}]$ \\
              $\Bs \to D_s^\mp \, [K_{2}^*(1980)^\pm\to K^\pm \, \rho(770)^{0}]$ \\
               $\Bs \to D_s^\mp \, [K_{2}^*(1980)^\pm\to \pi^\pm \, K_2^{*}(1430)^{0}]$ \\
              $\Bs \to D_s^\mp \, [K_{2}^*(1980)^\pm\to K^\pm \, f_2(1270)]$ \\
               $\Bs \to K^\pm [ D_{s1}(2536)^\mp\to  D_s^\mp \,  \, \rho(770)^{0}]$ \\
               $\Bs \to K^\pm [ D_{s1}(2536)^\mp\to  D_s^\mp \,  \, f_0(500)^0]$ \\
        \hline
	  \end{tabular}
    \end{subtable}%
    \hfill
    \begin{subtable}{.4\linewidth}
      \centering
        \caption{\small Quasi-two-body decays}
        \begin{tabular}{ll}
        \hline
              $\Bs \to f_0(500)^{0} (D_s^\mp K^\pm)_S$ \\
              $\Bs \to f_0(500)^{0} (D_s^\mp K^\pm)_P$ \\
              $\Bs \to f_0(980)^{0} (D_s^\mp K^\pm)_S$ \\
              $\Bs \to f_0(980)^{0} (D_s^\mp K^\pm)_P$ \\
              $\Bs \to \rho(770)^{0} (D_s^\mp K^\pm)_S$ \\
              $\Bs[S,P,D] \to \rho(770)^{0} \, (D_s^\mp K^\pm)_P$\\
              $\Bs \to  f_2(1270)^0 (D_s^\mp K^\pm)_S$ \\
              $\Bs \to  K^{*}(892)^{0} \, (D_s^\mp \pi^\pm)_S$ \\
              $\Bs[S,P,D]  \to  K^{*}(892)^{0} \, (D_s^\mp \pi^\pm)_P$ \\
              $\Bs \to  K_0^{*}(1430)^{0} \, (D_s^\mp \pi^\pm)_S$ \\
              $\Bs \to  K_2^{*}(1430)^{0} \, (D_s^\mp \pi^\pm)_S$ \\

              $\Bs \to (D_s^\mp K^\pm)_{S} \, (\pi^\pm \pi^\mp)_{S}$ \\
        \hline
              \\
              \\
              \\
              \\
              \\
              \\
              \\
              \\
              \\
              \\
              \\
              \\
              \\
              \\
              \\
              \\
              \\
              \\
	        \end{tabular}
    \end{subtable}
\end{table}

The full list of intermediate-state amplitudes considered for the model building procedure is given in Table~\ref{tab:ampsDsKpipiLASSO}.
In addition to those, the presence of unknown states decaying into $D_s^- \pi^+ \pi^-$, $D_s^- K^+ \pi^-$, $D_s^- \pi^+$ or $D_s^- K^+$ is investigated.
These states are typically fitted with huge decay widths (effectively indistinguishable from a non-resonant state) and small decay fractions. Therefore,  such states (if existent in the relevant mass range) are not resolvable with the current data sample.

\begin{figure}[p]
		\includegraphics[width=0.32\textwidth, height = !]{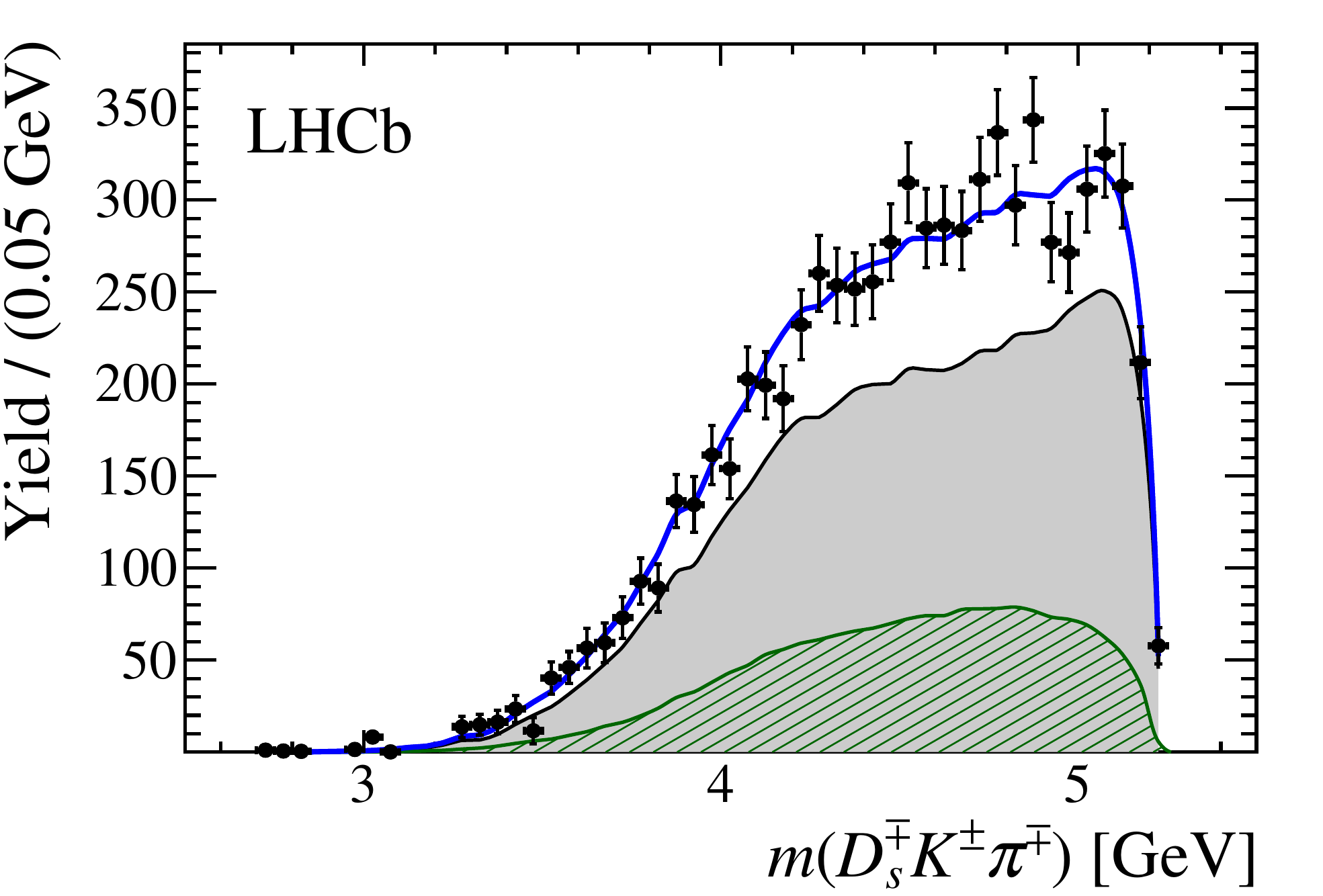} 	
		\includegraphics[width=0.32\textwidth, height = !]{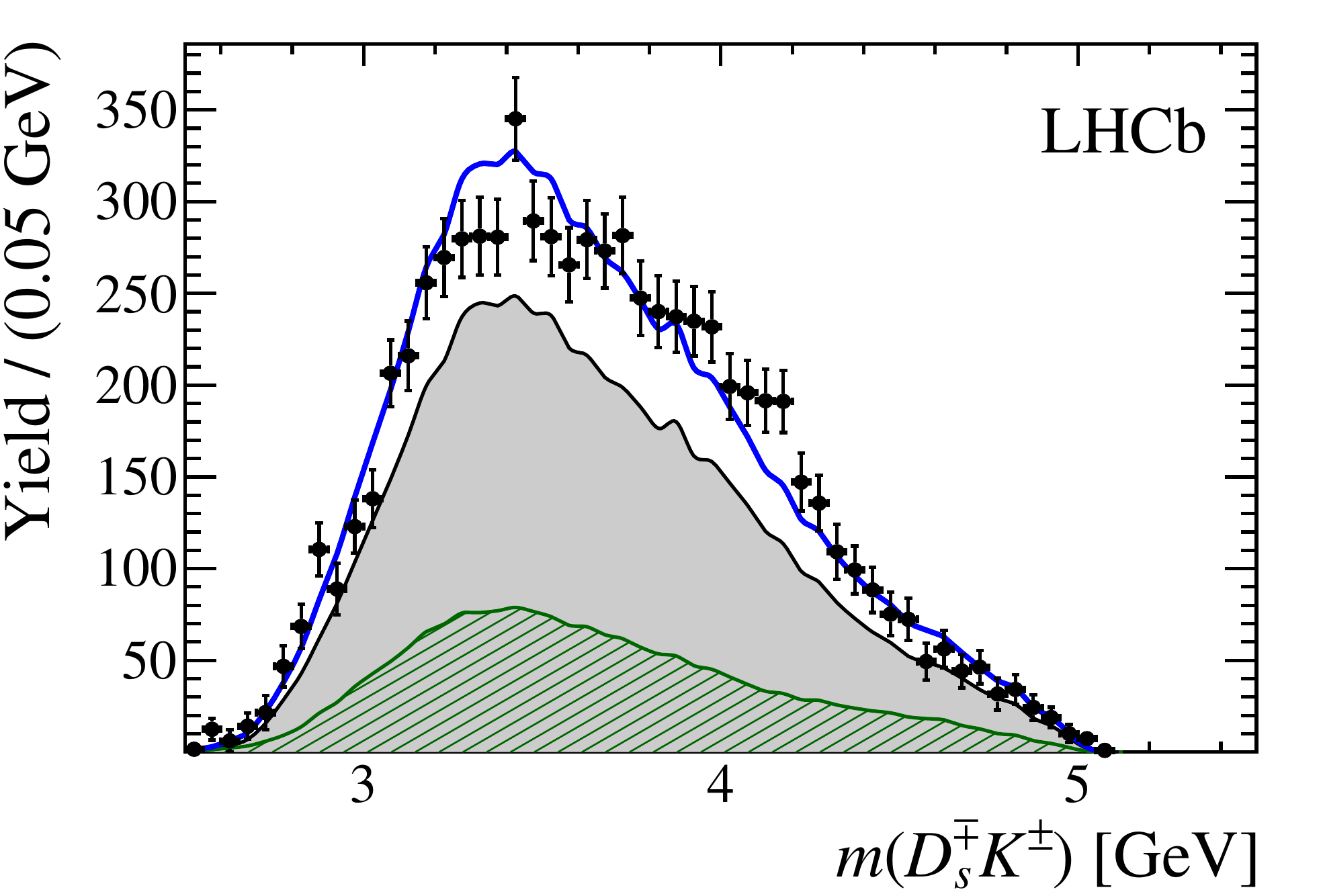}
		\includegraphics[width=0.32\textwidth, height = !]{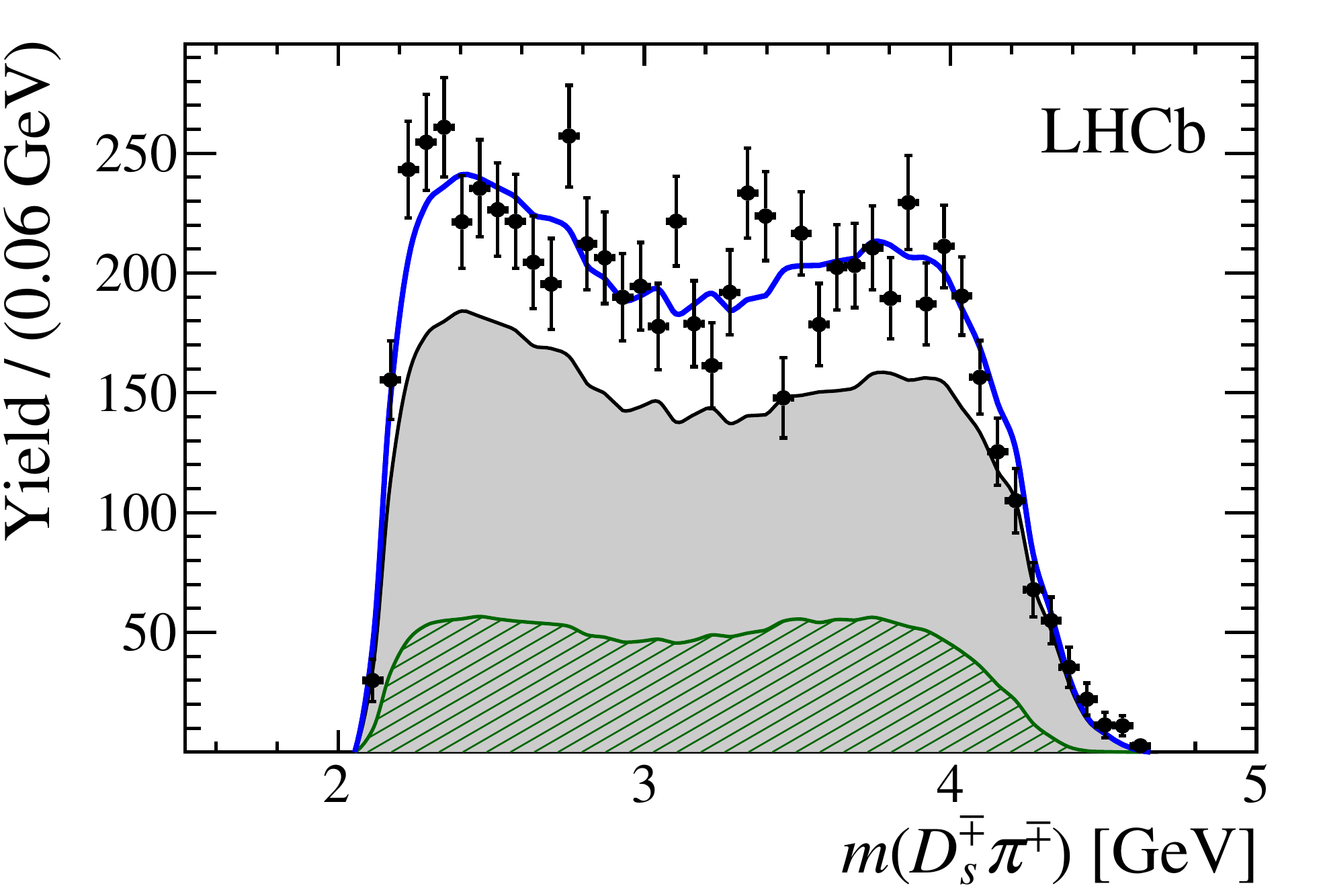}
		
		\includegraphics[width=0.32\textwidth, height = !]{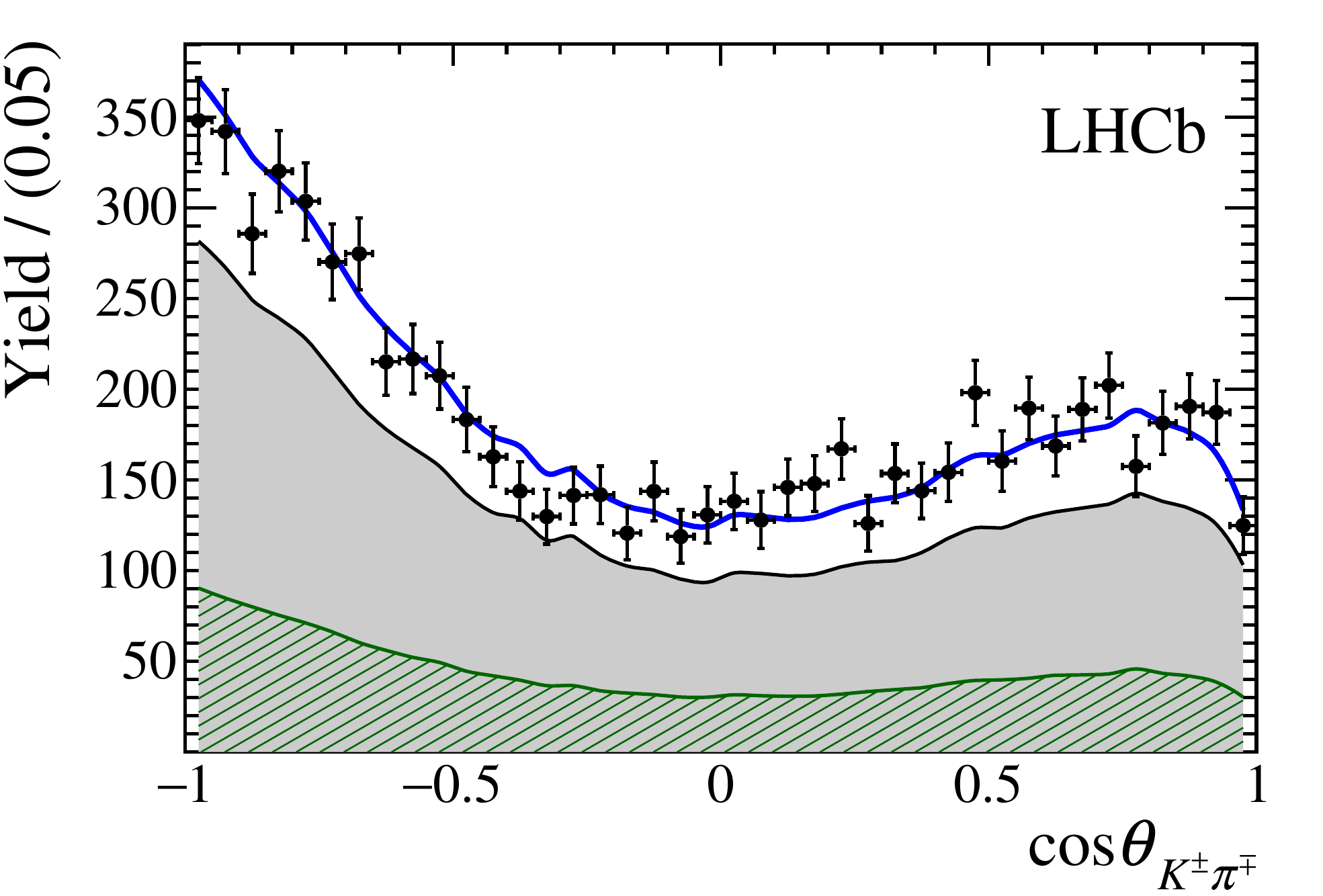}
		\includegraphics[width=0.32\textwidth, height = !]{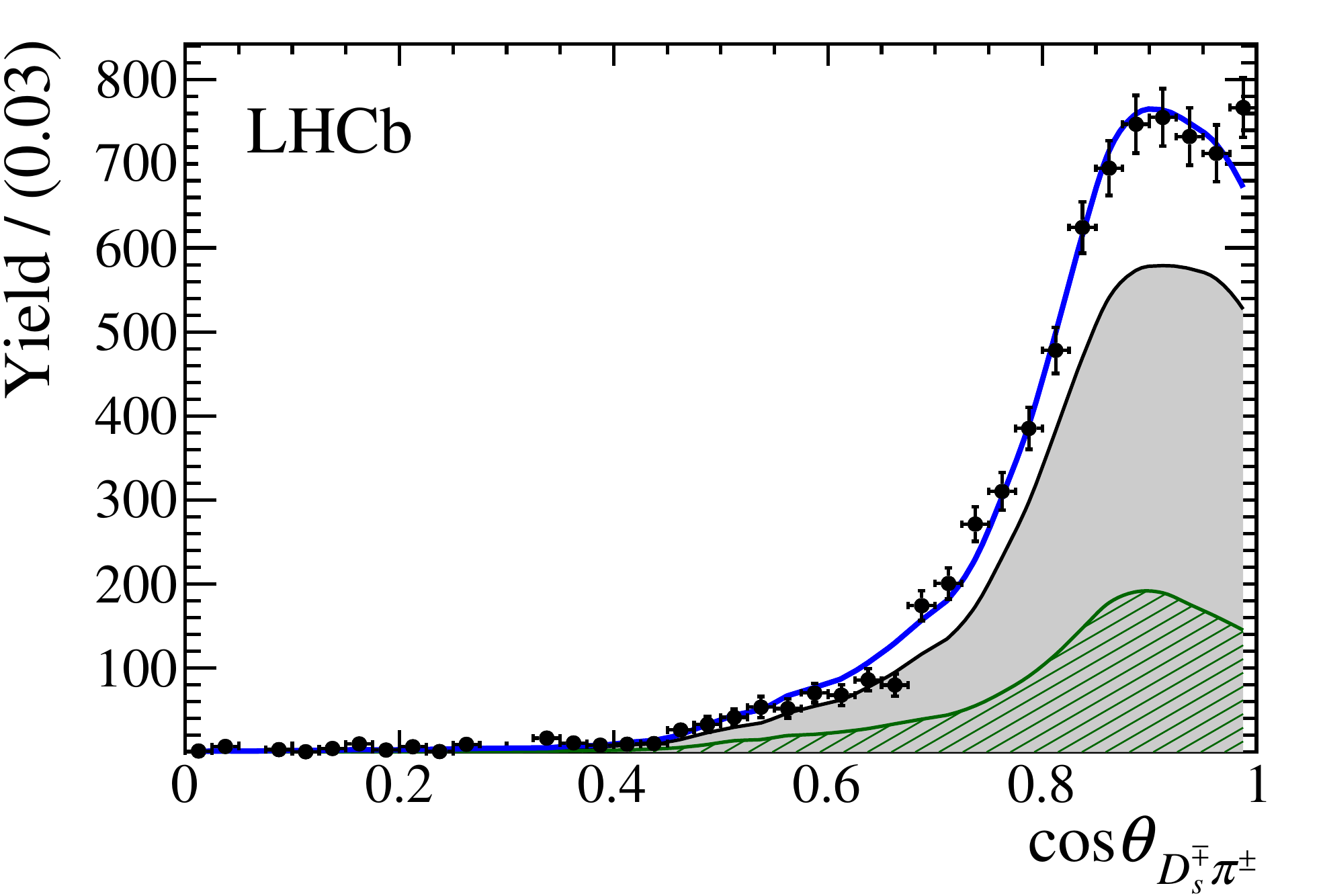}
        \includegraphics[width=0.32\textwidth, height = !]{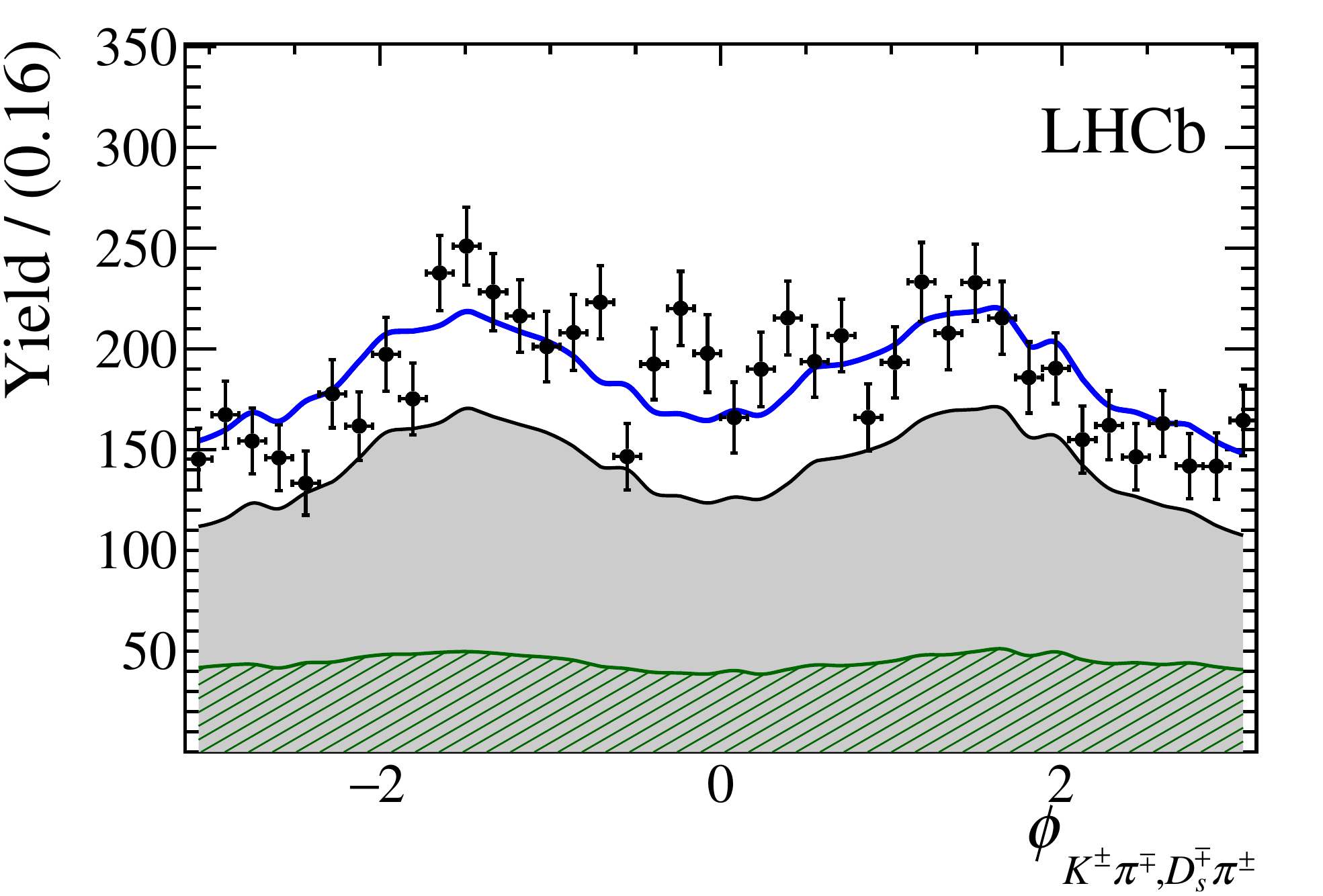}
		\caption{
		Invariant-mass  and angular distributions of background-subtracted \mbox{\signal} candidates (data points) and fit projections (blue solid line).
			Contributions from $b\to c$ and $b\to u$ decay amplitudes are overlaid, colour coded as in Fig.~\ref{fig:fullFit}.
		  } 		
		  \label{fig:fullFit3}
\end{figure}

\begin{figure}[p]
		\includegraphics[width=0.32\textwidth, height = !]{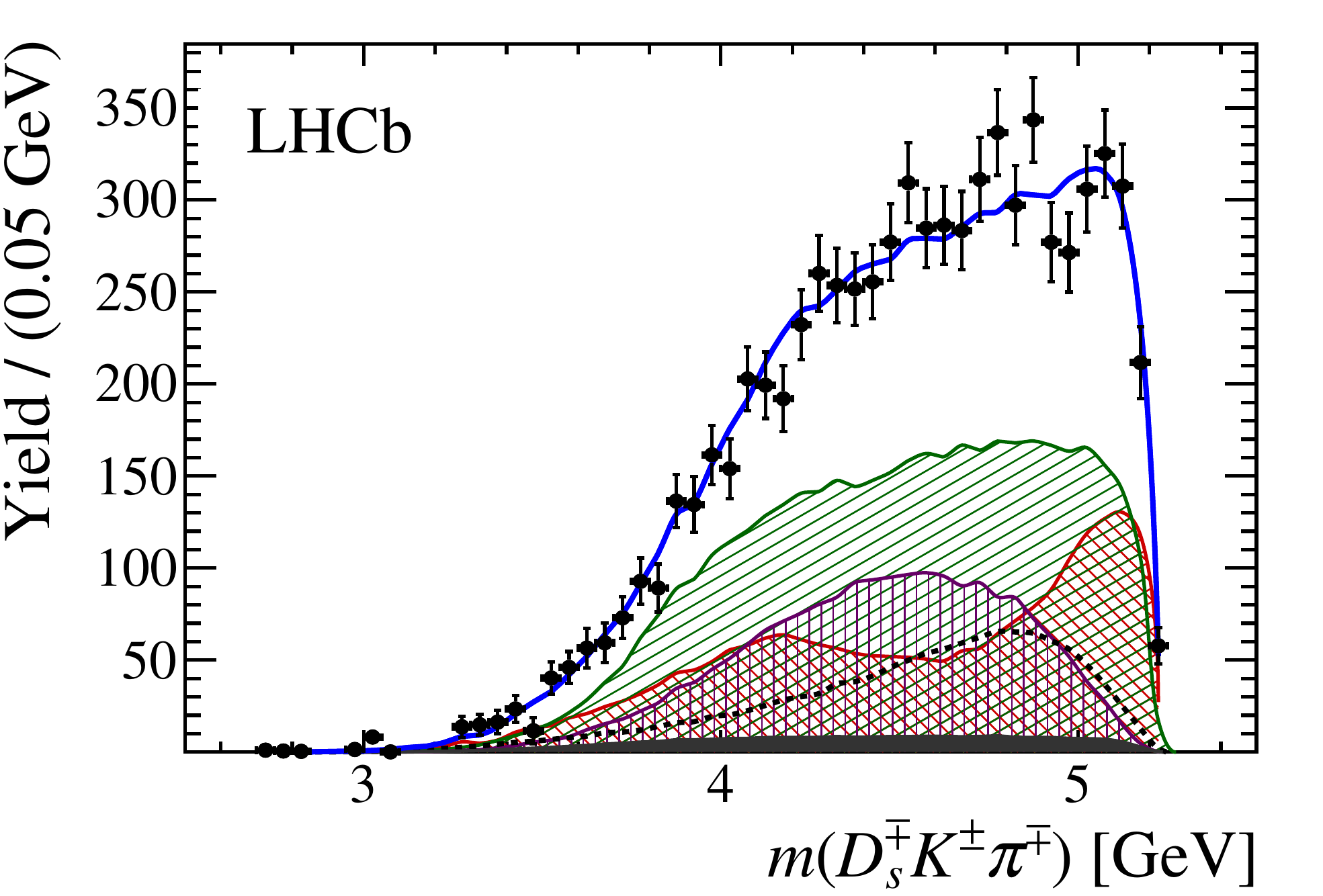} 	
		\includegraphics[width=0.32\textwidth, height = !]{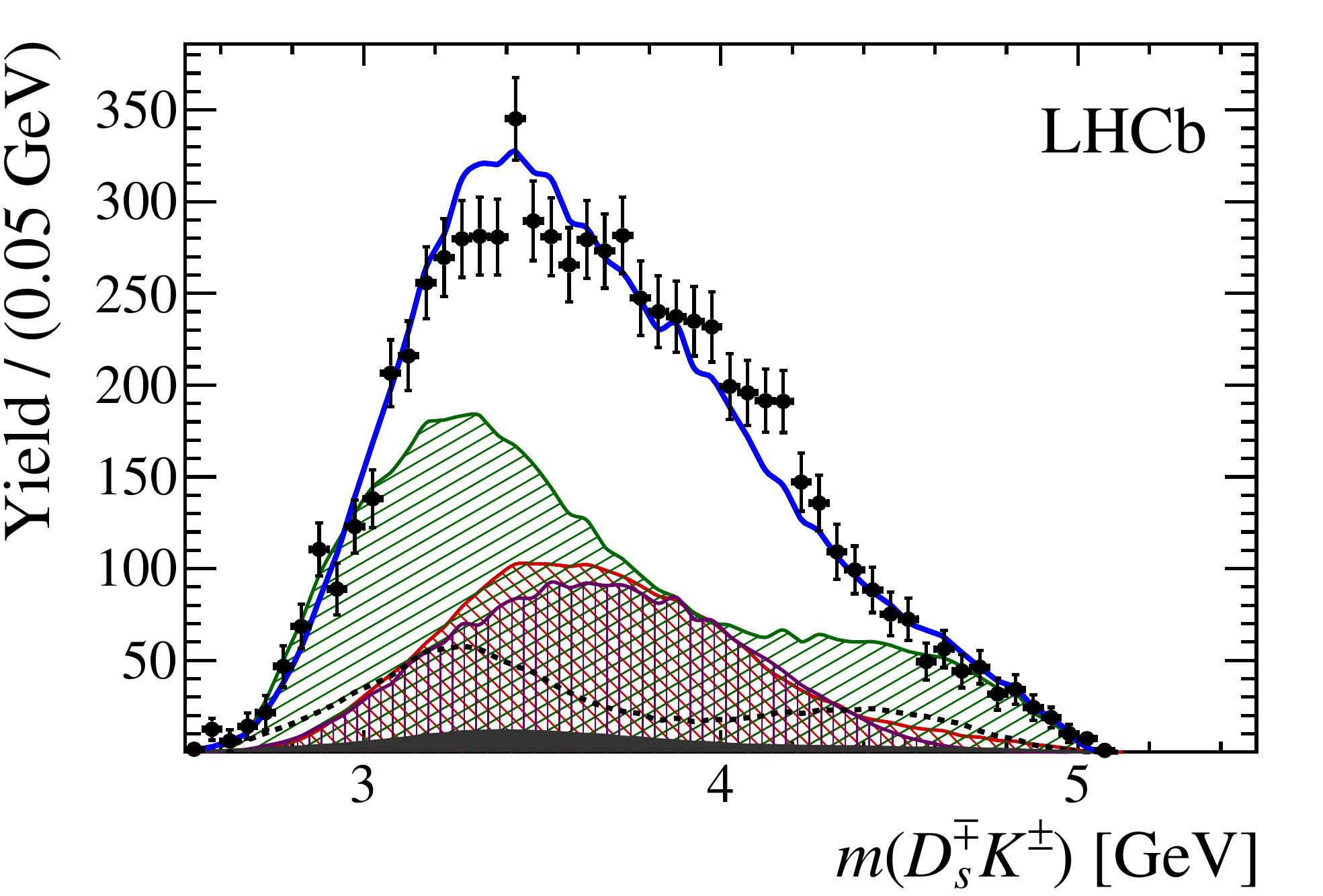}
		\includegraphics[width=0.32\textwidth, height = !]{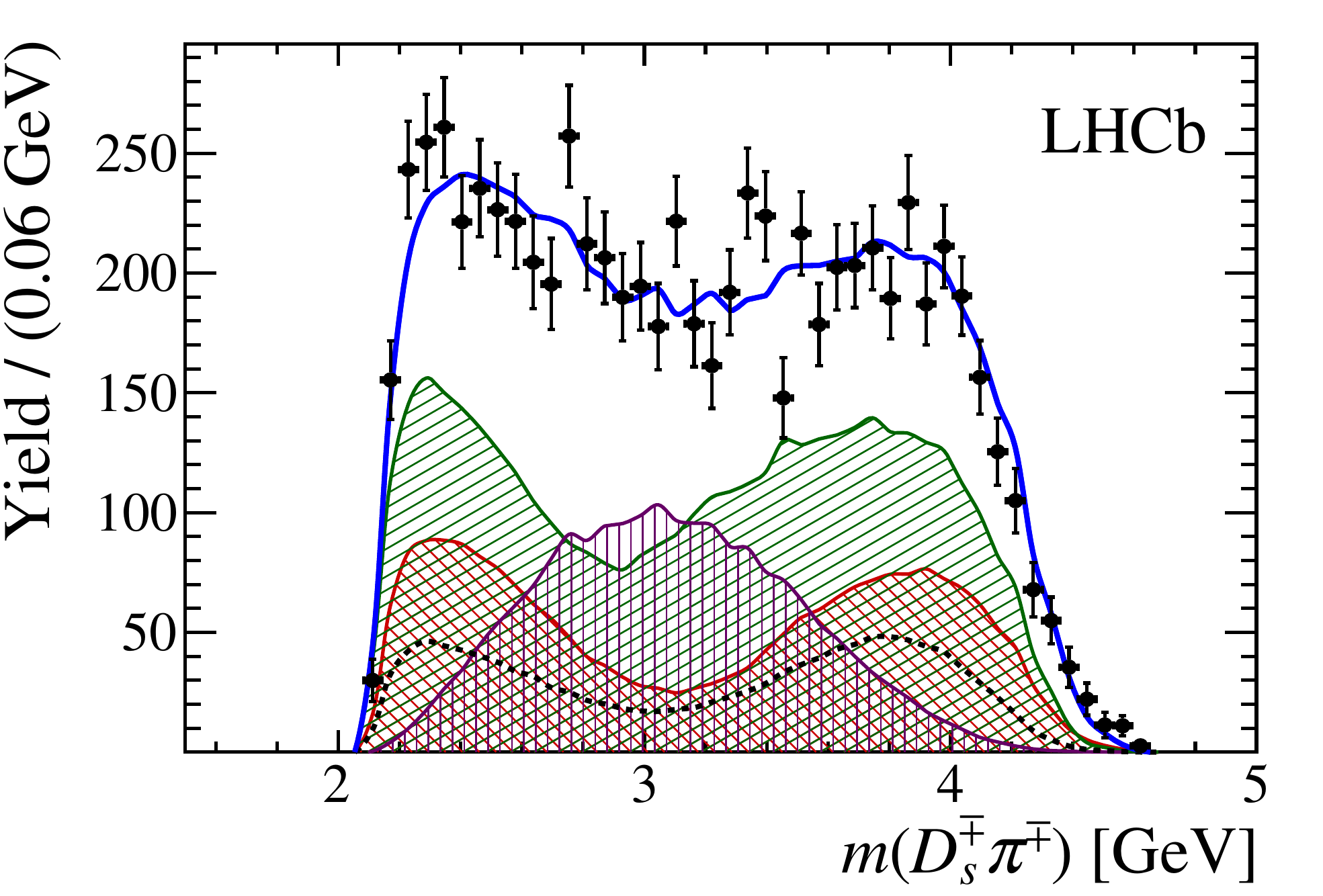}
		
		\includegraphics[width=0.32\textwidth, height = !]{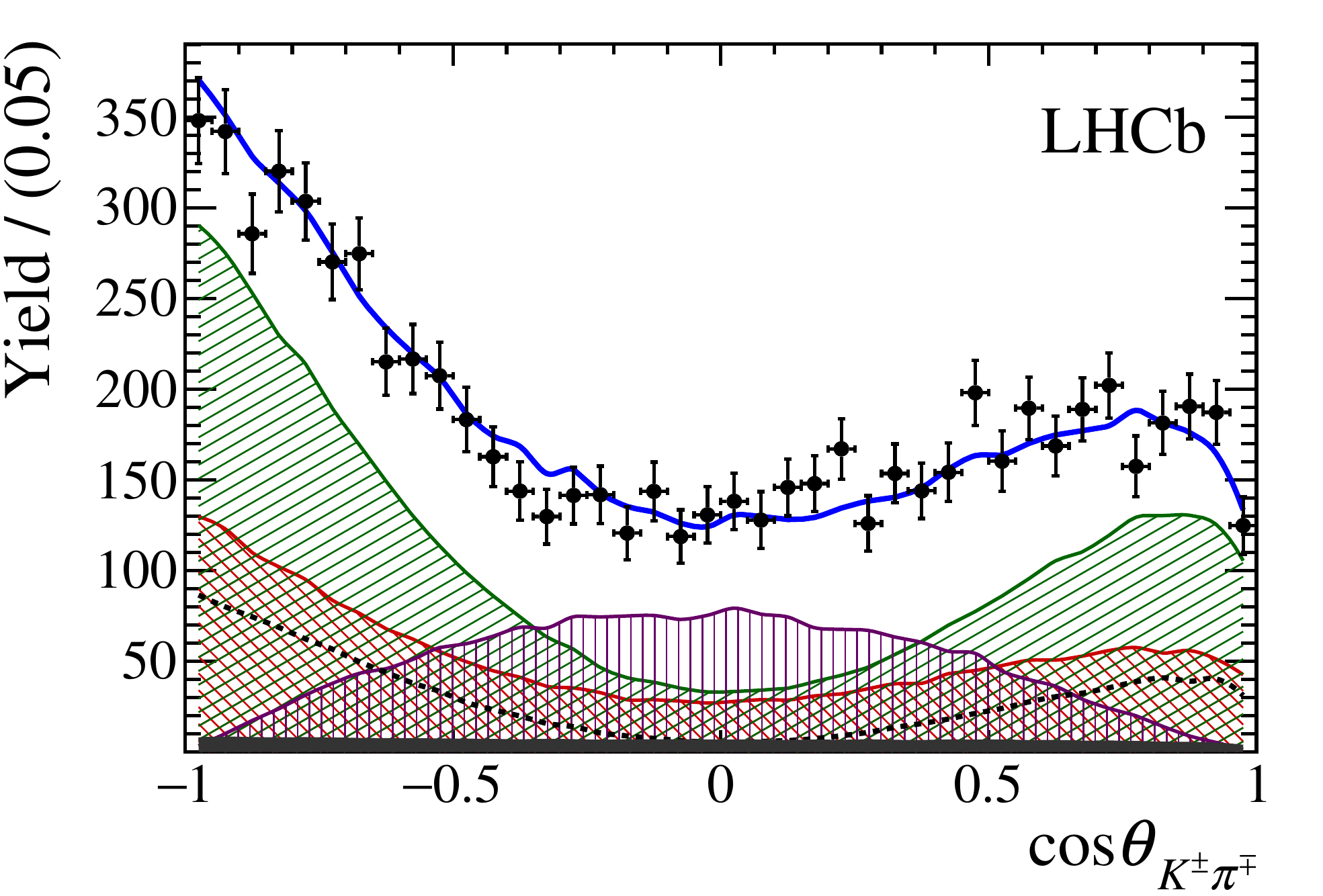}
		\includegraphics[width=0.32\textwidth, height = !]{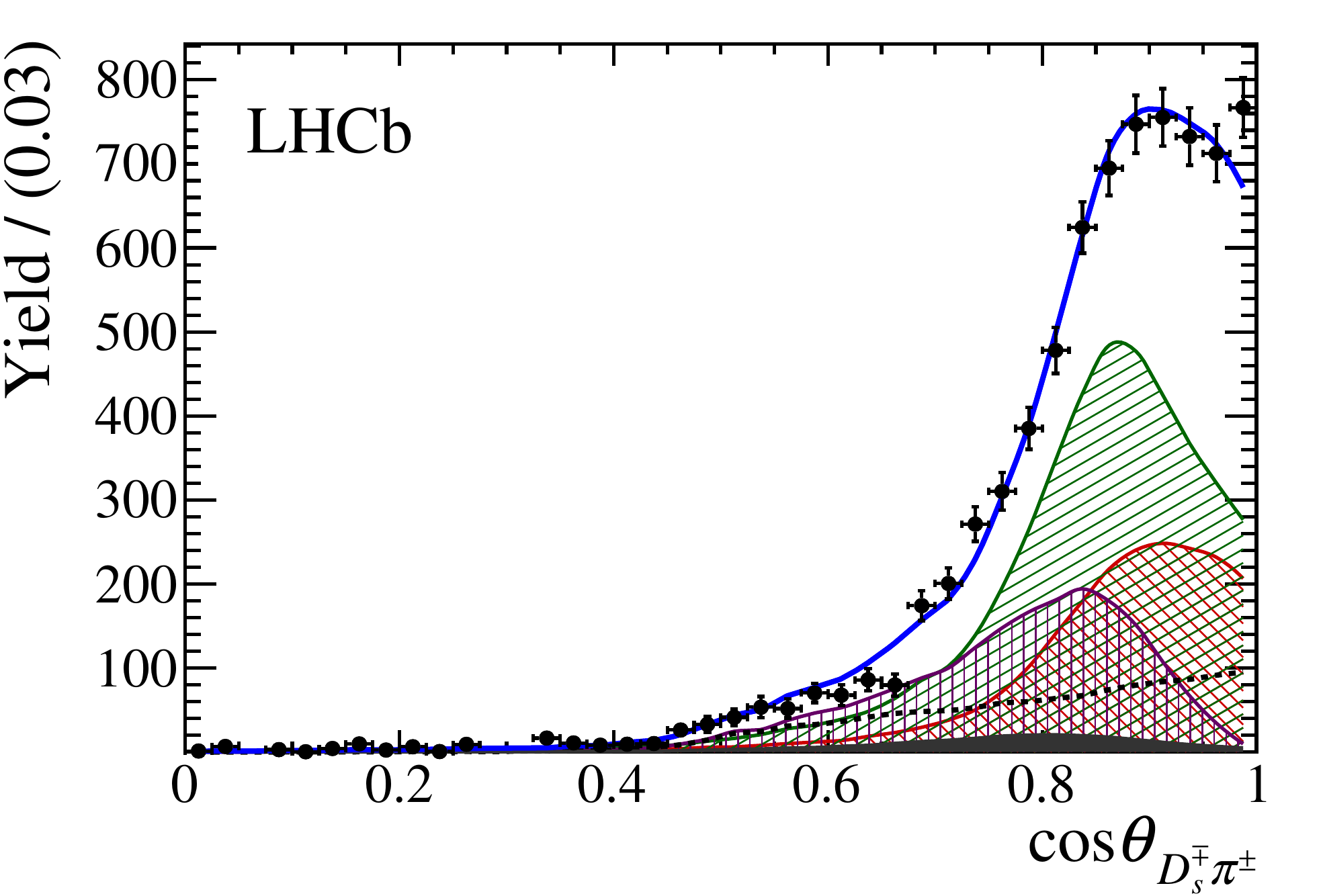}
        \includegraphics[width=0.32\textwidth, height = !]{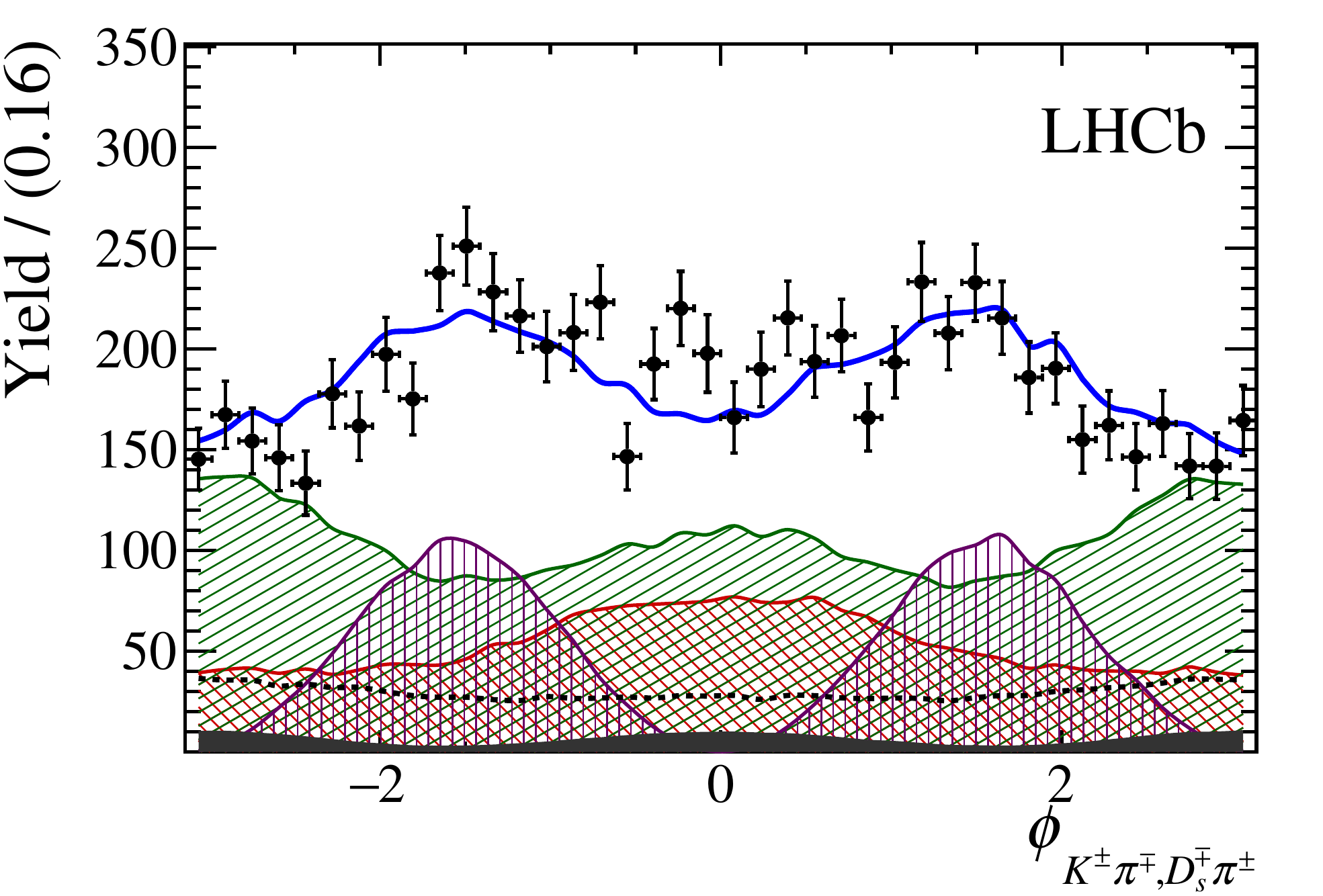}

		\caption{Invariant-mass  and angular distributions of background-subtracted \mbox{\signal} candidates (data points) and fit projections (blue solid line). Incoherent contributions from intermediate-state components are overlaid, colour coded as in Fig.~\ref{fig:fullFit2}.
		  } 		
		\label{fig:fullFit4}
\end{figure}

\clearpage
Tables~\ref{tab:fullResult} and~\ref{tab:fullResultCascade} list the moduli and phases of the complex amplitude coefficients
obtained by fitting the baseline model to the background-subtracted \signal signal candidates.
The systematic uncertainties are summarised in Table~\ref{tab:sigSys3}.
The fit results are used to derive the amplitude ratio, $r_i$, and
strong-phase difference, $\delta_i$, for a given decay channel, defined as
\begin{align*}
    r_i &= r \, \left\vert \frac{ a_i^u }{ a_i^c } \right\vert \, \sqrt{\frac{\int \vert A^c(\phsPoint) \vert^2 \, \text d \Phi_4(\phsPoint)}{\int \vert A^u(\phsPoint) \vert^2 \, \text d \Phi_4(\phsPoint)} }, &    \\
    \delta_i &= \delta - \arg(a_i^c) + \arg(a_i^u)
    - \arg\left(\int A^c(\phsPoint)^* \, A^u(\phsPoint) \, \text d \Phi_4(\phsPoint)\right). &
\end{align*}
The values are given in Table~\ref{tab:perAmp_r_delta}.
Tables~\ref{tab:intFracDsKpipi} ($b\to c$ amplitudes) and~\ref{tab:intFracDsKpipi2}  ($b\to u$ amplitudes)  list the interference fractions ordered by magnitude, for the baseline model.
Figures~\ref{fig:fullFit3} and~\ref{fig:fullFit4} show additional fit projections of the baseline model.
Here, the acoplanarity angle,
${\phi}_{K^+\pi^-,D_s^-\pi^+}$, is defined as the angle between the two decay planes formed by
the $K^+\pi^-$ system and the $D_s^- \pip$ system
in the $\Bs$ rest frame; boosting into the rest frames of the two-body systems defining these decay planes,
the two helicity variables $\cos \theta_{K^+\pi^-}$ and $\cos \theta_{D_s^- \pip}$
are defined as the cosine of the angle
of the $K^+$ or $D_s^-$ momentum with the $\Bs$ flight direction.

The decay fractions for several alternative models are summarised in \mbox{Tables~\ref{tab:altModelsDsKpipi} and~\ref{tab:altModelsDsKpipi2}}.

\begin{table}[h]
\centering
\caption{
Moduli and phases of the amplitude coefficients for decays via $b \to c$ and $b \to u$ quark-level transitions.
The uncertainties are statistical and systematic.
}
\resizebox{\linewidth}{!}{
	\renewcommand{\arraystretch}{1.1}

\begin{tabular}{l
r@{$\,\pm\,$}c@{$\,\pm\,$}l
r@{$\,\pm\,$}c@{$\,\pm\,$}l
r@{$\,\pm\,$}c@{$\,\pm\,$}l
r@{$\,\pm\,$}c@{$\,\pm\,$}l
}
\hline
\hline
\multicolumn{1}{c}{Decay channel} & \multicolumn{6}{c}{$A^{c}(\phsPoint)$} & \multicolumn{6}{c}{$A^{u}(\phsPoint)$}  \\
 & \multicolumn{3}{c}{$\vert a_i^c \vert$}  & \multicolumn{3}{c}{$\arg(a_i^c) [\degrees]$}  & \multicolumn{3}{c}{$\vert a_i^u \vert$} & \multicolumn{3}{c}{$\arg(a_i^u) [\degrees]$} \\
\hline
$\Bs \to \Dsmp \, K_1(1270)^\pm  $ & 0.50 & 0.05 & 0.05 & 24 & 4 & 6 & 0.51 & 0.27 & 0.25 & $-129$ & 18 & 26 \\
$\Bs \to \Dsmp \,  K_1(1400)^\pm $ &  \multicolumn{3}{c}{1.0} & \multicolumn{3}{c}{0.0} & \multicolumn{3}{c}{1.0} & \multicolumn{3}{c}{0.0}  \\
$\Bs \to \Dsmp \,  K^{*}(1410)^\pm $ & 0.45 & 0.05 & 0.05 & 53 & 6 & 6 & 0.81 & 0.20 & 0.15 & 2 & 13 & 11 \\
$\Bs \to \Dsmp \,  K(1460)^\pm  $ & \multicolumn{3}{c}{} & \multicolumn{3}{c}{} &0.78 & 0.20 & 0.16 & $-92$ & 10 & 15 \\
$\Bs \to ( \Dsmp \, \pipm)_{P} \, K^{*}(892)^0 $ & 0.40 & 0.04 & 0.04 & $-33$ & 5 & 4 & 1.21 & 0.26 & 0.22 &$-7$ & 8 & 17 \\
$\Bs \to ( \Dsmp \, K^\pm)_{P} \, \rho(770)^0 $ & 0.12 & 0.03 & 0.03 & 109 & 9 & 14 & \multicolumn{3}{c}{} & \multicolumn{3}{c}{} \\

\hline
\hline
\end{tabular}
 }
\label{tab:fullResult}

\end{table}

\begin{table}[h]
\centering
	\renewcommand{\arraystretch}{1.1}

\caption{
Moduli and phases of the amplitude coefficients for cascade decays.
The amplitude coefficients are defined relative to the respective three-body production amplitude coefficients in Table~\ref{tab:fullResult}
and are shared among $b \to c$ and $b \to u$ transitions.
The uncertainties are statistical and systematic.
}
\begin{tabular}{l
r@{$\,\pm\,$}c@{$\,\pm\,$}l
r@{$\,\pm\,$}c@{$\,\pm\,$}l
}
\hline
\hline
\multicolumn{1}{c}{Decay channel}& \multicolumn{3}{c}{$\vert a_i \vert$}  & \multicolumn{3}{c}{$\arg(a_i) [\degrees]$}  \\
\hline
$K_1(1270)^\pm \to K^\pm \, \rho(770)^0  $ &  \multicolumn{3}{c}{1.0} &  \multicolumn{3}{c}{0.0}  \\
$K_1(1270)^\pm \to K^{*}(892)^0 \pi^\pm$ & 0.90 & 0.09 & 0.10 & 46 & 7 & 4   \\
$K_1(1270)^\pm \to K^{*}_{0}(1430)^0 \pipm $ & 0.46 & 0.05 & 0.03 & 120 & 5 & 6   \\
$K^{*}(1410)^\pm \to K^{*}(892)^0 \, \pipm  $ &  \multicolumn{3}{c}{1.0} &  \multicolumn{3}{c}{0.0}  \\
$K^{*}(1410)^\pm \to K^\pm \, \rho(770)^0 $ & 0.66 & 0.05 & 0.05 & $-165$ & 6 & 5    \\

\hline
\hline
\end{tabular}
\label{tab:fullResultCascade}

\end{table}

\clearpage
\begin{sidewaystable}[h]
\centering
\caption{Systematic uncertainties on the fit parameters of the full time-dependent amplitude fit to \signal data in units of the statistical standard deviations.
The different contributions are: 1) fit bias, 2) background subtraction, 3) correlation of observables, 4) time acceptance, 5) resolution, 6) decay-time bias, 7) nuisance asymmetries, 8) $\Delta m_s$,
9) phase-space acceptance, 10) acceptance factorisation,
11) lineshape models, 12) masses and widths of resonances,
13) form factor.
}

\resizebox{\linewidth}{!}{
	\renewcommand{\arraystretch}{1.1}
\begin{tabular}{l  c  c  c  c  c  c  c  c  c  c  c  c  c  c    }
\hline
\hline
Parameter & 1 & 2 & 3 & 4 & 5 & 6 & 7 & 8& 9 & 10 & 11 & 12 & 13 &   Total  \\
\hline
$\Bs \to \Dsmp \, K_1(1270)^\pm  \, \vert a_i^c \vert $ & 0.09 & 0.32 & 0.17 & 0.01 & 0.36 & 0.05 & 0.00 & 0.02 & 0.60 & 0.11 & 0.31 & 0.07 & 0.48 &   0.99 \\
$\Bs \to \Dsmp \,  K_1(1270)^\pm  \, \arg(a_i^c)$ & 0.10 & 0.37 & 0.12 & 0.03 & 0.13 & 0.05 & 0.05 & 0.02 & 0.40 & 0.31 & 0.98 & 0.21 & 0.58 &   1.33 \\
$\Bs \to \Dsmp \,  K_1(1270)^\pm  \, \vert a_i^u \vert$ & 0.17 & 0.26 & 0.14 & 0.04 & 0.41 & 0.07 & 0.01 & 0.03 & 0.35 & 0.18 & 0.60 & 0.02 & 0.26 &   0.94 \\
$\Bs \to \Dsmp \,  K_1(1270)^\pm  \, \arg(a_i^u)$ & 0.12 & 0.77 & 0.21 & 0.08 & 0.39 & 0.05 & 0.11 & 0.04 & 0.66 & 0.55 & 0.52 & 0.07 & 0.54 &   1.46 \\
$ K_1(1270)^\pm \to K^{*}(892)^0 \, \pipm  \, \vert a_i \vert$ & 0.01 & 0.32 & 0.27 & 0.01 & 0.31 & 0.01 & 0.00 & 0.01 & 0.17 & 0.36 & 0.55 & 0.15 & 0.55 &   1.03 \\
$K_1(1270)^\pm \to K^{*}(892)^0 \, \pipm  \, \arg(a_i)$ & 0.14 & 0.11 & 0.11 & 0.01 & 0.13 & 0.02 & 0.00 & 0.00 & 0.19 & 0.14 & 0.33 & 0.07 & 0.34 &   0.59 \\
$K_1(1270)^\pm \to K^{*}_{0}(1430)^0 \, \pipm  \, \vert a_i \vert $ & 0.10 & 0.48 & 0.14 & 0.00 & 0.05 & 0.01 & 0.00 & 0.00 & 0.10 & 0.25 & 0.42 & 0.05 & 0.17 &   0.73 \\
$ K_1(1270)^\pm \to K^{*}_{0}(1430)^0 \, \pipm  \, \arg(a_i) $ & 0.24 & 0.33 & 0.53 & 0.00 & 0.14 & 0.01 & 0.00 & 0.00 & 0.27 & 0.46 & 0.16 & 0.08 & 0.85 &   1.22 \\
$\Bs \to \Dsmp \, K^{*}(1410)^\pm  \, \vert a_i^c \vert$ & 0.27 & 0.35 & 0.14 & 0.02 & 0.29 & 0.04 & 0.00 & 0.01 & 0.87 & 0.32 & 0.09 & 0.10 & 0.20 &   1.10 \\
$\Bs \to \Dsmp \,  K^{*}(1410)^\pm  \, \arg(a_i^c)$ & 0.32 & 0.29 & 0.10 & 0.01 & 0.40 & 0.02 & 0.00 & 0.01 & 0.23 & 0.21 & 0.46 & 0.06 & 0.64 &   1.04 \\
$\Bs \to \Dsmp \,  K^{*}(1410)^\pm  \, \vert a_i^u \vert$ & 0.17 & 0.16 & 0.17 & 0.03 & 0.25 & 0.07 & 0.07 & 0.03 & 0.38 & 0.32 & 0.26 & 0.05 & 0.34 &   0.77 \\
$\Bs \to \Dsmp \,  K^{*}(1410)^\pm \, \arg(a_i^u)$ & 0.21 & 0.42 & 0.41 & 0.05 & 0.13 & 0.01 & 0.01 & 0.01 & 0.07 & 0.26 & 0.21 & 0.05 & 0.47 &   0.87 \\
$ K^{*}(1410)^\pm \to \Kpm \, \rho(770)^0  \, \vert a_i \vert$ & 0.58 & 0.37 & 0.26 & 0.00 & 0.19 & 0.02 & 0.10 & 0.00 & 0.43 & 0.45 & 0.31 & 0.04 & 0.06 &   1.03 \\
$ K^{*}(1410)^\pm \to \Kpm \, \rho(770)^0  \, \arg(a_i)$ & 0.05 & 0.42 & 0.24 & 0.00 & 0.19 & 0.04 & 0.06 & 0.00 & 0.58 & 0.03 & 0.22 & 0.05 & 0.19 &   0.84 \\
$\Bs \to \Dsmp \,  K(1460)^\pm \, \vert a_i^u \vert$ & 0.02 & 0.22 & 0.28 & 0.05 & 0.13 & 0.04 & 0.01 & 0.02 & 0.22 & 0.18 & 0.63 & 0.04 & 0.19 &   0.82 \\
$\Bs \to \Dsmp \,  K(1460)^\pm  \, \arg(a_i^u)$ & 0.14 & 0.51 & 0.10 & 0.06 & 0.30 & 0.01 & 0.19 & 0.02 & 0.39 & 0.48 & 0.11 & 0.10 & 1.15 &   1.46 \\
$\Bs \to ( \Dsmp \, \pipm)_{P} \, \, K^{*}(892)^0 \, \vert a_i^c \vert$ & 0.06 & 0.37 & 0.11 & 0.01 & 0.13 & 0.05 & 0.01 & 0.02 & 0.09 & 0.32 & 0.58 & 0.09 & 0.62 &   1.00 \\
$\Bs \to ( \Dsmp \, \pipm)_{P} \, \, K^{*}(892)^0 \, \arg(a_i^c)$ & 0.03 & 0.60 & 0.29 & 0.03 & 0.19 & 0.04 & 0.02 & 0.02 & 0.22 & 0.10 & 0.43 & 0.06 & 0.29 &   0.90 \\
$\Bs \to ( \Dsmp \, \pipm)_{P} \, \, K^{*}(892)^0 \, \vert a_i^u \vert$ & 0.08 & 0.26 & 0.15 & 0.05 & 0.20 & 0.02 & 0.01 & 0.02 & 0.18 & 0.26 & 0.62 & 0.04 & 0.33 &   0.86 \\
$\Bs \to ( \Dsmp \, \pipm)_{P} \, \, K^{*}(892)^0 \, \arg(a_i^u)$ & 0.20 & 0.75 & 0.66 & 0.05 & 0.54 & 0.06 & 0.15 & 0.04 & 0.36 & 0.68 & 0.31 & 0.13 & 1.37 &   1.98 \\
$\Bs \to ( \Dsmp \, \Kpm)_{P} \, \, \rho(770)^0 \, \vert a_i^c \vert$ & 0.11 & 0.46 & 0.14 & 0.01 & 0.12 & 0.03 & 0.00 & 0.00 & 0.32 & 0.05 & 0.71 & 0.06 & 0.37 &   1.01 \\
$\Bs \to ( \Dsmp \, \Kpm)_{P} \, \, \rho(770)^0 \, \arg(a_i^c)$ & 0.18 & 0.66 & 0.66 & 0.01 & 0.23 & 0.01 & 0.00 & 0.01 & 0.20 & 0.22 & 1.12 & 0.06 & 0.47 &   1.59 \\
\hline
\hline
\end{tabular}
 }
\label{tab:sigSys3}
\end{sidewaystable}
\clearpage

\begin{table}[h]
\centering
    \caption{Amplitude ratio and strong-phase difference for a given decay channel. The uncertainties are statistical and systematic.}
 \label{tab:perAmp_r_delta}
\renewcommand{\arraystretch}{1.2}
\begin{tabular}{l
r@{$\,\pm\,$}c@{$\,\pm\,$}l
r@{$\,\pm\,$}c@{$\,\pm\,$}l
}
\hline
\hline
\multicolumn{1}{c}{Decay channel} & \multicolumn{3}{c}{$r_i$} & \multicolumn{3}{c}{$\delta_i [\degrees]$}  \\
\hline
$\Bs \to \Dsmp \, K_1(1270)^\pm $ &  0.31 & 0.20 & 0.16    &  $-109$ & 22 & 28   \\

$\Bs \to \Dsmp \, K_1(1400)^\pm $ &   0.30 & 0.03 & 0.02   &  44 & 10 & 4    \\

$\Bs \to \Dsmp \, K^{*}(1410)^\pm $ &  0.54 & 0.18 & 0.15   & $-7$ & 17  & 14 \\

$\Bs \to ( \Dsmp \, \pipm)_{P} \, \, K^{*}(892)^0 $ &  0.92 & 0.24 & 0.20 &    69 & 15 & 18 \\

\hline
\hline
\end{tabular}
\end{table}

\begin{table}[h]
\centering
\caption{Interference fractions (ordered by magnitude) of the $b\to c$ intermediate-state amplitudes
   included in the baseline model. Only the statistical uncertainties are given.}
\resizebox{\linewidth}{!}{
\tiny
	\renewcommand{\arraystretch}{1.25}

\begin{tabular}{l l r }
\hline
\hline
\multicolumn{1}{c}{Decay channel $i$} & \multicolumn{1}{c}{Decay channel $j$} & \multicolumn{1}{c}{$I_{ij}[\%]$}  \\
\hline
$\Bs \to \Dsmp \, ( K_1(1400)^\pm \to K^{*}(892)^0 \, \pipm )$ & $\Bs \to ( \Dsmp \, \pipm)_{P} \, \, K^{*}(892)^0$ & $-15.4$ $\pm$ 3.1 \\
$\Bs \to \Dsmp \, ( K_1(1270)^\pm \to K^{*}(892)^0 \, \pipm )$ & $\Bs \to \Dsmp \, ( K_1(1400)^\pm \to K^{*}(892)^0 \, \pipm )$ & $-12.2$ $\pm$ 5.8 \\
$\Bs \to \Dsmp \, ( K_1(1270)^\pm \to K^{*}(892)^0 \, \pipm )$ & $\Bs \to ( \Dsmp \, \pipm)_{P} \, \, K^{*}(892)^0$ & $-10.2$ $\pm$ 1.2 \\
$\Bs \to \Dsmp \, ( K_1(1400)^\pm \to K^{*}(892)^0 \, \pipm )$ & $\Bs \to \Dsmp \, ( K_1(1270)^\pm \to K^\pm\, \rho(770)^0 )$ & 5.8 $\pm$ 0.5 \\
$\Bs \to \Dsmp \, ( K^{*}(1410)^\pm \to K^{*}(892)^0 \, \pipm )$ & $\Bs \to \Dsmp \, ( K^{*}(1410)^\pm \to K^\pm\, \rho(770)^0 )$ & 4.1 $\pm$ 0.3 \\
$\Bs \to ( \Dsmp \, K^\pm)_{P} \, \, \rho(770)^0$ & $\Bs \to \Dsmp \, ( K_1(1270)^\pm \to K^\pm\, \rho(770)^0 )$ & 3.1 $\pm$ 0.5 \\
$\Bs \to \Dsmp \, ( K_1(1400)^\pm \to K^{*}(892)^0 \, \pipm )$ & $\Bs \to ( \Dsmp \, K^\pm)_{P} \, \, \rho(770)^0$ & 1.3 $\pm$ 0.3 \\
$\Bs \to \Dsmp \, ( K_1(1270)^\pm \to K^{*}_{0}(1430) \, \pipm )$ & $\Bs \to \Dsmp \, ( K_1(1270)^\pm \to K^\pm\, \rho(770)^0 )$ & $-1.3$ $\pm$ 0.9 \\
$\Bs \to ( \Dsmp \, K^\pm)_{P} \, \, \rho(770)^0$ & $\Bs \to \Dsmp \, ( K_1(1270)^\pm \to K^{*}_{0}(1430) \, \pipm )$ & $-0.5$ $\pm$ 0.3 \\
$\Bs \to ( \Dsmp \, \pipm)_{P} \, \, K^{*}(892)^0$ & $\Bs \to ( \Dsmp \, K^\pm)_{P} \, \, \rho(770)^0$ & $-0.2$ $\pm$ 0.1 \\
$\Bs \to \Dsmp \, ( K_1(1270)^\pm \to K^{*}(892)^0 \, \pipm )$ & $\Bs \to \Dsmp \, ( K_1(1270)^\pm \to K^\pm\, \rho(770)^0 )$ & $-0.1$ $\pm$ 0.5 \\
$\Bs \to ( \Dsmp \, \pipm)_{P} \, \, K^{*}(892)^0$ & $\Bs \to \Dsmp \, ( K_1(1270)^\pm \to K^\pm\, \rho(770)^0 )$ & $-0.1$ $\pm$ 0.1 \\
$\Bs \to \Dsmp \, ( K_1(1400)^\pm \to K^{*}(892)^0 \, \pipm )$ & $\Bs \to \Dsmp \, ( K_1(1270)^\pm \to K^{*}_{0}(1430) \, \pipm )$ & $-0.1$ $\pm$ 0.0 \\
$\Bs \to \Dsmp \, ( K_1(1270)^\pm \to K^{*}(892)^0 \, \pipm )$ & $\Bs \to ( \Dsmp \, K^\pm)_{P} \, \, \rho(770)^0$ & 0.0 $\pm$ 0.1 \\
$\Bs \to \Dsmp \, ( K_1(1400)^\pm \to K^{*}(892)^0 \, \pipm )$ & $\Bs \to \Dsmp \, ( K^{*}(1410)^\pm \to K^\pm\, \rho(770)^0 )$ & $0.0$ $\pm$ 0.0 \\
$\Bs \to ( \Dsmp \, \pipm)_{P} \, \, K^{*}(892)^0$ & $\Bs \to \Dsmp \, ( K_1(1270)^\pm \to K^{*}_{0}(1430) \, \pipm )$ & 0.0 $\pm$ 0.0 \\
$\Bs \to \Dsmp \, ( K^{*}(1410)^\pm \to K^\pm\, \rho(770)^0 )$ & $\Bs \to \Dsmp \, ( K_1(1270)^\pm \to K^\pm\, \rho(770)^0 )$ & $0.0$ $\pm$ 0.0 \\
$\Bs \to \Dsmp \, ( K_1(1400)^\pm \to K^{*}(892)^0 \, \pipm )$ & $\Bs \to \Dsmp \, ( K^{*}(1410)^\pm \to K^{*}(892)^0 \, \pipm )$ & $0.0$ $\pm$ 0.0 \\
$\Bs \to ( \Dsmp \, \pipm)_{P} \, \, K^{*}(892)^0$ & $\Bs \to \Dsmp \, ( K^{*}(1410)^\pm \to K^\pm\, \rho(770)^0 )$ & 0.0 $\pm$ 0.0 \\
$\Bs \to \Dsmp \, ( K_1(1270)^\pm \to K^{*}(892)^0 \, \pipm )$ & $\Bs \to \Dsmp \, ( K_1(1270)^\pm \to K^{*}_{0}(1430) \, \pipm )$ & $0.0$ $\pm$ 0.0 \\
$\Bs \to \Dsmp \, ( K^{*}(1410)^\pm \to K^{*}(892)^0 \, \pipm )$ & $\Bs \to \Dsmp \, ( K_1(1270)^\pm \to K^{*}_{0}(1430) \, \pipm )$ & $0.0$ $\pm$ 0.0 \\
$\Bs \to \Dsmp \, ( K_1(1270)^\pm \to K^{*}(892)^0 \, \pipm )$ & $\Bs \to \Dsmp \, ( K^{*}(1410)^\pm \to K^\pm\, \rho(770)^0 )$ & 0.0 $\pm$ 0.0 \\
$\Bs \to \Dsmp \, ( K^{*}(1410)^\pm \to K^{*}(892)^0 \, \pipm )$ & $\Bs \to \Dsmp \, ( K_1(1270)^\pm \to K^\pm\, \rho(770)^0 )$ & 0.0 $\pm$ 0.0 \\
$\Bs \to \Dsmp \, ( K^{*}(1410)^\pm \to K^{*}(892)^0 \, \pipm )$ & $\Bs \to ( \Dsmp \, \pipm)_{P} \, \, K^{*}(892)^0$ & $0.0$ $\pm$ 0.0 \\
$\Bs \to \Dsmp \, ( K^{*}(1410)^\pm \to K^\pm\, \rho(770)^0 )$ & $\Bs \to \Dsmp \, ( K_1(1270)^\pm \to K^{*}_{0}(1430) \, \pipm )$ & $0.0$ $\pm$ 0.0 \\
$\Bs \to \Dsmp \, ( K^{*}(1410)^\pm \to K^{*}(892)^0 \, \pipm )$ & $\Bs \to ( \Dsmp \, K^\pm)_{P} \, \, \rho(770)^0$ & 0.0 $\pm$ 0.0 \\
$\Bs \to \Dsmp \, ( K_1(1270)^\pm \to K^{*}(892)^0 \, \pipm )$ & $\Bs \to \Dsmp \, ( K^{*}(1410)^\pm \to K^{*}(892)^0 \, \pipm )$ & $0.0$ $\pm$ 0.0 \\
$\Bs \to \Dsmp \, ( K^{*}(1410)^\pm \to K^\pm\, \rho(770)^0 )$ & $\Bs \to ( \Dsmp \, K^\pm)_{P} \, \, \rho(770)^0$ & 0.0 $\pm$ 0.0 \\
\hline
\hline
\end{tabular}

 }
\label{tab:intFracDsKpipi}
\end{table}
\begin{table}[h]
\centering
\caption{Interference fractions (ordered by magnitude) of the $b\to u$ intermediate-state amplitudes
   included in the baseline model. Only the statistical uncertainties are given.}
\resizebox{\linewidth}{!}{
\tiny
	\renewcommand{\arraystretch}{1.25}

\begin{tabular}{l l r }
\hline
\hline
\multicolumn{1}{c}{Decay channel $i$} & \multicolumn{1}{c}{Decay channel $j$} & \multicolumn{1}{c}{$I_{ij}[\%]$}  \\
\hline
$\Bs \to \Dsmp \, ( K_1(1270)^\pm \to K^{*}(892)^0 \, \pipm )$ & $\Bs \to ( \Dsmp \, \pipm)_{P} \, \, K^{*}(892)^0$ & 9.5 $\pm$ 2.6 \\
$\Bs \to \Dsmp \, ( K_1(1270)^\pm \to K^{*}(892)^0 \, \pipm )$ & $\Bs \to \Dsmp \, ( K_1(1400)^\pm \to K^{*}(892)^0 \, \pipm )$ & 9.2 $\pm$ 3.2 \\
$\Bs \to \Dsmp \, ( K (1460)^\pm \to K^{*}(892)^0 \, \pipm )$ & $\Bs \to ( \Dsmp \, \pipm)_{P} \, \, K^{*}(892)^0$ & $-5.5$ $\pm$ 0.9 \\
$\Bs \to \Dsmp \, ( K^{*}(1410)^\pm \to K^{*}(892)^0 \, \pipm )$ & $\Bs \to \Dsmp \, ( K^{*}(1410)^\pm \to  K^\pm\, \rho(770)^0 )$ & 4.1 $\pm$ 0.8 \\
$\Bs \to \Dsmp \, ( K_1(1400)^\pm \to K^{*}(892)^0 \, \pipm )$ & $\Bs \to ( \Dsmp \, \pipm)_{P} \, \, K^{*}(892)^0$ & $-3.7$ $\pm$ 5.9 \\
$\Bs \to \Dsmp \, ( K_1(1400)^\pm \to K^{*}(892)^0 \, \pipm )$ & $\Bs \to \Dsmp \, ( K_1(1270)^\pm \to  K^\pm\, \rho(770)^0 )$ & $-1.2$ $\pm$ 0.6 \\
$\Bs \to \Dsmp \, ( K_1(1270)^\pm \to K^{*}_{0}(1430) \, \pipm )$ & $\Bs \to \Dsmp \, ( K_1(1270)^\pm \to  K^\pm\, \rho(770)^0 )$ & $-0.4$ $\pm$ 0.3 \\
$\Bs \to ( \Dsmp \, \pipm)_{P} \, \, K^{*}(892)^0$ & $\Bs \to \Dsmp \, ( K_1(1270)^\pm \to  K^\pm\, \rho(770)^0 )$ & 0.1 $\pm$ 0.3 \\
$\Bs \to \Dsmp \, ( K^{*}(1410)^\pm \to K^{*}(892)^0 \, \pipm )$ & $\Bs \to ( \Dsmp \, \pipm)_{P} \, \, K^{*}(892)^0$ & $-0.1$ $\pm$ 0.0 \\
$\Bs \to ( \Dsmp \, \pipm)_{P} \, \, K^{*}(892)^0$ & $\Bs \to \Dsmp \, ( K^{*}(1410)^\pm \to  K^\pm\, \rho(770)^0 )$ & $0.0$ $\pm$ 0.0 \\
$\Bs \to \Dsmp \, ( K_1(1270)^\pm \to K^{*}(892)^0 \, \pipm )$ & $\Bs \to \Dsmp \, ( K_1(1270)^\pm \to  K^\pm\, \rho(770)^0 )$ & $0.0$ $\pm$ 0.2 \\
$\Bs \to ( \Dsmp \, \pipm)_{P} \, \, K^{*}(892)^0$ & $\Bs \to \Dsmp \, ( K_1(1270)^\pm \to K^{*}_{0}(1430) \, \pipm )$ & $0.0$ $\pm$ 0.0 \\
$\Bs \to \Dsmp \, ( K (1460)^\pm \to K^{*}(892)^0 \, \pipm )$ & $\Bs \to \Dsmp \, ( K_1(1270)^\pm \to  K^\pm\, \rho(770)^0 )$ & $0.0$ $\pm$ 0.0 \\
$\Bs \to \Dsmp \, ( K^{*}(1410)^\pm \to K^{*}(892)^0 \, \pipm )$ & $\Bs \to \Dsmp \, ( K_1(1270)^\pm \to  K^\pm\, \rho(770)^0 )$ & 0.0 $\pm$ 0.0 \\
$\Bs \to \Dsmp \, ( K_1(1400)^\pm \to K^{*}(892)^0 \, \pipm )$ & $\Bs \to \Dsmp \, ( K (1460)^\pm \to K^{*}(892)^0 \, \pipm )$ & $0.0$ $\pm$ 0.0 \\
$\Bs \to \Dsmp \, ( K_1(1270)^\pm \to K^{*}(892)^0 \, \pipm )$ & $\Bs \to \Dsmp \, ( K (1460)^\pm \to K^{*}(892)^0 \, \pipm )$ & $0.0$ $\pm$ 0.0 \\
$\Bs \to \Dsmp \, ( K_1(1400)^\pm \to K^{*}(892)^0 \, \pipm )$ & $\Bs \to \Dsmp \, ( K^{*}(1410)^\pm \to K^{*}(892)^0 \, \pipm )$ & $0.0$ $\pm$ 0.0 \\
$\Bs \to \Dsmp \, ( K (1460)^\pm \to K^{*}(892)^0 \, \pipm )$ & $\Bs \to \Dsmp \, ( K^{*}(1410)^\pm \to  K^\pm\, \rho(770)^0 )$ & 0.0 $\pm$ 0.0 \\
$\Bs \to \Dsmp \, ( K_1(1400)^\pm \to K^{*}(892)^0 \, \pipm )$ & $\Bs \to \Dsmp \, ( K^{*}(1410)^\pm \to  K^\pm\, \rho(770)^0 )$ & 0.0 $\pm$ 0.0 \\
$\Bs \to \Dsmp \, ( K^{*}(1410)^\pm \to  K^\pm\, \rho(770)^0 )$ & $\Bs \to \Dsmp \, ( K_1(1270)^\pm \to  K^\pm\, \rho(770)^0 )$ & 0.0 $\pm$ 0.0 \\
$\Bs \to \Dsmp \, ( K_1(1400)^\pm \to K^{*}(892)^0 \, \pipm )$ & $\Bs \to \Dsmp \, ( K_1(1270)^\pm \to K^{*}_{0}(1430) \, \pipm )$ & 0.0 $\pm$ 0.0 \\
$\Bs \to \Dsmp \, ( K_1(1270)^\pm \to K^{*}(892)^0 \, \pipm )$ & $\Bs \to \Dsmp \, ( K^{*}(1410)^\pm \to  K^\pm\, \rho(770)^0 )$ & $0.0$ $\pm$ 0.0 \\
$\Bs \to \Dsmp \, ( K^{*}(1410)^\pm \to K^{*}(892)^0 \, \pipm )$ & $\Bs \to \Dsmp \, ( K_1(1270)^\pm \to K^{*}_{0}(1430) \, \pipm )$ & $0.0$ $\pm$ 0.0 \\
$\Bs \to \Dsmp \, ( K (1460)^\pm \to K^{*}(892)^0 \, \pipm )$ & $\Bs \to \Dsmp \, ( K^{*}(1410)^\pm \to K^{*}(892)^0 \, \pipm )$ & $0.0$ $\pm$ 0.0 \\
$\Bs \to \Dsmp \, ( K_1(1270)^\pm \to K^{*}(892)^0 \, \pipm )$ & $\Bs \to \Dsmp \, ( K_1(1270)^\pm \to K^{*}_{0}(1430) \, \pipm )$ & $0.0$ $\pm$ 0.0 \\
$\Bs \to \Dsmp \, ( K (1460)^\pm \to K^{*}(892)^0 \, \pipm )$ & $\Bs \to \Dsmp \, ( K_1(1270)^\pm \to K^{*}_{0}(1430) \, \pipm )$ & $0.0$ $\pm$ 0.0 \\
$\Bs \to \Dsmp \, ( K_1(1270)^\pm \to K^{*}(892)^0 \, \pipm )$ & $\Bs \to \Dsmp \, ( K^{*}(1410)^\pm \to K^{*}(892)^0 \, \pipm )$ & 0.0 $\pm$ 0.0 \\
$\Bs \to \Dsmp \, ( K^{*}(1410)^\pm \to  K^\pm\, \rho(770)^0 )$ & $\Bs \to \Dsmp \, ( K_1(1270)^\pm \to K^{*}_{0}(1430) \, \pipm )$ & 0.0 $\pm$ 0.0 \\
\hline
\hline

\end{tabular}

 }
\label{tab:intFracDsKpipi2}
\end{table}

\clearpage

\clearpage

\begin{table}[h]
\caption{Decay fractions in percent for several alternative amplitude models \mbox{(Alt.~1 - Alt.~6).}
Resonance parameters and the  observables $r, \kappa, \delta, \gamma - 2 \beta_s$ are also given.}
\resizebox{\linewidth}{!}{
	\renewcommand{\arraystretch}{1.1}
	\footnotesize
\begin{tabular}{l l    r  r  r  r  r  r  }
\hline
\hline
& &  \multicolumn{1}{c}{Alt.1}  & \multicolumn{1}{c}{Alt.2}  & \multicolumn{1}{c}{Alt.3}  & \multicolumn{1}{c}{Alt.4}  & \multicolumn{1}{c}{Alt.5}  & \multicolumn{1}{c}{Alt.6}  \\
\hline
\multirow{18}{*}{$b \to c$}  & $\Bs \to \Dsmp \, ( K_1(1270)^\pm \to K^{*}(892)^0 \, \pipm )$  & 12.2 & 12.0 & 13.9 & 11.5 & 14.7 & 12.8 \\
 & $\Bs \to \Dsmp \, ( K_1(1270)^\pm[D] \to K^{*}(892)^0 \, \pipm )$   &   & 0.4 &   &   &  &   \\
 & $\Bs \to \Dsmp \, ( K_1(1270)^\pm \to \Kpm\, \rho(770)^0 )$  & 16.2 & 14.6 & 21.7 & 15.6 & 18.6 & 16.5 \\
 & $\Bs \to \Dsmp \, ( K_1(1270)^\pm \to \Kpm\, \rho(1450)^0 )$   &   & 0.3 &   &   &   &   \\
 & $\Bs \to \Dsmp \, ( K_1(1270)^\pm \to K^{*}_{0}(1430)^0 \, \pipm )$  & 3.9 & 3.5 & 3.7 & 3.6 & 3.0 & 3.6 \\
 & $\Bs \to \Dsmp \, ( K_1(1400)^\pm \to K^{*}(892)^0 \, \pipm )$  & 43.1 & 61.1 & 59.8 & 57.7 & 69.5 & 60.6 \\
 & $\Bs \to \Dsmp \, ( K_1(1400)^\pm \to \Kpm\, \rho(770)^0 )$   &    &    & 0.5 &    &    &    \\
 & $\Bs \to \Dsmp \, ( K^{*}(1410)^\pm \to K^{*}(892)^0 \, \pipm )$  & 12.9 & 14.0 & 13.6 & 13.8 & 24.2 & 13.5 \\
 & $\Bs \to \Dsmp \, ( K^{*}(1410)^\pm \to \Kpm\, \rho(770)^0 )$  & 4.9 & 5.8 & 4.9 & 5.3 & 3.6 & 5.2 \\
 & $\Bs \to \Dsmp \, ( K(1460)^\pm \to K^{*}(892)^0 \, \pipm )$   & 1.7 &    &    & 1.3 &    &    \\
 & $\Bs \to \Dsmp \, ( K(1460)^\pm \to \Kpm\, \rho(770)^0 )$   &    &    &    & 0.2 &    &    \\
 & $\Bs \to \Dsmp \, ( K(1460)^\pm \to \Kpm\, f_0(500)^0 )$   &    &    &    & 0.6 &    &    \\
 & $\Bs \to \Dsmp \, ( K^{*}(1680)^\pm \to K^{*}(892)^0 \, \pipm )$   &    &    &    &    & 5.5 &    \\
 & $\Bs \to \Dsmp \, ( K^{*}(1680)^\pm \to \Kpm\, \rho(770)^0 )$   &    &    &    &    & 1.9 &    \\
 & $\Bs \to \Dsmp \, ( K_2(1770) \to K^{*}(892)^0 \, \pipm )$   &    &    &    &    &    & 0.4 \\
 & $\Bs \to ( \Dsmp \, \pipm)_{P} \, \, K^{*}(892)^0$  & 8.8 & 9.3 & 9.5 & 7.7 & 10.2 & 9.2 \\
 & $\Bs \to ( \Dsmp \, \Kpm)_{P} \, \, \rho(770)^0$  & 1.3 & 0.7 & 0.1 & 0.6 & 0.2 & 0.5 \\
\multirow{18}{*}{$b \to u$}  & $\text{Sum}$  & 105.1 & 121.7 & 127.6 & 118.0 & 151.5 & 122.1 \\
\hline
 & $\Bs \to \Dsmp \, ( K_1(1270)^\pm \to K^{*}(892)^0 \, \pipm )$  & 4.5 & 2.4 & 1.1 & 4.9 & 2.0 & 3.4 \\
 & $\Bs \to \Dsmp \, ( K_1(1270)^\pm[D] \to K^{*}(892)^0 \, \pipm )$   &    & 0.1 &    &    &    &    \\
 & $\Bs \to \Dsmp \, ( K_1(1270)^\pm \to \Kpm\, \rho(770)^0 )$  & 6.0 & 2.9 & 1.6 & 6.6 & 2.6 & 4.4 \\
 & $\Bs \to \Dsmp \, ( K_1(1270)^\pm \to \Kpm\, \rho(1450)^0 )$   &    & 0.1 &    &    &    &    \\
 & $\Bs \to \Dsmp \, ( K_1(1270)^\pm \to K^{*}_{0}(1430)^0 \, \pipm )$  & 1.4 & 0.7 & 0.3 & 1.5 & 0.4 & 1.0 \\
 & $\Bs \to \Dsmp \, ( K_1(1400)^\pm \to K^{*}(892)^0 \, \pipm )$  & 23.3 & 18.9 & 26.2 & 10.3 & 28.2 & 23.0 \\
 & $\Bs \to \Dsmp \, ( K_1(1400)^\pm \to \Kpm\, \rho(770)^0 )$   &    &    & 0.2 &    &    &    \\
 & $\Bs \to \Dsmp \, ( K^{*}(1410)^\pm \to K^{*}(892)^0 \, \pipm )$  & 14.8 & 13.7 & 12.7 & 7.2 & 2.4 & 11.5 \\
 & $\Bs \to \Dsmp \, ( K^{*}(1410)^\pm \to \Kpm\, \rho(770)^0 )$  & 5.6 & 5.7 & 4.6 & 2.8 & 0.4 & 4.4 \\
 & $\Bs \to \Dsmp \, ( K(1460)^\pm \to K^{*}(892)^0 \, \pipm )$  & 12.9 & 10.8 & 12.6 & 8.1 & 12.1 & 11.9 \\
 & $\Bs \to \Dsmp \, ( K(1460)^\pm \to \Kpm\, \rho(770)^0 )$   &    &    &    & 1.0 &    &    \\
 & $\Bs \to \Dsmp \, ( K(1460)^\pm \to \Kpm\, f_0(500)^0 )$   &    &    &    & 3.7 &    &    \\
 & $\Bs \to \Dsmp \, ( K^{*}(1680)^\pm \to K^{*}(892)^0 \, \pipm )$   &    &    &    &    & 5.6 &    \\
 & $\Bs \to \Dsmp \, ( K^{*}(1680)^\pm \to \Kpm\, \rho(770)^0 )$   &    &    &    &    & 2.0 &    \\
 & $\Bs \to \Dsmp \, ( K_2(1770) \to K^{*}(892)^0 \, \pipm )$   &    &    &    &    &    & 0.9 \\
 & $\Bs \to ( \Dsmp \, \pipm)_{P} \, \, K^{*}(892)^0$  & 32.5 & 33.4 & 31.3 & 38.3 & 27.4 & 30.7 \\
 & $\text{Sum}$  & 103.9 & 88.7 & 90.7 & 84.4 & 83.0 & 91.4 \\
\hline
 & $m_{K_1(1400)} \, [\text{MeV}]$  & 1401 & 1406 & 1404 & 1401 & 1397 & 1408 \\
 & $\Gamma_{K_1(1400)} \, [\text{MeV}]$  & 189 & 193 & 192 & 196 & 199 & 193 \\
 & $m_{K^{*}(1410)} \, [\text{MeV}]$  & 1434 & 1437 & 1436 & 1442 & 1419 & 1434 \\
 & $\Gamma_{K^{*}(1410)} \, [\text{MeV}]$  & 391 & 400 & 386 & 382 & 453 & 395 \\
 & $r$ &  0.50 & 0.55 & 0.59 & 0.48 & 0.54 & 0.58 \\
 & $\kappa$  & 0.64 & 0.75 & 0.74 & 0.77 & 0.71 & 0.75 \\
 & $\delta \, [\degrees]$ & $-4$ & $-15$ & $-14$ & $-24$ & $-19$ & $-18$ \\
 & $\gamma - 2 \beta_{s} \, [\degrees]$  & 40 & 44 & 42 & 50 & 52 & 51 \\
\hline
\hline
\end{tabular}
 }
\label{tab:altModelsDsKpipi}
\end{table}

\begin{table}[h]
\centering
\caption{Decay fractions in percent for several alternative amplitude models \mbox{(Alt.~7 - Alt.~12).}
Resonance parameters and the observables $r, \kappa, \delta, \gamma - 2 \beta_s$ are also given.}
\resizebox{\linewidth}{!}{
	\renewcommand{\arraystretch}{1.1}
	\footnotesize
\begin{tabular}{l l  r  r  r  r  r  r  }
\hline
\hline
&  & \multicolumn{1}{c}{Alt.7}  & \multicolumn{1}{c}{Alt.8}  & \multicolumn{1}{c}{Alt.9}  & \multicolumn{1}{c}{Alt.10}  & \multicolumn{1}{c}{Alt.11}  & \multicolumn{1}{c}{Alt.12}  \\
\hline
\multirow{18}{*}{$b \to c$}  & $\Bs \to \Dsmp \, ( K_1(1270)^\pm \to K^{*}(892)^0 \, \pipm )$ & 13.3 & 12.8 & 15.5 & 24.0 & 11.2 & 13.3 \\
 & $\Bs \to \Dsmp \, ( K_1(1270)^\pm \to \Kpm\, \rho(770)^0 )$ & 17.5 & 17.5 & 19.1 & 14.9 & 17.7 & 19.9 \\
 & $\Bs \to \Dsmp \, ( K_1(1270)^\pm \to K^{*}_{0}(1430)^0 \, \pipm )$ & 3.3 & 4.0 & 4.3 & 2.8 & 3.2 & 3.7 \\
 & $\Bs \to \Dsmp \, ( K_1(1400)^\pm \to K^{*}(892)^0 \, \pipm )$ & 66.0 & 93.1 & 55.5 & 77.9 & 44.7 & 63.5 \\
 & $\Bs \to \Dsmp \, ( K^{*}(1410)^\pm \to K^{*}(892)^0 \, \pipm )$ & 12.5 & 16.5 & 13.3 & 13.7 & 15.1 & 12.7 \\
 & $\Bs \to \Dsmp \, ( K^{*}(1410)^\pm \to \Kpm\, \rho(770)^0 )$ & 5.4 & 5.9 & 6.4 & 4.7 & 5.8 & 5.1 \\
 & $\Bs \to ( \Dsmp \, \pipm)_{S} \, \, K^{*}(892)^0$ &     &     &     & 5.1 &     &     \\
 & $\Bs \to ( \Dsmp \, \pipm)_{P} \, \, K^{*}(892)^0$ & 11.4 & 19.8 & 8.5 &     & 5.5 & 10.8 \\
 & $\Bs[P] \to ( \Dsmp \, \pipm)_{P} \, \, K^{*}(892)^0$ &     & 1.7 &     &     &     &     \\
 & $\Bs[D] \to ( \Dsmp \, \pipm)_{P} \, \, K^{*}(892)^0$ &     & 4.7 &     &     &     &     \\
 & $\Bs \to ( \Dsmp \, \Kpm)_{S} \, \, f_0(500)^0 $&     &     &     &     & 1.8 &     \\
 & $\Bs \to ( \Dsmp \, \Kpm)_{S} \, \, f_0(980)^0$ &     &     &     &     & 1.5 &     \\
 & $\Bs \to ( \Dsmp \, \Kpm)_{S} \, \, f_2(1270)^0$ &     &     &     &     & 0.1 &     \\
 & $\Bs \to ( \Dsmp \, \Kpm)_{S} \, \, \rho(770)^0$ & 0.3 &     &     &     &     &     \\
 & $\Bs \to ( \Dsmp \, \Kpm)_{P} \, \, \rho(770)^0$ &     & 1.6 & 3.5 & 1.3 & 0.5 &     \\
 & $\Bs[P] \to ( \Dsmp \, \Kpm)_{P} \, \, \rho(770)^0$ &     &     & 0.2 &     &     &     \\
 & $\Bs[D] \to ( \Dsmp \, \Kpm)_{P} \, \, \rho(770)^0$ &     &     & 0.9 &     &     &     \\
\multirow{18}{*}{$b \to u$}  & $\text{Sum}$ & 129.6 & 177.5 & 127.2 & 144.4 & 107.1 & 129.0 \\
\hline
 & $\Bs \to \Dsmp \, ( K_1(1270)^\pm \to K^{*}(892)^0 \, \pipm )$ & 3.3 & 5.2 & 1.1 & 10.6 & 2.5 & 2.5 \\
 & $\Bs \to \Dsmp \, ( K_1(1270)^\pm \to \Kpm\, \rho(770)^0 )$ & 4.4 & 7.2 & 1.4 & 6.6 & 3.9 & 3.7 \\
 & $\Bs \to \Dsmp \, ( K_1(1270)^\pm \to K^{*}_{0}(1430)^0 \, \pipm )$ & 0.8 & 1.6 & 0.3 & 1.2 & 0.7 & 0.7 \\
 & $\Bs \to \Dsmp \, ( K_1(1400)^\pm \to K^{*}(892)^0 \, \pipm )$ & 19.8 & 35.8 & 27.8 & 7.7 & 17.1 & 22.4 \\
 & $\Bs \to \Dsmp \, ( K^{*}(1410)^\pm \to K^{*}(892)^0 \, \pipm )$ & 14.2 & 9.8 & 9.9 & 11.7 & 4.0 & 14.5 \\
 & $\Bs \to \Dsmp \, ( K^{*}(1410)^\pm \to \Kpm\, \rho(770)^0 )$ & 6.1 & 3.5 & 4.8 & 4.0 & 1.5 & 5.8 \\
 & $\Bs \to \Dsmp \, ( K(1460)^\pm \to K^{*}(892)^0 \, \pipm )$ & 11.7 & 2.6 & 13.2 & 8.8 & 13.6 & 12.3 \\
 & $\Bs \to ( \Dsmp \, \pipm)_{S} \, \, K^{*}(892)^0$ &     &     &     & 22.3 &     &     \\
 & $\Bs \to ( \Dsmp \, \pipm)_{P} \, \, K^{*}(892)^0$ & 25.8 & 68.5 & 33.6 &     & 43.1 & 29.3 \\
 & $\Bs[P] \to ( \Dsmp \, \pipm)_{P} \, \, K^{*}(892)^0$ &     & 5.8 &     &     &     &     \\
 & $\Bs[D] \to ( \Dsmp \, \pipm)_{P} \, \, K^{*}(892)^0$ &     & 16.3 &     &     &     &     \\
 & $\Bs \to ( \Dsmp \, \Kpm)_{S} \, \, f_0(500)^0$ &     &     &     &     & 0.7 &     \\
 & $\Bs \to ( \Dsmp \, \Kpm)_{S} \, \, f_0(980)^0$ &     &     &     &     & 0.6 &     \\
 & $\Bs \to ( \Dsmp \, \Kpm)_{S} \, \, f_2(1270)^0$ &     &     &     &     & 0.0 &     \\
 & $\Bs \to ( \Dsmp \, \Kpm)_{S} \, \, \rho(770)^0$ & 0.4 &     &     &     &     &     \\
 & $\Bs \to ( \Dsmp \, \Kpm)_{P} \, \, \rho(770)^0$ &     &     & 4.4 &     &     &     \\
 & $\Bs[P] \to ( \Dsmp \, \Kpm)_{P} \, \, \rho(770)^0$ &     &     & 0.3 &     &     &     \\
 & $\Bs[D] \to ( \Dsmp \, \Kpm)_{P} \, \, \rho(770)^0$ &     &     & 1.1 &     &     &     \\
 & $\text{Sum}$ & 86.5 & 156.3 & 97.9 & 73.0 & 87.8 & 91.0 \\
\hline
 & $m_{K_1(1400)} \, [\text{MeV}]$ & 1405 & 1398 & 1404 & 1365 & 1406 & 1406 \\
 & $\Gamma_{K_1(1400)} \, [\text{MeV}]$ & 193 & 247 & 188 & 203 & 184 & 190 \\
 & $m_{K^{*}(1410)} \, [\text{MeV}]$ & 1430 & 1443 & 1447 & 1440 & 1427 & 1434 \\
 & $\Gamma_{K^{*}(1410)} \, [\text{MeV}]$ & 406 & 432 & 419 & 373 & 406 & 399 \\
 & $r$ & 0.57 & 0.75 & 0.58 & 0.45 & 0.54 & 0.57 \\
 & $\kappa$ & 0.73 & 0.81 & 0.73 & 0.70 & 0.75 & 0.73 \\
 & $\delta \, [\degrees]$ & $-16$ & $-18$ & $-12$ & $-14$ & $-19$ & $-14$ \\
 & $\gamma - 2 \beta_{s} \, [\degrees]$ & 41 & 45 & 44 & 52 & 53 & 41 \\
\hline
\hline
\end{tabular}
 }
\label{tab:altModelsDsKpipi2}
\end{table}

\clearpage
\section{Interpretation of the \CP coefficients}
\label{a:CPcoeff}
The statistical and systematic correlation matrices of the \CP coefficients obtained from the phase-space integrated fit are given in Table~\ref{tab:statCorr} and Table~\ref{tab:systCorr}.

\begin{figure}[h]
  \centering
  \includegraphics[width=0.6\textwidth, height=!]{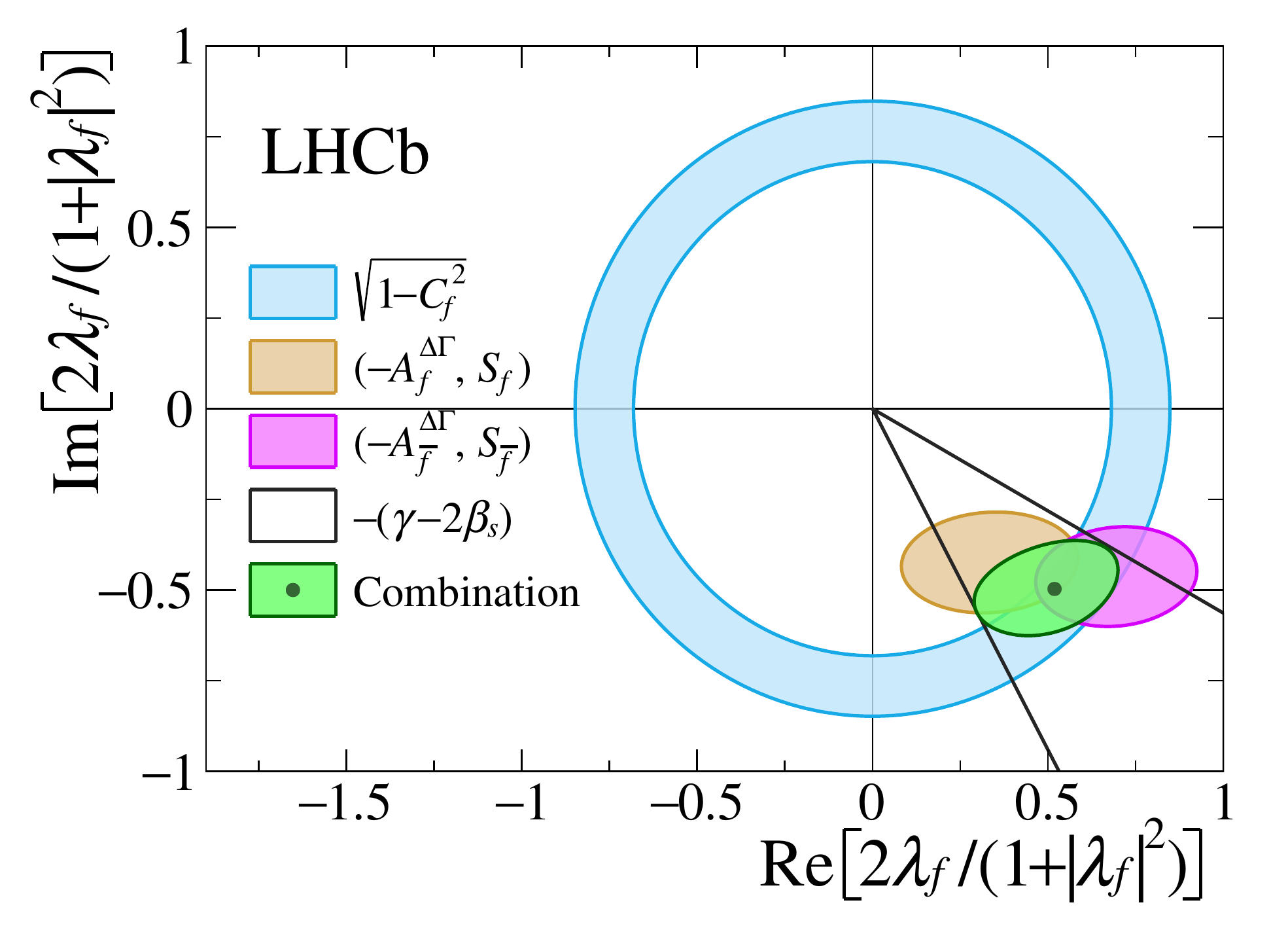}
  \caption{
Visualisation of how the \CP coefficients contribute towards the overall
constraint on the weak phase, $\gamma - 2\beta_s$.
The difference between the phase of
$(-A_f^{\Delta\Gamma},S_f)$
and
$(-A_{\bar{f}}^{\Delta\Gamma},S_{\bar{f}})$
is proportional to the strong phase $\delta$, which is close to $0\degrees$
and thus not indicated
in the figure.}
  \label{fig:dskpipi_markup}
\end{figure}

Figure~\ref{fig:dskpipi_markup} visualises the measured \CP coefficients
in the complex plane,
where
\mbox{$\lambda_f \equiv (q/p) \, (\int A^u(\phsPoint) \, \dphs / \int A^c(\phsPoint) \, \dphs)$.}
The points determined by
$(-A_f^{\Delta\Gamma},S_f)$
and
$(-A_{\bar{f}}^{\Delta\Gamma},S_{\bar{f}})$
are proportional to
$r \, \kappa \, e^{\pm \delta - (\gamma-2\beta_s)}$,
whilst an additional constraint on $r$ arises from $C_f$.

\setcounter{table}{0}
\setcounter{equation}{0}

\renewcommand{\thetable}{C.\arabic{table}}
\renewcommand{\theequation}{C.\arabic{equation}}

\begin{table}[h]
\centering
\caption{Statistical correlation of the \CP coefficients.}
	\renewcommand{\arraystretch}{1.1}
\begin{tabular}{l | r r r r r }
\hline
\hline
 & \multicolumn{1}{c}{$C_f$} & \multicolumn{1}{c}{$A^{\Delta\Gamma}_f$} & \multicolumn{1}{c}{$A^{\Gamma\Delta}_{\bar f}$} & \multicolumn{1}{c}{$S_f$} & \multicolumn{1}{c}{$S_{\bar f}$}  \\
\hline
$C_f$ &  1.000 & 0.135 & 0.168 & 0.040 & 0.042 \\
$A^{\Delta\Gamma}_f$ &  & 1.000 & 0.504 & $-0.082$ &  $-0.052$ \\
$A^{\Delta\Gamma}_{\bar{f}}$ &  & &  1.000 &  $-0.036$ & $-0.113$ \\
$S_f$ &  &&& 1.000 & 0.014 \\
$S_{\bar{f}}$ & &&&& 1.000 \\
\hline
\hline
\end{tabular}
\label{tab:statCorr}

\caption{Systematic correlation of the \CP coefficients.}
	\renewcommand{\arraystretch}{1.1}
\begin{tabular}{l | r r r r r }
\hline
\hline
 & \multicolumn{1}{c}{$C_f$} & \multicolumn{1}{c}{$A^{\Delta\Gamma}_f$} & \multicolumn{1}{c}{$A^{\Gamma\Delta}_{\bar f}$} & \multicolumn{1}{c}{$S_f$} & \multicolumn{1}{c}{$S_{\bar f}$}  \\
\hline
$C_f$ &  1.000  &  0.039  &  0.042  &   $-0.136$   &    $-0.006$ \\
$A^{\Delta\Gamma}_f$ &  & 1.000 &  0.208    & $-0.031$  &   $-0.023$ \\
$A^{\Delta\Gamma}_{\bar{f}}$ &  & &  1.000 &  $-0.021$  &   $-0.035$  \\
$S_f$ &  &&& 1.000 &  $-0.278$ \\
$S_{\bar{f}}$ & &&&& 1.000 \\
\hline
\hline
\end{tabular}
\label{tab:systCorr}
\end{table}

\clearpage

\addcontentsline{toc}{section}{References}
\setboolean{inbibliography}{true}
\bibliographystyle{LHCb}
\ifx\mcitethebibliography\mciteundefinedmacro
\PackageError{LHCb.bst}{mciteplus.sty has not been loaded}
{This bibstyle requires the use of the mciteplus package.}\fi
\providecommand{\href}[2]{#2}

\newpage
\centerline
{\large\bf LHCb collaboration}
\begin
{flushleft}
\small
R.~Aaij$^{31}$,
C.~Abell{\'a}n~Beteta$^{49}$,
T.~Ackernley$^{59}$,
B.~Adeva$^{45}$,
M.~Adinolfi$^{53}$,
H.~Afsharnia$^{9}$,
C.A.~Aidala$^{84}$,
S.~Aiola$^{25}$,
Z.~Ajaltouni$^{9}$,
S.~Akar$^{64}$
% ...

{\footnotesize \it

% ...
$ ^{9}$Universit{\'e} Clermont Auvergne, CNRS/IN2P3, LPC, Clermont-Ferrand, France\\
% ...
$ ^{25}$INFN Sezione di Milano, Milano, Italy\\
% ...
$ ^{31}$Nikhef National Institute for Subatomic Physics, Amsterdam, Netherlands\\
% ...
$ ^{45}$Instituto Galego de F{\'\i}sica de Altas Enerx{\'\i}as (IGFAE), Universidade de Santiago de Compostela, Santiago de Compostela, Spain\\
% ...
$ ^{49}$Physik-Institut, Universit{\"a}t Z{\"u}rich, Z{\"u}rich, Switzerland\\
% ...
$ ^{53}$H.H. Wills Physics Laboratory, University of Bristol, Bristol, United Kingdom\\
% ...
$ ^{59}$Oliver Lodge Laboratory, University of Liverpool, Liverpool, United Kingdom\\
% ...
$ ^{64}$University of Cincinnati, Cincinnati, OH, United States\\
% ...
$ ^{67}$Syracuse University, Syracuse, NY, United States\\
% ...
$ ^{84}$University of Michigan, Ann Arbor, United States, associated to $^{67}$\\
\bigskip
% ...
\medskip
}
\end{flushleft} 

\end{document}